\documentclass[sn-mathphys,Numbered]{sn-jnl}% Math and Physical Sciences Reference Style
\usepackage{graphicx}%
\usepackage{multirow}%
\usepackage{amsmath,amssymb,amsfonts}%
\usepackage{amsthm}%
\usepackage{mathrsfs}%
\usepackage[title]{appendix}%
\usepackage{xcolor}%
\usepackage{textcomp}%
\usepackage{manyfoot}%
\usepackage{booktabs}%
\usepackage{algorithm}%
\usepackage{algorithmicx}%
\usepackage{algpseudocode}%
\usepackage{listings}%
\usepackage{float}
\usepackage{subfig}
\usepackage{placeins} 
\usepackage{xcolor}
\usepackage{color}

\theoremstyle{thmstyleone}%
\theoremstyle{thmstyletwo}%

\theoremstyle{thmstylethree}%

\begin{document}

\title[Accuracy Assessment of Discontinuous Galerkin Spectral Element 
       Method in Simulating Supersonic Free Jets]
       {Accuracy Assessment of Discontinuous Galerkin Spectral Element 
       Method in Simulating Supersonic Free Jets}


\author*[1]{\fnm{Diego F.} \sur{Abreu}}\email{mecabreu@yahoo.com.br}

\author[2]{\fnm{João Luiz F.} \sur{Azevedo}}\email{joaoluiz.azevedo@gmail.com}

\author[3]{\fnm{Carlos} \sur{Junqueira-Junior}}\email{junior.junqueira@ensam.eu}

\affil*[1]{\orgdiv{Graduate Program in Space Sciences and Technologies}, 
           \orgname{Instituto Tecnológico de Aeronáutica, DCTA/ITA},
           \orgaddress{\street{Praça Marechal Eduardo Gomes, 50},
           \city{São José dos Campos},
           \postcode{12228--900},
           \state{SP},
           \country{Brazil}}}

\affil[2]{\orgdiv{Aerodynamics Division}, 
          \orgname{Instituto de Aeronáutica e Espaço, DCTA/IAE/ALA},
          \orgaddress{
          \city{São José dos Campos},
          \postcode{12228--904},
          \state{SP},
          \country{Brazil}}}

\affil[3]{\orgdiv{DynFluid Laboratory},
          \orgname{Arts et Métiers Institute of Technology, CNAM},
                   %HESAM University},
          \orgaddress{\street{151 Boulevard de l'Hôpital},
          \city{Paris},
          \postcode{75013},
          \state{Île-de-France},
          \country{France}}}


\abstract{The study performs large-eddy simulations of supersonic free jet
flows using the Discontinuous Galerkin Spectral Element Method (DGSEM). The
main objective of the present work is to assess the resolution requirements 
for adequate simulation of such flows with the DGSEM approach. The study 
looked at the influence of the mesh and the spatial discretization accuracy 
on the simulation results. The present analysis involves four simulations, 
incorporating three different numerical meshes and two different orders of 
spatial discretization accuracy. The numerical meshes are generated with 
distinct mesh topologies and refinement levels. Detailed descriptions of the 
grid generation and refinement procedures are presented. The study compares 
flow property profiles and power spectral densities of velocity components with
experimental data. The results show a consistent improvement in the computed 
data as the simulation resolution increases. This investigation revealed a 
trade-off between mesh and polynomial refinement, striking a balance between
computational cost and the accuracy of large-eddy simulation results for 
turbulent flow analyses.}




\maketitle

%-------------------------------------------------------------------------------
\section{Introduction}
%-------------------------------------------------------------------------------

Free jet flows exist in many industrial applications, including 
aerospace engineering, heat transfer, and metalworking. In aerospace
engineering, for example, the jet flows are the main responsible for generating
thrust from small jet aircraft to large launch vehicles. In the operation of 
jet flows, the interaction of the high-velocity flow of the jet with the 
ambient air generates a highly turbulent flow, which induces high levels of
pressure fluctuations, responsible for the production of noise and structural
loads. The development of the next generation of vehicles and engines requires
high-fidelity information on the jet flows in the early stages of the design 
process to achieve the performance requirements and allow the right design 
of the structures in the surroundings of the flow. The assessment of the 
characteristics of the jet flows can be obtained from a test bench in physical 
experiments \cite{Woodmansee2004, BridgesWernet2008, MorrisZaman2010} and 
employing numerical simulations \cite{BogeyBailly2010, Debonis2017, Bres2017, 
Junior2018, Abreu2022}.

The choice of the approach to obtain a particular characteristic depends on 
the analysis to be performed. Mean properties, such as thrust and mean 
velocity field, can be obtained with Reynolds-Averaged Navier-Stokes (RANS) 
simulations\cite{Zhangetal2015}, while determining velocity and pressure
fluctuations, to estimate the noise produced by the jet at takeoff, demand 
more sophisticated numerical formulations or the realization of physical
experiments. The present work is aligned with the efforts of academia to
produce high-fidelity numerical methodologies for the simulation of 
supersonic jet flows. The large-eddy simulations (LES), which are based on
the scale separation of the flow, is the methodology chosen due to its
capacity to provide high-frequency flow field information with lower costs 
when compared to the direct numerical simulation (DNS)\cite{Pope2000} or the
realization of physical experiments. The use of large-eddy simulations to 
solve jet flows started some decades ago\cite{BodonyLele2008,
BogeyMarsdenBailly2011}. The limited availability of computational resources
restricted the complexity of the geometries and the Reynolds number that could
be resolved. The majority of the simulations were resolving subsonic regimes
with the nozzle absent from the computational domain or presented with a 
simplified geometry. The increase in computational resources allowed the
development of simulations with higher Reynolds number flows and more complex 
geometries \cite{Yang2020, PradhanGhosh2023, Deng2022, Noah2021}. 

The influence of nozzle-exit conditions was analyzed with the imposition of
different boundary layer profiles\cite{BogeyBailly2010}, with a tripped
boundary condition\cite{BogeyMarsdenBailly2011} and with a highly disturbed
boundary layer\cite{BogeyMarsden2016}. The simulations employed a high-order
finite difference method to solve the subsonic jet flows with the nozzle 
represented by a constant radius pipe. The flow field and noise generated by
subsonic jet flows were investigated using a compact high-order finite
difference method on unstructured meshes\cite{faranosovetal2013}. Subsonic 
heated jet flows were simulated through a high-order framework with a central 
difference scheme on structured meshes\cite{Debonis2017} to investigate 
turbulent temperature fluctuations and heat flux. Multiple strategies to 
develop the flow inside the nozzle to improve the nozzle outlet condition for 
a subsonic jet flow were investigated using a finite volume framework on
unstructured meshes\cite{Bresetal2018}. High-order frameworks using 
discontinuous Galerkin have been used to simulate subsonic jet 
flows\cite{Lorteauetal2018, Lindbladetal2023} in order to perform acoustic 
analyses. Diverse options of numerical methods have proven capable of
simulating subsonic jet flows with accurate flow fields and noise predictions.

The employment of large-eddy simulations for analyzing supersonic jets 
is less diverse than the subsonic applications. The noise generated by
supersonic jet flows has been simulated with finite volume methods on
unstructured grids\cite{Mendezetal2012, Bres2017, Langenaisetal2019}. The
acoustic characteristics of supersonic jet flows have also been simulated with
finite difference methods\cite{Bogey2021}. The finite difference method has
also been examined to identify the effect of the choice of subgrid-scale (SGS)
models\cite{Junior2020} on the results of the LES of supersonic jet flow. 
More recently, simulations using the discontinuous Galerkin schemes are being
employed to predict the noise generated by supersonic jet 
flows\cite{ShenMiller2020, ChauhanMassa2022}. 

\begin{figure*}[htb!]
    \subfloat[Velocity contours.]{
	\includegraphics[trim = 0mm 0mm 100mm 0mm, clip, width=0.48\linewidth]
    {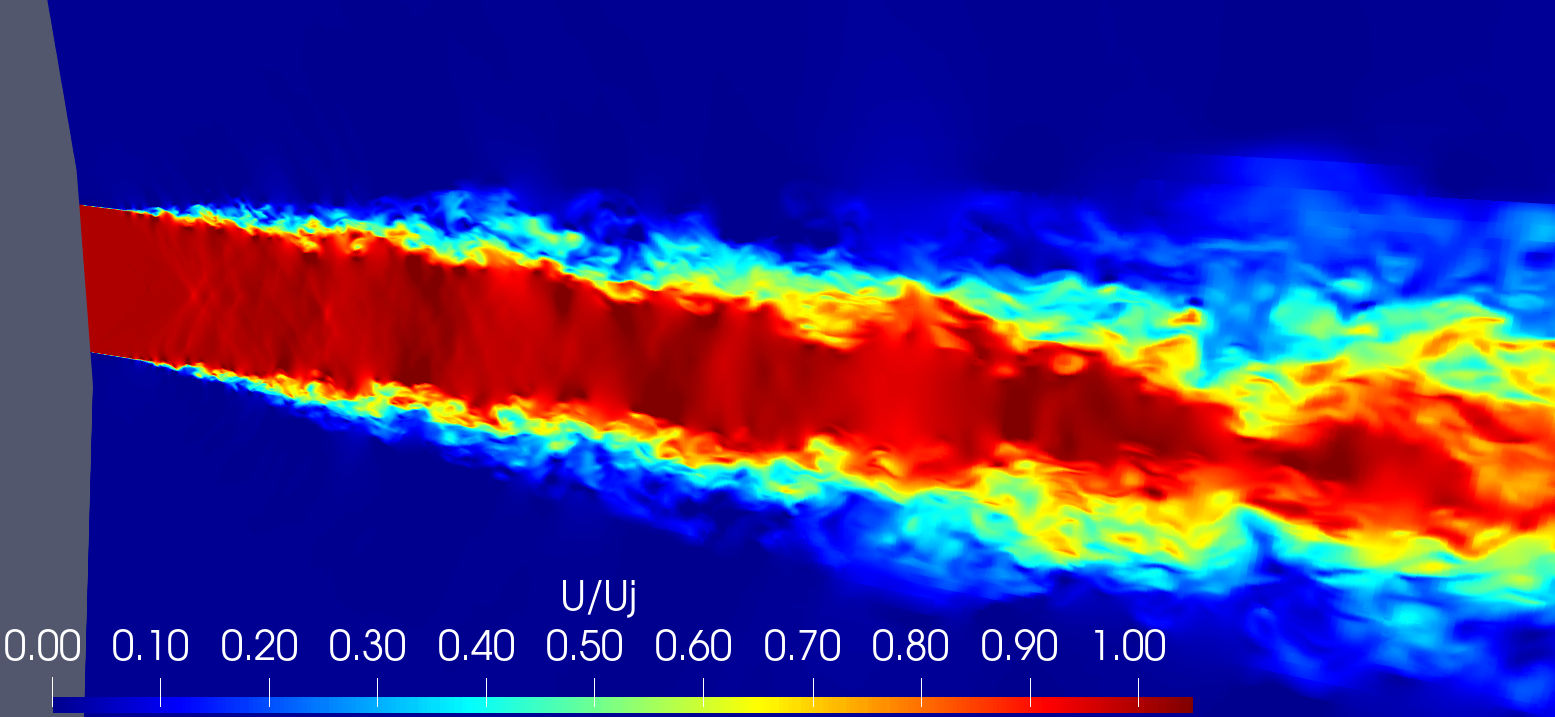}
	}
\subfloat[Pressure contours with isosurfaces of Q-criterion colored by
          vorticity magnitude.]{
	\includegraphics[trim = 0mm 0mm 100mm 0mm, clip, width=0.48\linewidth]
    {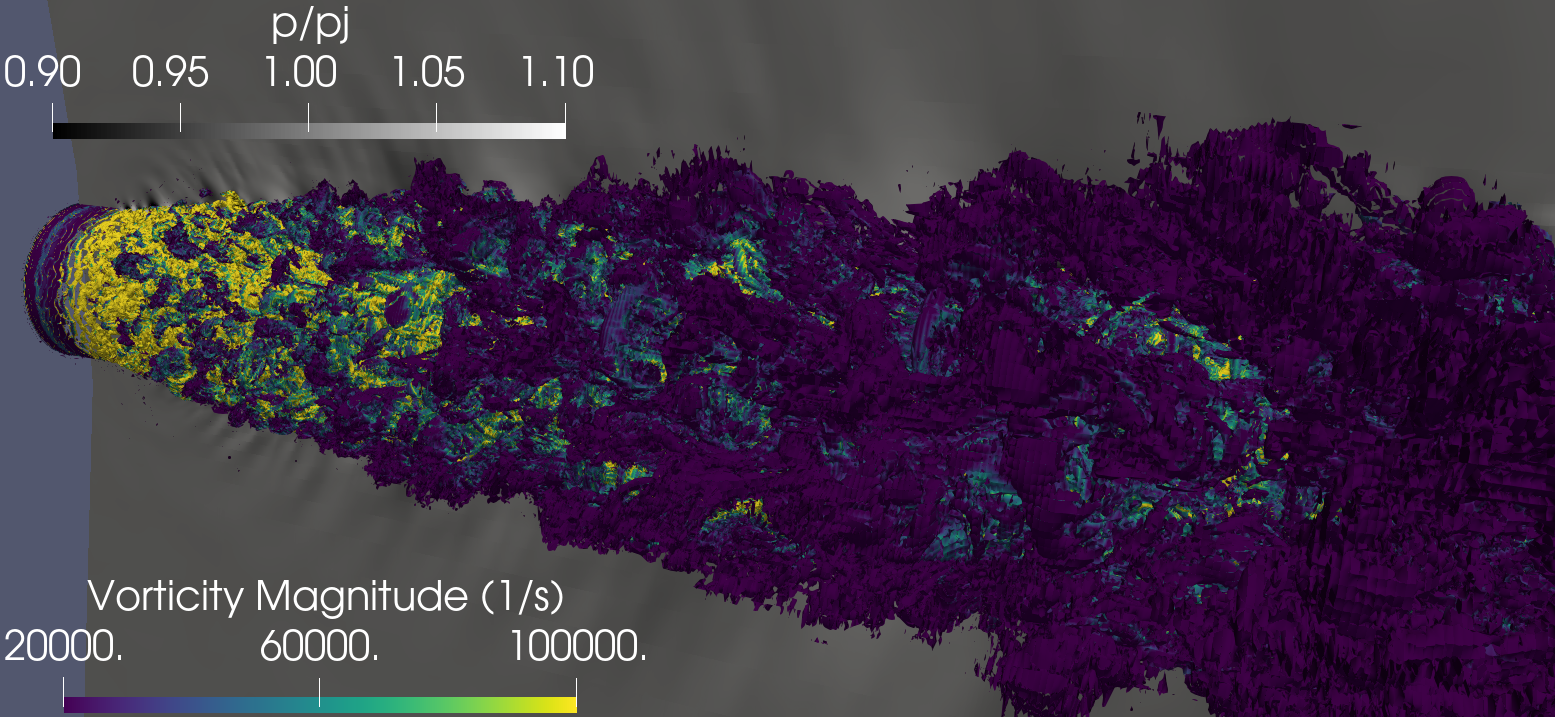}
	}
\caption{{Visualization of the jet flow of interest in the
         present work.}}
\label{fig:jet_presentation}
\end{figure*}

In the present work, a numerical investigation is performed to assess the 
accuracy of the discontinuous Galerkin spectral element method 
(DGSEM)\cite{Kopriva2010, Hindenlang2012} in the simulation of supersonic jet
flows. {For visual representation purposes, 
Fig.~\ref{fig:jet_presentation} presents 2-D cuts of the supersonic flow
assessed in the article, depicting velocity contours and Q-criterion 
isosurfaces colored by vorticity magnitude over pressure contours in grayscale 
in the background}. The 
presence of shock waves and the possible need to model the jet
nozzle introduces additional challenges to the simulation of a supersonic jet
flows compared to subsonic cases. The DGSEM scheme is a high-order framework 
on unstructured meshes with enhanced computational efficiency obtained by
collocating the interpolation and integration points and using a tensor 
product nodal basis functions inside the hexahedron elements. In the present
effort, the version of the DGSEM approach, as implemented in the open-source
FLEXI numerical framework\cite{Kraisetal2021}, is used. The numerical 
investigation comprises three numerical meshes with two topologies and two
polynomial degrees for the spatial discretization in order to identify the 
mesh and polynomial \textit{hp} refinement requirements to obtain 
high-quality results with the numerical scheme.

This work is organized with the governing equations of the LES simulations and
the numerical formulation of the discontinuous Galerkin spectral element 
method outlined in Sec.~\ref{sec.gov_eq}. The jet flow problem is presented in
Sec.~\ref{sec.jetflow} with the definition of the physical domain and the 
mesh generation strategy. Information on the simulation settings are provided
in Sec.~\ref{sec.numerics}. The results of the simulations are presented in
Sec.~\ref{sec.results}, and the concluding remarks close the work in 
Sec.~\ref{sec.conclusions}.

%-------------------------------------------------------------------------------
\section{Numerical Formulation}
%-------------------------------------------------------------------------------
\label{sec.gov_eq}

\subsection{Governing Equations}

The large-eddy simulation is a modeling strategy built on the scale separation 
of turbulent motions, performed by a spatial filtering 
process\cite{Garnieretal2009}. The definition of the numerical mesh is one 
form of spatial filter, which can be referred to as an implicit filter. The
simulation resolves the turbulent eddies that are compatible with the mesh
size. In contrast, the turbulent eddies that are shorter than the mesh size 
cannot be simulated, and therefore, they must be modeled. The modeling of 
the small eddies is performed by a subgrid-scale model (SGS).

The filtered Navier-Stokes equations, in conservative form, are given by
\begin{equation}
\frac{\partial \mathbf{Q} }{\partial t} + \nabla \cdot \mathbf{F} =0,
\label{eq.filterNS}
\end{equation}
where $\mathbf{Q}$ is the vector of filtered conserved variables, and
$\mathbf{F}$ is the flux vector. The flux vector is given by
$\mathbf{F}=\mathbf{F}^e-\mathbf{F}^v$, where the superscripts $e$ and $v$
denotes the advective, or the Euler, and the viscous fluxes. The vector of
conserved variables and the vectors of the numerical fluxes are given by
\begin{equation}
\mathbf{Q}= \left[ \begin{array}{c} 
                    \bar{\rho} \\ \bar{\rho}\tilde{u}_1 \\ \bar{\rho}\tilde{u}_2 
                    \\ \bar{\rho} \tilde{u}_3 \\ \bar{\rho}\check{E}
                   \end{array} \right] \hspace*{1 cm} 
\mathbf{F}_i^e= \left[ \begin{array}{c} 
                    \bar{\rho} \tilde{u}_i \\ \bar{\rho} \tilde{u}_1 \tilde{u}_i
                    + \delta_{1i} \bar{p}
                    \\ \bar{\rho} \tilde{u}_2 \tilde{u}_i + \delta_{2i}\bar{p} 
                    \\ \bar{\rho} \tilde{u}_3 \tilde{u}_i + \delta_{3i}\bar{p} 
                    \\ (\bar{\rho} \check{E} + \bar{p}) \tilde{u}_i
                   \end{array} \right]  \hspace*{1 cm} 
\mathbf{F}_i^v= \left[ \begin{array}{c} 
                    0 \\ \bar{\tau}_{1i} \\ \bar{\tau}_{2i} \\ \bar{\tau}_{3i} 
                    \\ \tilde{u}_j \bar{\tau}_{ij} - \bar{q}_i
                   \end{array} \right], \hspace*{0.1 cm} \mbox{for } i = 1,2,3. 
\label{eq:num_fluxes}
\end{equation}
The Favre averaged velocity components are represented by the velocity vector
$\tilde{\boldsymbol{u}}$ or $(\tilde{u}_1, \tilde{u}_2, \tilde{u}_3)$,
$\bar{\rho}$ is the filtered density, $\bar{p}$ is the filtered pressure, and
$\bar{\rho} \check{E}$ is the filtered total energy per unit volume, which is
defined according to the definition proposed by Vreman in its ``system 
I''\cite{Vremanetal1995},
\begin{equation}
\bar{\rho} \check{E} = \frac{\bar{p}}{\gamma - 1} + \frac{1}{2}\bar{\rho}
\tilde{u}_i\tilde{u}_i.
\label{eq:rhoe}
\end{equation}
The Kronecker delta is represented by $\delta$. The filtered pressure, Favre
averaged temperature and filtered density are correlated using the ideal gas
equation of state $\bar{p}= \bar{\rho} R \tilde{T}$, where $R$ is the specific
gas constant with a value of $287.058$~J/(kg$\cdot$K).

The viscous fluxes present the filtered stress tensor $\bar{\tau}$ and the heat
flux vector $\bar{q}$, which are calculated by
\begin{equation}
\bar{\tau}_{ij}=(\mu + \mu_{SGS}) \left(\frac{\partial \tilde{u}_i}
{\partial x_j} + \frac{\partial \tilde{u}_j}{\partial x_i} \right) 
 - \frac{2}{3} (\mu + \mu_{SGS}) \left(\frac{\partial \tilde{u}_k}
 {\partial x_k} \right) \delta_{ij},
\label{eq:bartau}
\end{equation}
\begin{equation}
\bar{q}_i=-(k+k_{SGS})\frac{\partial \tilde{T}}{\partial x_i}.
\label{eq:barq}
\end{equation}
Sutherland's law, given by 
\begin{equation}
\mu(\tilde{T}) \approx \mu_0 \left( \frac{\tilde{T}}{T_0} \right)^{3/2} 
                        \frac{T_0+s}{\tilde{T}+s},
\label{eq:sutherland}                        
\end{equation}
is employed to estimate the dynamic viscosity coefficient $\mu$. The reference
values are $\mu_0=1.716 \times 10^{-5}$ N$\cdot$s/m$^2$ and $T_0=273$ K, and 
$s=111$ K is the Sutherland constant\cite{White2006}. The thermal conductivity
coefficient of the fluid, $k$, is calculated by
\begin{equation}
k=\frac{\mu c_p}{Pr},
\label{eq:kq}
\end{equation}
where the specific heat at constant pressure is calculated by 
$c_p=R\gamma/(\gamma-1)$. The Prandtl number $Pr$ is equal to $0.72$ and the
heat capacity $\gamma$ is equal to $1.4$. The terms $\mu_{SGS}$ and $k_{SGS}$
are the subgrid-scale dynamic viscosity and thermal conductivity coefficients.

In the present work, the subgrid-scale contribution is provided by the
Smagorinsky model\cite{Smagorinsky1963}, in which the SGS viscosity coefficient
is calculated according to
\begin{equation}
\mu_{SGS}=\bar{\rho}C_s^2\Delta^2 | \tilde{S} |.
\label{eq:musgs}
\end{equation}
The calculation of the SGS viscosity coefficient requires the estimate of the 
filter size $\Delta$, the magnitude of the filtered strain rate $| \tilde{S} |$, 
which is calculated by
\begin{equation}
| \tilde{S} | = (2 \tilde{S}_{ij} \tilde{S}_{ij})^{0.5},
\label{eq:magtildeS}
\end{equation}
\begin{equation}
\tilde{S}_{ij}=\frac{1}{2}\left( \frac{\partial \tilde{u}_i}
{\partial x_j} 
+ \frac{\partial \tilde{u}_j}{\partial x_i} \right),
\label{eq:tildeSij}
\end{equation}
and the Smagorinsky model constant $C_s$, with a value of 
$0.148$\cite{Garnieretal2009}. The filter size $\Delta$ is obtained by the
cubic root of the element volume. The SGS thermal conductivity coefficient is
calculated by
\begin{equation}
k_{SGS}=\frac{\mu_{SGS} c_p}{Pr_{SGS}}
\label{eq:ksgs}
\end{equation}
where $Pr_{SGS}$ is the SGS Prandtl number with a value of $0.9$.

\subsection{Numerical Methods}
\label{sec.dgsem}

The filtered Navier-Stokes equations, Eq.~(\ref{eq.filterNS}), are solved with
a nodal discontinuous Galerkin (DG) scheme. The specific form of the DG scheme
used in the present work is termed the discontinuous Galerkin spectral element
method (DGSEM). It follows the description and implementation in Refs. 
\cite{Kopriva2010,Hindenlang2012}. The scheme was applied and validated with
numerous problems\cite{GassnerBeck2013, Gassner2014, Becketal2014, 
FladGassner2017}. The DGSEM used is implemented in the FLEXI numerical 
framework\cite{Kraisetal2021}. The numerical scheme is employed on a discretized
physical domain of non-overlapping hexahedral elements. The elements from the
physical domain are mapped onto a reference unit cube element 
$\Omega=[-1,1]^3$. The choice of hexahedral elements is due to computational
efficiency and the simplified implementation. The filtered Navier-Stokes
equations, presented in Eq.~(\ref{eq.filterNS}), need to be mapped to the
reference domain leading to 
\begin{equation}
J \frac{\partial \mathbf{Q}}{\partial t} + \nabla_{\xi} \cdot 
\mathcal{F} = 0.
\label{eq:mapfilterNS}
\end{equation}
The symbol $\nabla_{\xi}$ represent the divergence operator to the reference
element coordinates, $\boldsymbol{\xi}=(\xi,\eta,\zeta)^T$, $J= \arrowvert
\partial \boldsymbol{x} / \partial \boldsymbol{\xi} \arrowvert$ is the Jacobian
of the coordinate transformation, and $\mathcal{F}$ is the contravariant flux
vector which takes into account inviscid and viscous terms.

The solution in each element is approximated by a nodal polynomial interpolation
of the form presented in 
\begin{equation}
\mathbf{Q}(\xi) \approx \sum_{p,q,r=0}^N \mathbf{Q}_h
(\xi_p,\eta_q,\zeta_r,t)\phi_{pqr}(\boldsymbol{\xi}).
\label{eq:num_sol}
\end{equation}
At each interpolation node in the reference element, the vector of conserved
variables is represented by $\mathbf{Q}_h(\xi_p,\eta_q,\zeta_r,t)$ and
$\phi_{pqr} (\boldsymbol{\xi})$ is the interpolating polynomial. For hexahedral
elements, the interpolating polynomial is a tensor product of one-dimensional
Lagrange polynomials $\mathscr{L}$ in each space direction, given by
\begin{equation}
\phi_{pqr}(\boldsymbol{\xi})
=\mathscr{L}_p(\xi)\mathscr{L}_q(\eta)\mathscr{L}_r(\zeta) 
\hspace{10pt} , \hspace{10pt} \mathscr{L}_p(\xi)= \prod_{\substack{i=0 \\ i \ne p}}^{N_p} 
\frac{\xi-\xi_i}{\xi_p-\xi_i}.
\label{eq:interp_poly}
\end{equation}
The definition presented for $\mathscr{L}_p(\xi)$ is equivalent for the other 
two directions. In a similar process, the components of the contravariant 
fluxes $\mathcal{F}$ are approximated by interpolation at the same nodal 
points of the solution; the example in the $\xi$-component is presented in 
\begin{equation}
\mathcal{F} \left( \mathbf{Q}(\boldsymbol{\xi}) \right) \approx 
\sum_{p,q,r=0}^N \boldsymbol{\mathcal{F}}_{\xi,h}
(\xi_p,\eta_q,\zeta_r,t)\phi_{pqr}(\boldsymbol{\xi}).
\label{eq:cont_flux}
\end{equation}

The discontinuous Galerkin formulation is obtained by multiplying 
Eq.~(\ref{eq:mapfilterNS}) by the test function $\psi=\psi(\xi)$ and 
integrating over the reference element $\Omega$. The strong formulation of 
the discontinuous Galerkin scheme is obtained when the second term in 
Eq.~(\ref{eq:mapfilterNS}) is integrated by parts two consecutive times,
resulting in 
\begin{equation}
\int_\Omega J \frac{\partial \mathbf{Q}}{\partial t} \psi d 
\boldsymbol{\xi} + \int_{\partial \Omega} \left( \nabla_\xi \cdot 
\boldsymbol{\mathcal{F}}^e \right) \psi d \boldsymbol{\xi} 
+ \int_{\partial \Omega} \left( \boldsymbol{\mathcal{F}}^{e,\star}
- \boldsymbol{\mathcal{F}}^e \right) \psi d \boldsymbol{S}_\xi
- \int_\Omega \left( \nabla_\xi \cdot \boldsymbol{\mathcal{F}}^v
\right) \psi d \boldsymbol{\xi} = 0.
\label{eq:DGstrongmapfilterNS}
\end{equation}
The boundaries of the element $\Omega$ are defined by $\partial \Omega$, and
$\boldsymbol{\mathcal{F}^{e,\star}}$ is the numerical interface flux. The 
second and third terms in Eq.~(\ref{eq:DGstrongmapfilterNS}) represents the
inviscid part of the flux, $\hat{\boldsymbol{\mathcal{F}}}^e$. One alternative
to obtain the discrete form of the inviscid part of the flux is by substituting
the integrals by Gaussian quadratures. For simplicity, only the $\xi$-component
is presented in
\begin{align}
 \left( \boldsymbol{\hat{\mathcal{F}}}^{e}_{\xi,h} \right)_{ijk}
 = \left[ \boldsymbol{\mathcal{F}}^{e,\star}_{\xi,h} (1,\eta_j,\zeta_k;
 \boldsymbol{n}) 
 - \boldsymbol{\mathcal{F}}^{e}_{\xi,h} (1,\eta_j,\zeta_k) \right]
 \nonumber
 \\
 - \left[ \boldsymbol{\mathcal{F}}^{e,\star}_{\xi,h} (-1,\eta_j,\zeta_k;
 \boldsymbol{n}) 
 - \boldsymbol{\mathcal{F}}^{e}_{\xi,h} (-1,\eta_j,\zeta_k) \right]
 \nonumber
 \\
 + \sum_{m=0}^N\mathcal{D}_{im} \boldsymbol{\mathcal{F}}^{e}_{\xi,h}
 (\xi_m,\eta_j,\zeta_k).
 \label{eq:disc_inv_cont_flux}
\end{align}
The subscript $h$ indicates a discrete approach, $\boldsymbol{n}$ is the 
surface normal directed to the exterior of the element, and
$\mathcal{D}_{im}=\partial \mathscr{L}_j/\partial \xi |_{\xi_i}$ is the
derivative matrix. In Eq.~(\ref{eq:disc_inv_cont_flux}), the first two terms
are associated with the surface integrals, and the third term is the discrete
form of the volume integral.

One strategy to enhance the stability of the method is by employing a split
formulation to the inviscid fluxes $\boldsymbol{\mathcal{F}}^{e}$. The
approach is based on the work developed by Pirozzoli\cite{Pirozzoli2011}, and
it was adapted for the DGSEM formulation\cite{Gassner2016}. It was shown that
the split form is capable of providing improvements to the method, especially 
for high-order interpolations. The formulation relies on summation-by-parts
(SBP) property, which is satisfied when the Legendre-Gauss-Lobatto (LGL) 
quadrature points are used as the interpolation nodes of the approximate 
solution. The volumetric term of the discrete flux is approximated by
\begin{equation}
 \sum_{m=0}^N\mathcal{D}_{im} \boldsymbol{\mathcal{F}}^{e}_{\xi,h}
 (\xi_m,\eta_j,\zeta_k) \approx 2 \sum_{m=0}^N\mathcal{D}_{im} 
 \boldsymbol{\mathcal{F}}^{\#}_{\xi} 
 (\mathbf{Q}_{ijk},\mathbf{Q}_{mjk}),
 \label{eq:vol_split}
\end{equation}
where $\boldsymbol{\mathcal{F}}^{\#} _\xi(\mathbf{Q}_{ijk},\mathbf{Q}_{mjk})$
is the split form of the flux. The construction of the split form of the flux
uses quadratic or cubic splitting of the terms in Eq.\@ (\ref{eq:num_fluxes}). 
In this work, the split form used is presented in
\begin{equation}
 \mathbf{F}_{i,PI}^{\#}= \left[ \begin{array}{c} 
     \{\{\bar{\rho}\}\} \{\{\tilde{u}_i\}\} \\ 
     \{\{\bar{\rho}\}\} \{\{\tilde{u_1}\}\} \{\{\tilde{u}_i\}\} 
     + \delta_{1i} \{\{\bar{p}\}\} \\
     \{\{\bar{\rho}\}\} \{\{\tilde{u_2}\}\} \{\{\tilde{u}_i\}\} 
     + \delta_{2i} \{\{\bar{p}\}\} \\
     \{\{\bar{\rho}\}\} \{\{\tilde{u_3}\}\} \{\{\tilde{u}_i\}\} 
     + \delta_{3i}\{\{\bar{p}\}\} \\
     \{\{(\bar{\rho}\}\} \{\{H\}\}  \{\{\tilde{u}_i\}\}
                   \end{array} \right] \hspace{0.5 cm} i = 1,2,3.
\label{eq:splitform}
\end{equation}
The terms in $\{\{ \cdot \}\}$ represents the arithmetic average between the
quantities from $\mathbf{Q}_{ijk}$ and $\mathbf{Q}_{mjk}$. In Eq.\@
(\ref{eq:splitform}), the total enthalpy $H = \check{E}+\bar{p}/\bar{\rho}$ is 
used in the energy equation. The split form also needs to be employed in 
the numerical interface flux $\boldsymbol{\mathcal{F}}^{e,\star}$, which is
calculated by 
\begin{equation}
 \boldsymbol{\mathcal{F}}^{e,\star} = 
 \boldsymbol{\mathcal{F}}^{\#}_{\xi} 
 (\mathbf{Q}^+,\mathbf{Q}^-) -\epsilon_{stab}
 (\mathbf{Q}^+,\mathbf{Q}^-),
\label{eq:surf_split} 
\end{equation}
where the superscripts ``+'' and ``-'' indicate from what side of the interface
the information comes, and $\epsilon_{stab}$ is a stabilization term that comes 
from the Riemann solver. It is used to introduce some level of numerical
dissipation to the scheme. The split form was evaluated for the Taylor-Green 
vortex problem\cite{Gassner2016}, to a channel flow\cite{DauricioAzevedo2021}
and to a 2-D profile\cite{DauricioAzevedo2023}. In the three tests the
stability of the simulations were improved.

The Riemann solver used in the simulations is a Roe scheme with entropy fix
\cite{Harten1983}. The lifting scheme of Bassi and Rebay, BR2
\cite{BassiRebay1997}, is employed for the spatial discretization of viscous
flux terms. The standard Runge-Kutta time marching method with three stages 
and third order of accuracy is employed to promote the temporal evolution 
of the simulations\cite{Kopriva2009}. The finite-volume sub-cell shock 
capturing method\cite{Sonntag2017} is employed to handle the shock waves 
present in supersonic flows. The indicator used is based on the switching
function of Jameson-Schmidt-Turkel\cite{JST1981} and adapted to 3-D 
applications as a shock indicator\cite{Sonntag2017}. The operation of the 
shock-capturing method changes the formulation of the cells from the DGSEM to
a second-order reconstruction finite volume (FV) scheme with a minmod slope
limiter\cite{Hirsch1990b} with each interior node of the DGSEM scheme changing
to one finite volume sub-cell. The cell returns from FV formulation to the 
DGSEM when the indicator reduces below a specified threshold.

{\subsection{Numerical Framework}}

{The numerical methods described in the previous section were 
implemented in the FLEXI numerical framework \cite{Kraisetal2021,flexi}. The solver
is a high-order open-source tool created at the University of Stuttgart for
solving the compressible Navier-Stokes equations on CPU clusters. It was
implemented with Fortran language and parallelized using the Message
Passing Interface (MPI) standard. The framework \cite{flexi} is composed of a 
preprocessor, which is capable of converting linear meshes into high-order 
meshes, the solver itself, and, also, tools designed to convert the output files into other
extensions. The computational efficiency and the parallel performance of the
numerical framework was assessed for different problems in Refs. 
\cite{Kraisetal2021, Hindenlang2012, Becketal2014}.} 
\\

%-------------------------------------------------------------------------------
\section{Simulation Domain and Grid Generation}
%-------------------------------------------------------------------------------
\label{sec.jetflow}

Large-eddy simulations are performed to study a supersonic round jet flow.
The operating condition of the jet flow is the isothermal, perfectly expanded
supersonic flow with a Mach number of $1.4$ and a Reynolds number based on 
nozzle exit diameter of $1.58 \times 10^6$. The perfectly expanded supersonic 
condition is adequate for assessing numerical methods due to the weak shock 
that develops in the jet potential core. In this operating condition, the jet
static pressure and temperature match the values from the freestream. The flow
conditions are settled based on the experiments of Bridges and 
Wernet\cite{BridgesWernet2008}. In the present section, the generation of the
computational domain and the numerical meshes are assessed.

\subsection{Physical Geometry}
\label{sec.phys_domain}

The physical domain investigated in the numerical simulations is the external
control volume in which the flow leaving the nozzle develops into the jet flow
when in contact with ambient air. The nozzle geometry is absent from the
physical domain. It is represented by the inlet section circular surface of
the jet flow, in which the nozzle-exit condition is imposed. An axisymmetric
divergent shape is connected to the jet inlet section to complete the
physical domain. The choice for modeling only the external domain of the jet
flow was made to exploit the potential of simulating the jet flow with reduced
computational costs.

\begin{figure}[htb!]
\centering
	\includegraphics[width=0.6\linewidth]{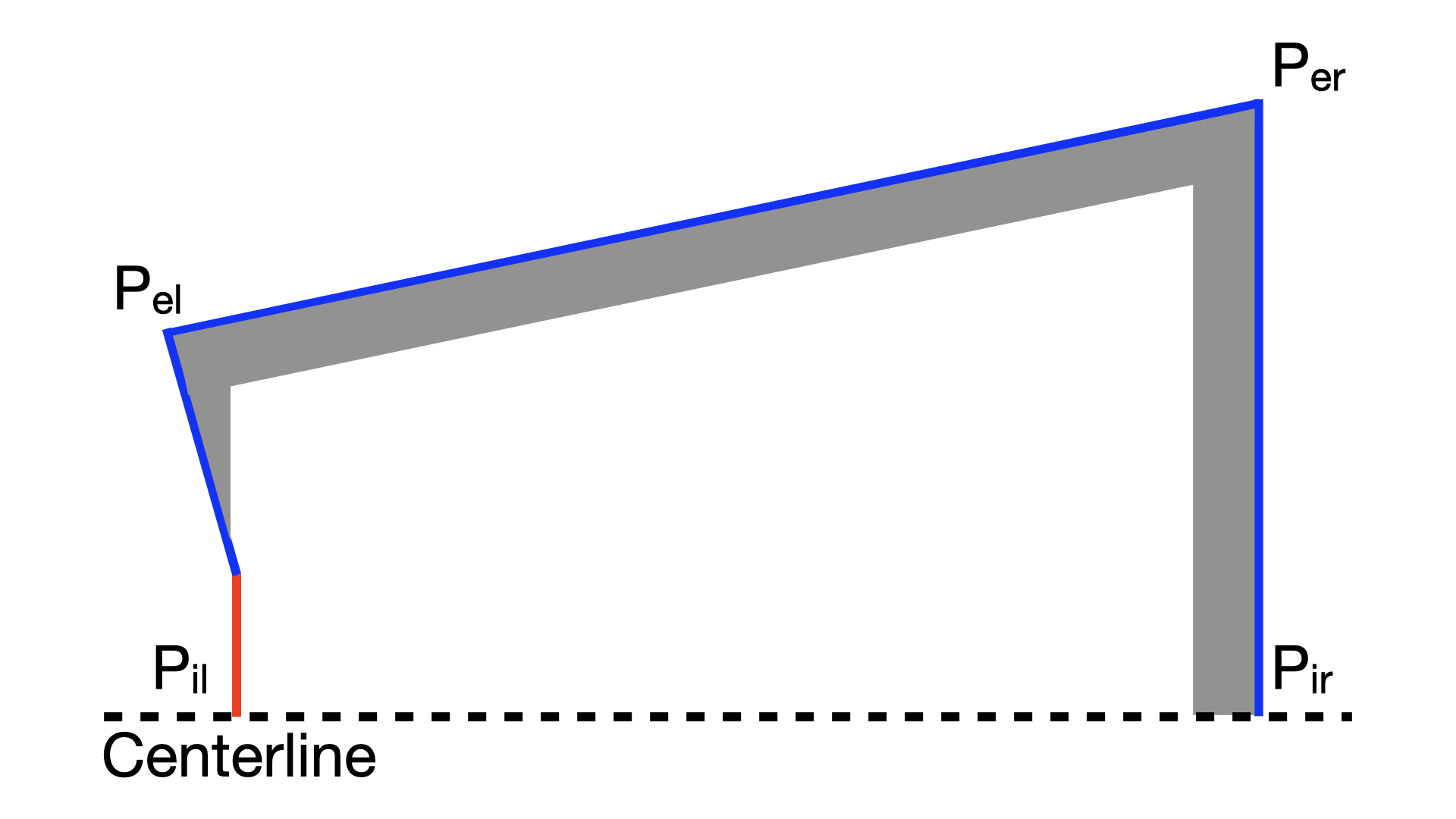}
    \caption{Schematic representation of an axisymmetric plane of the 
             physical domain used for the jet flow simulations.}
    \label{fig:geo_domain}
\end{figure}

The schematic representation of an axisymmetric plane of the physical domain
is presented in Fig.~\ref{fig:geo_domain} with the four reference points 
used to describe the domain's boundaries. The jet inlet diameter ($D_j$) 
is used as the reference length in the work. The computational domain has
a length of $40D_j$ in the centerline, from point $P_{il}$ to $P_{ir}$. Two
computational domains are generated to represent the physical domain. The first
computational domain, G1 geometry, has the coordinates $x/D_j=-0.95$ and
$r/D_j=5.65$ for $P_{el}$, and $x/D_j=36.6$ and $r/D_j=12.5$ for $P_{er}$. The
second computational domain, G2 geometry, has the coordinates $x/D_j=-0.95$ and
$r/D_j=8.0$ for $P_{el}$, and $x/D_j=40.0$ and $r/D_j=12.5$ for $P_{er}$.

\subsection{Numerical Meshes}

The FLEXI solver uses exclusively hexahedral grid elements for computational 
purposes\cite{Kraisetal2021}. Hence, the grid generation led to an integrated 
process to generate the computational domain and the numerical mesh. The 
computational domain comprises the three-dimensional version of the control
volume illustrated in Fig.~\ref{fig:geo_domain} with the dimensions provided 
in Sec. \ref{sec.phys_domain}. The geometries are generated with the 
open-source mesh generator Gmsh\cite{GeuzaineRemacle2009}.

\begin{figure*}[htb!]
\centering
\subfloat[Block topology for G1 geometry.]{
	\includegraphics[width=0.47\linewidth]
    {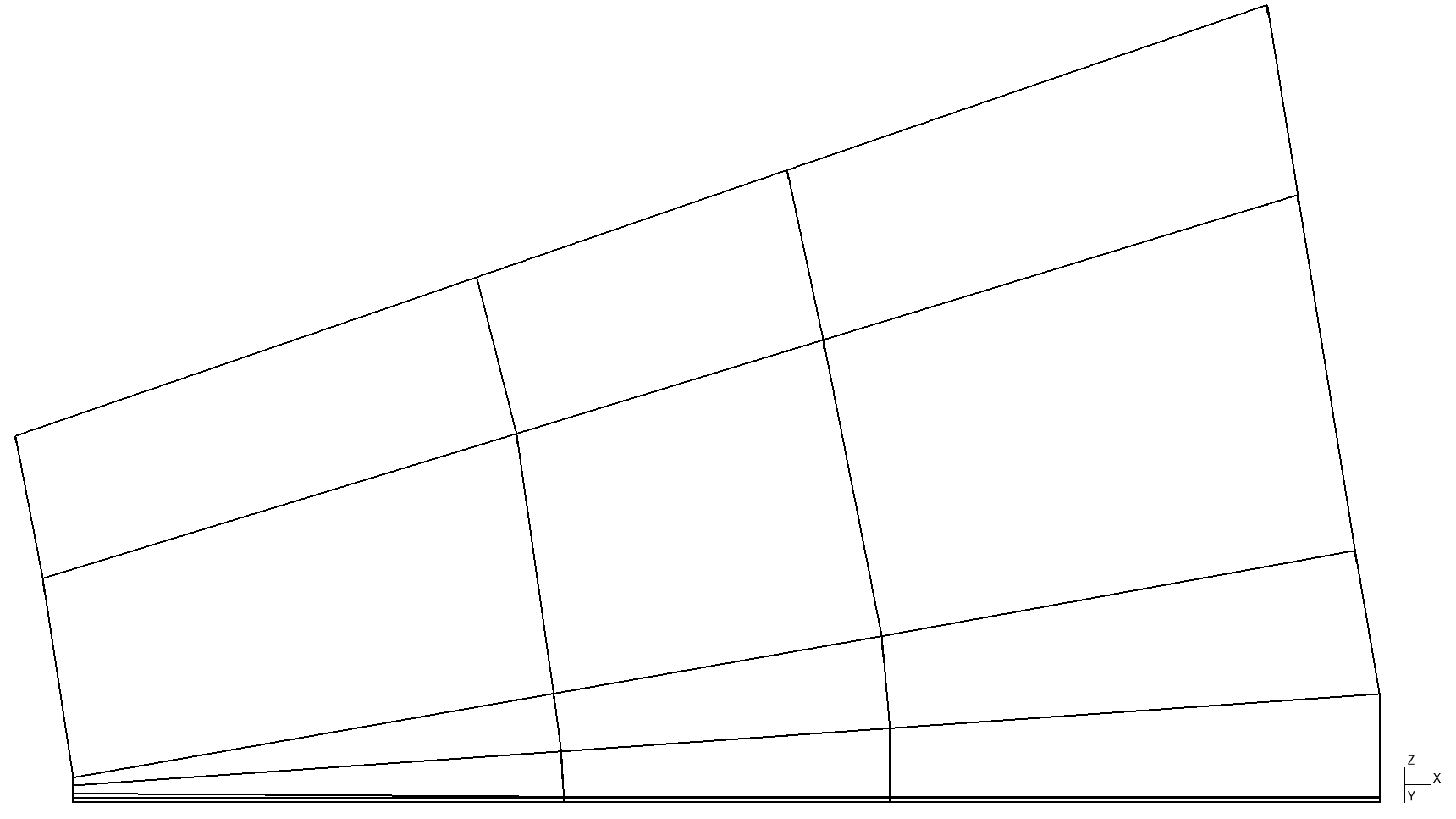}
    \label{fig:G1}
	}
\subfloat[Block topology for G2 geometry.]{
	\includegraphics[width=0.47\linewidth]
    {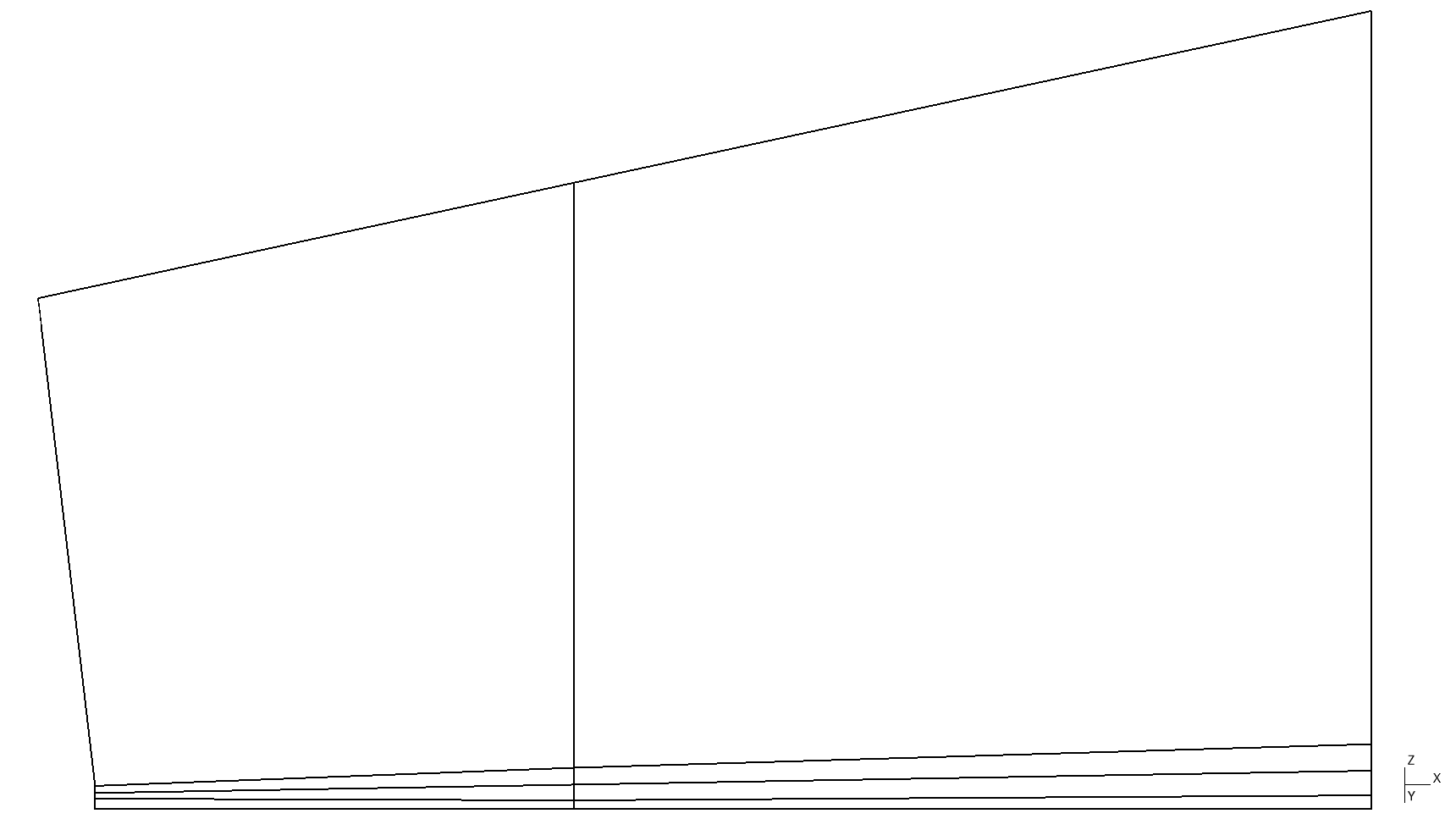}
    \label{fig:G2}
	}
\newline
\subfloat[Block topology in crossflow plane at ${x/D_j=0.0}$ for the G2
          geometry.]{
	\includegraphics[trim = 140mm 0mm 140mm 0mm, clip, width=0.25\linewidth]
    {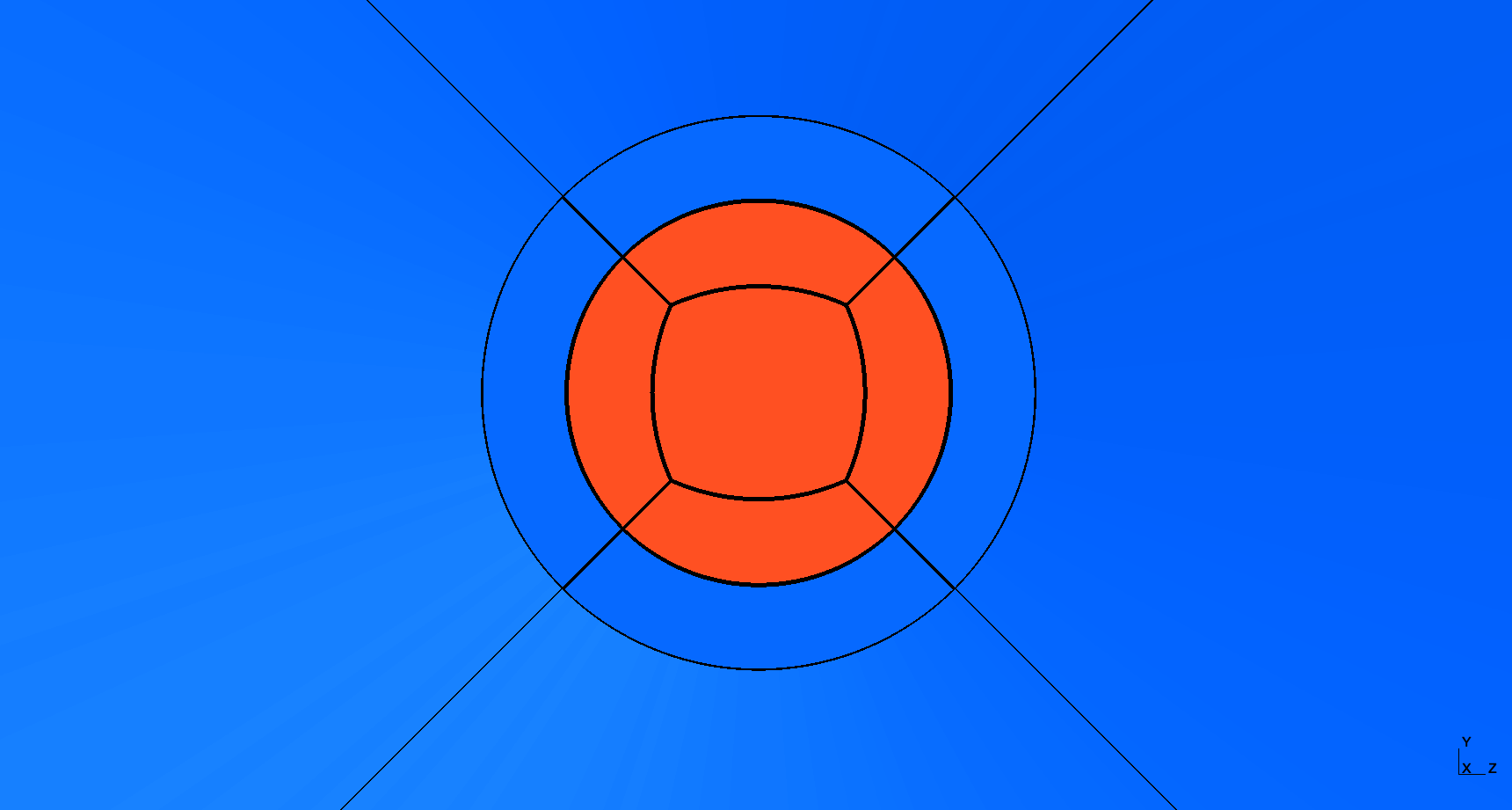}
    \label{fig:inlet_surf}
	}
\caption{Longitudinal and transversal cut planes presenting the block 
         topologies for the geometries used in the work.}
\label{fig.geo_cutplanes}
\end{figure*}

The mesh generation follows the multi-block strategy, where the computational
domain is divided into big blocks of 6-sided volumes. The multi-block mesh
generation with Gmsh requires that the block topology is defined in the 
computational domain. The generation of the G1 geometry and the first 
block topology follows the reference work\cite{Junior2018}, with adjustments
from the axisymmetric construction of the grid applied in the cited 
reference to the multi-block strategy employed in the present work. A cut 
plane through the centerline of the G1 geometry is presented in
Fig.~\ref{fig:G1} with the block distribution. After the simulations performed
with the G1 geometry and the meshes originated with the proposed topology, a 
new multi-block topology was generated utilizing the G2 geometry as the
reference domain. A cut plane through the centerline of the G2 geometry is
presented in Fig.~\ref{fig:G2} to show the differences between the topology
used in the two geometries. A view of the mesh blocks in a crossflow plane
is presented in Fig.~\ref{fig:inlet_surf}.

Three numerical meshes are generated to discretize the computational domain.
The first two numerical meshes are named M-1 and M-2. They are generated using 
the G1 geometry, and they have approximately $6 \times 10^6$ and 
$1.8 \times 10^6$ elements, respectively. When performing a simulation with 
the M-1 mesh and a second-order accurate spatial discretization, the total
number of degrees of freedom from the simulation is approximately the same 
as the simulation of the M-2 mesh with third-order accurate discretization.
In summary, the first two numerical grids are designed to investigate the
influence of the order of accuracy in the solution of a jet flow with a
constant number of degrees of freedom.

The third mesh generated has a different topology and a higher number of 
elements when compared to the first two meshes. It was generated to enclose 
a larger refinement in the region of the lipline close to the nozzle exit 
section and a smooth transition through the domain centerline with 
increasing distance from the jet inlet section. The local mesh refinement 
in the jet lipline and the axial mesh refinement are in agreement with other
numerical grids from different jet simulations\cite{BogeyBailly2010,
BogeyMarsden2016, Bres2017, Debonis2017}. This numerical mesh, named M-3 
mesh was designed to capture the jet within the section of $x/D_j=15.0$. 
After that point, the mesh coarsens at a higher rate to damp oscillations 
that could introduce instabilities to the simulation. The G2 geometry
is employed for the generation of the M-3 mesh. The numerical grid also has
higher coarsening rates in radial positions farther from the jet. Based on 
the results of the simulations using M-1 and M-2 meshes, the jet opening 
angle evaluated is approximately $2$ degrees.

\begin{figure*}[htb!]
\centering
\subfloat[Radial mesh refinement at $x/D_j=0$.]{
	\includegraphics[width=0.47\linewidth]{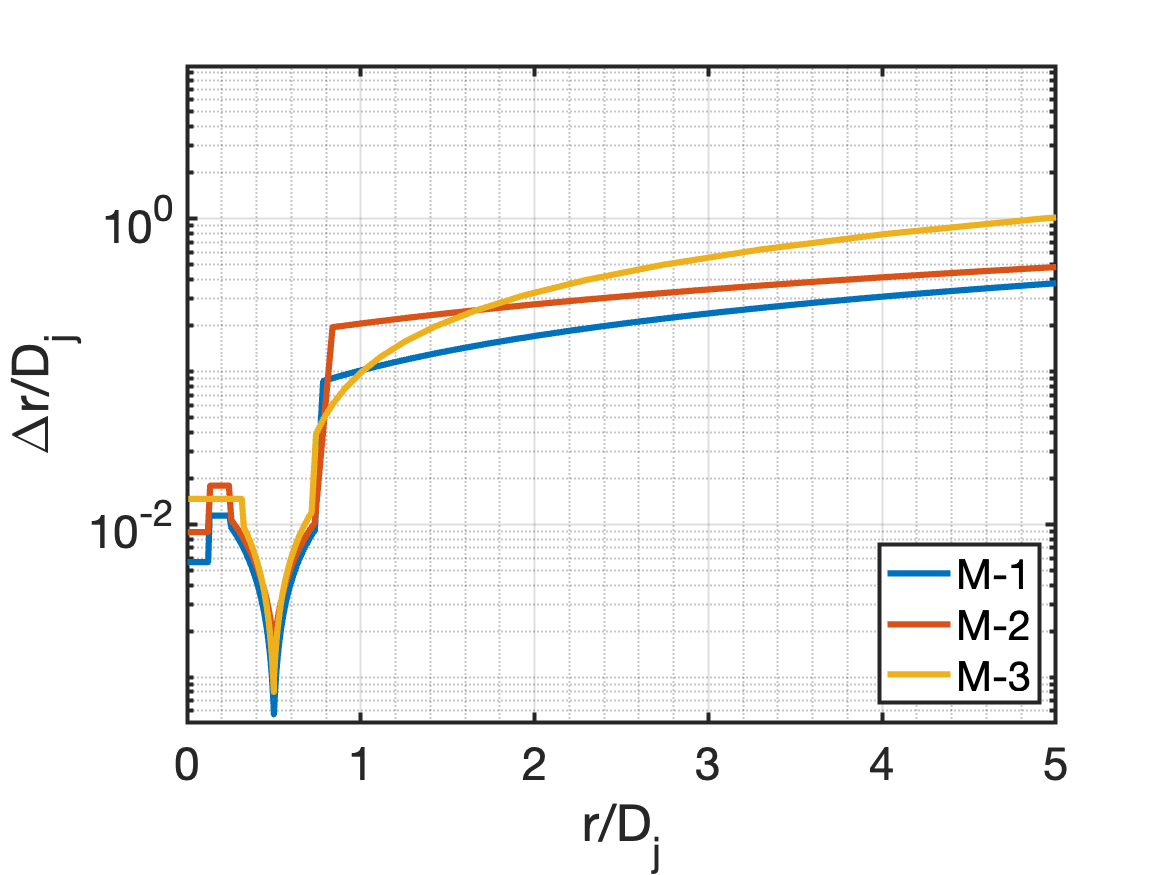}
	\label{fig.mref1}
	}
\subfloat[Radial mesh refinement at $x/D_j=15.0$.]{
	\includegraphics[width=0.47\linewidth]{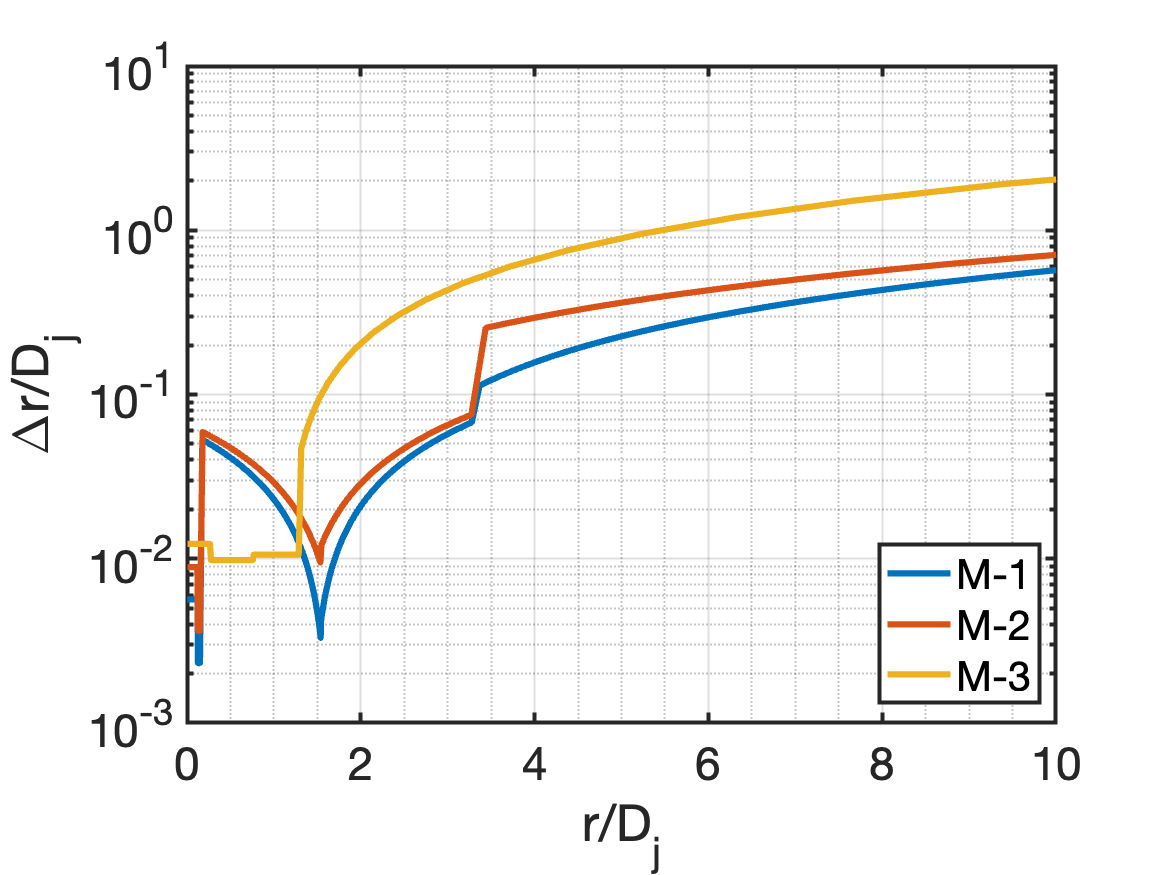}
	\label{fig.mref2}
	}
\\
\subfloat[Longitudinal mesh refinement at the jet centerline $r/D_j=0.0$.]{
	\includegraphics[width=0.47\linewidth]{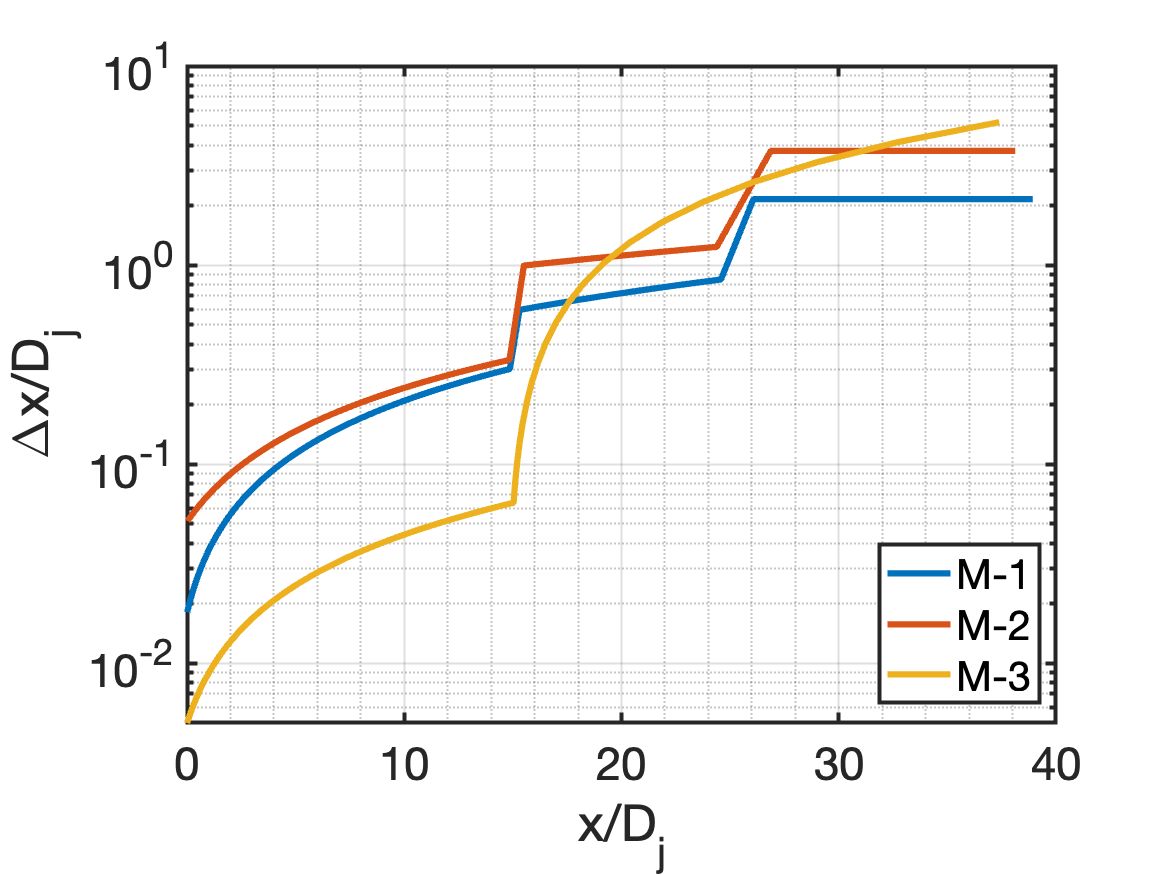}
	\label{fig.mref3}
	}
\caption{Distribution of grid spacing indicating radial and longitudinal 
         refinement for the three numerical meshes.}
\label{fig.M-3mesh}
\end{figure*}

\begin{figure*}[htb!]
\centering
\subfloat[M-1 mesh.]{
	\includegraphics[width=0.48\linewidth]
    {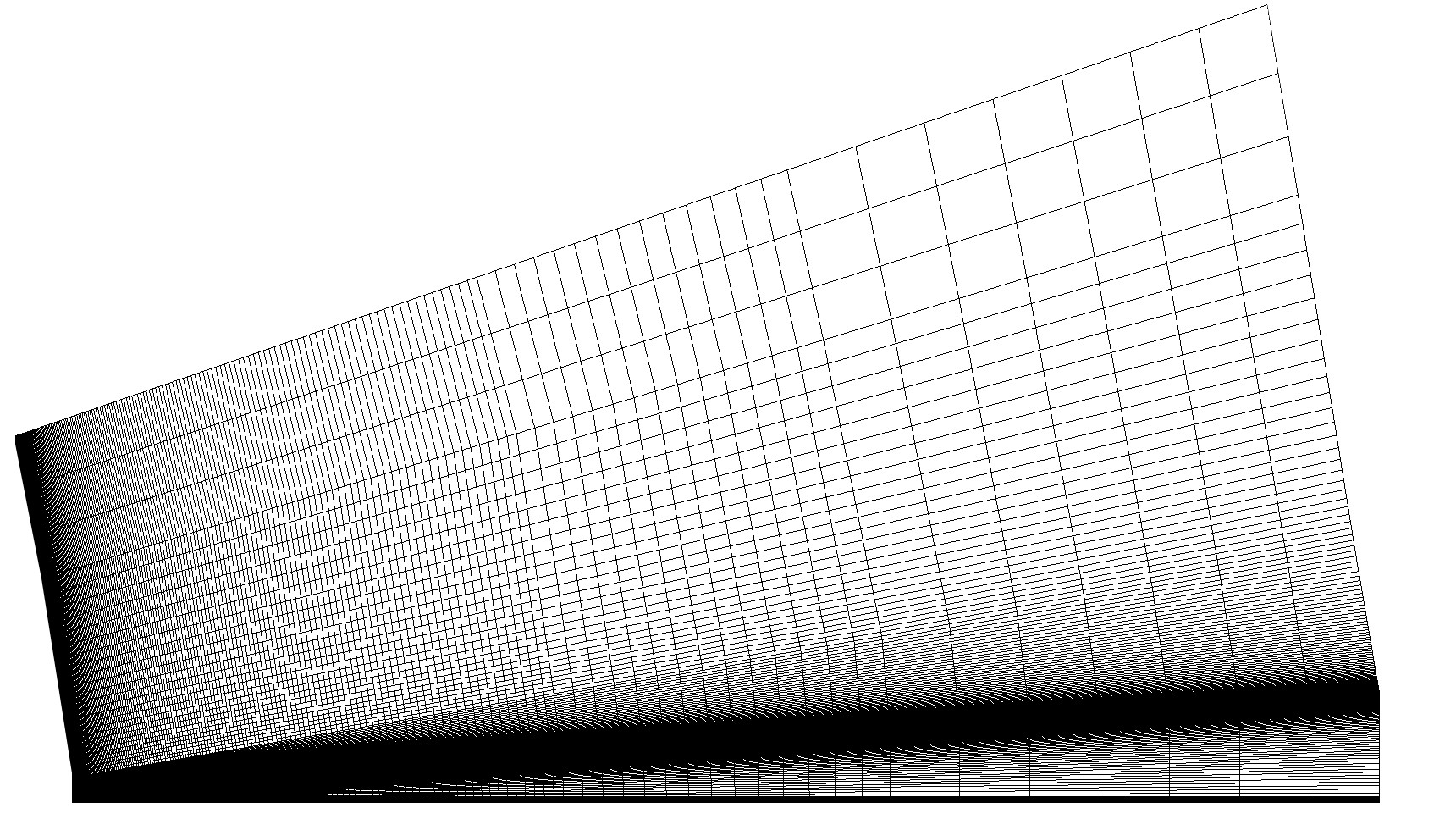}
	\label{fig.mesh1}
	}
\subfloat[M-2 mesh.]{
	\includegraphics[width=0.48\linewidth]
    {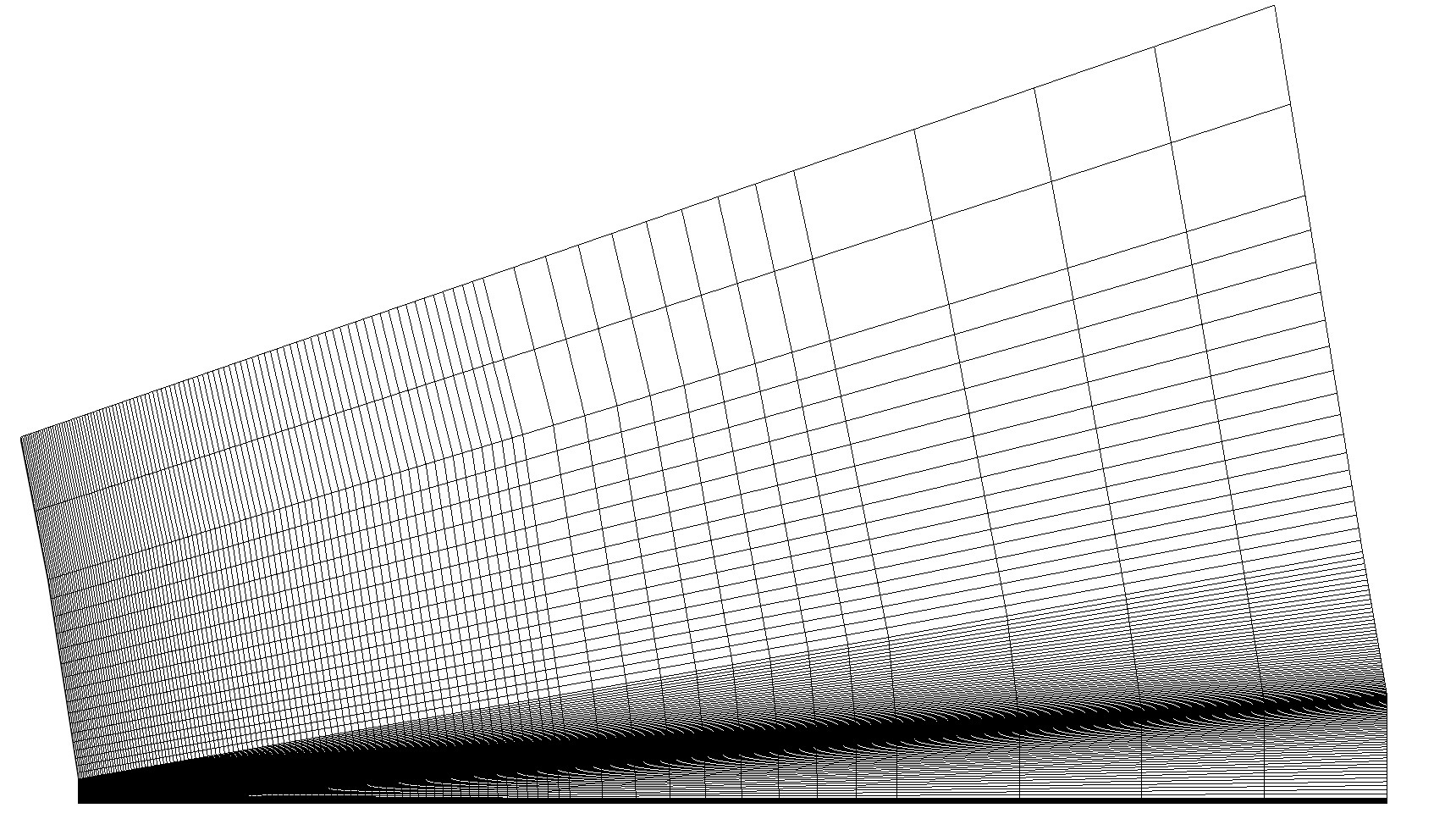}
	\label{fig.mesh2}
	}
\\
\subfloat[M-3 mesh.]{
	\includegraphics[width=0.48\linewidth]
    {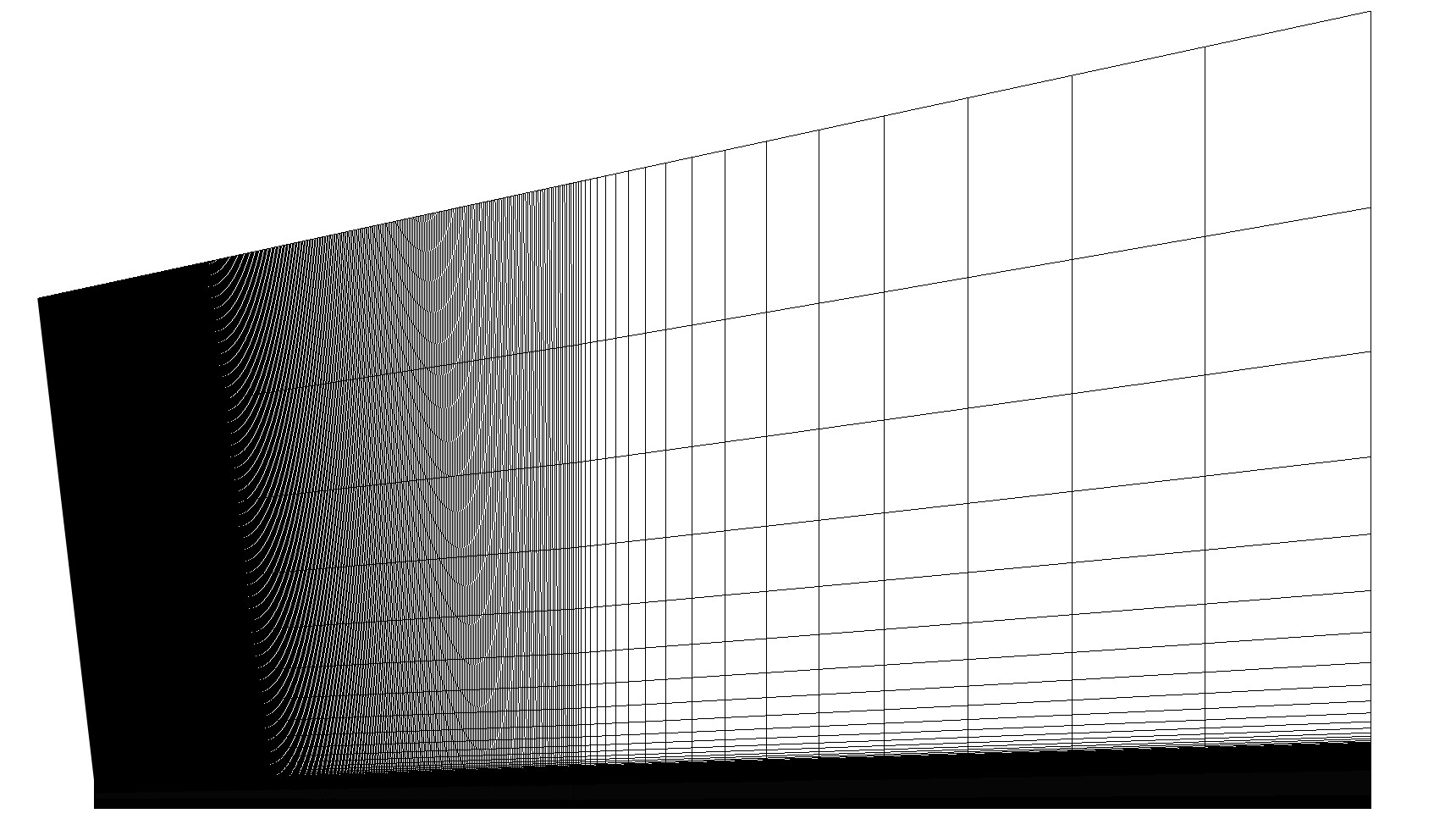}
	\label{fig.mesh3}
	}	
\caption{Visualization of the half-domain longitudinal cut planes for the 
        meshes used in the present work.}
\label{fig.meshes}
\end{figure*}

The radial mesh refinement in longitudinal positions $x/D_j=0.0$ and 
$x/D_j=15.0$ are presented in Figs.~\ref{fig.mref1} and \ref{fig.mref2}, and
longitudinal distributions of mesh elements are presented in
Fig.~\ref{fig.mref3}. In the azimuthal direction, the M-1 mesh has one 
element every $2$ degree, the M-2 mesh has one element every $3$ degree and 
M-3 mesh has the same distribution as the M-1 mesh. The three numerical 
meshes' half-domain longitudinal cut planes are presented in 
Fig.~\ref{fig.meshes}. A summary of the information from the numerical
geometries and meshes is presented in Tab.~\ref{tab:mesh} 

\begin{table*}[htb!]
\centering
\caption{Characteristics of the three numerical meshes used in the work.}
\begin{tabular}{ c c c c } \hline
Numerical & Computational & Elements  & Topology \\ 
Meshes & Domain & ($10^{6}$) & \\ \hline
M-1 & G1 & $6.2$ & Legacy \\
M-2 & G1 & $1.8$ & Legacy \\
M-3 & G2 & $15.4$ & Improved \\ 
\hline
\end{tabular}
\label{tab:mesh}
\end{table*}

%-------------------------------------------------------------------------------
\section{Numerical Procedure}
%-------------------------------------------------------------------------------
\label{sec.numerics}

The definition of the computational domain and generation of the numerical 
mesh is the starting point for the performance of the numerical simulations.
It is necessary to specify the boundary conditions to be used in the
domain's boundaries. The definition of the polynomial degree to represent
the numerical solution influences the order of accuracy of the numerical 
simulation and the limits of stability of the computational methods. The
specification of the boundary conditions, the definition of the simulation 
settings, and the calculation procedure for the statistical
properties are presented in the present section.

\subsection{Boundary Conditions}

The jet inflow, $(\cdot)_{j}$, and far-field, $(\cdot)_{ff}$, are the two
boundary surfaces present in the computational domain. The surfaces are
represented by the red and blue lines, respectively, in 
Fig.~\ref{fig:geo_domain}. The boundary conditions imposed for both surfaces
are weakly enforced solutions to Riemann problems in which the flow state 
outside of the domain is specified. The Riemann solver applied in the boundary 
condition enforcement is the same one employed to calculate the numerical 
fluxes between element interfaces. The difference between the boundary 
conditions rests on the properties of the reference state. An inviscid profile
represents the jet inflow condition. The inviscid profile represents a uniform
flow with the same properties prescribed independent of radial and azimuthal
positions. The properties imposed in the jet inlet condition match the 
specified Mach number and Reynolds number based on the jet inlet diameter of 
the experimental reference\cite{BridgesWernet2008}. The flow is characterized 
as a perfectly expanded and isothermal, meaning that $p_{j}/p_{ff}=1$ and
$T_{j}/T_{ff}=1$, where $p$ stands for pressure and $T$ for temperature. The
Mach number of the jet inlet is $M_{j}=1.4$, and the Reynolds number is 
$Re_{j} = 1.58 \times 10^6$. A small longitudinal velocity component with
$M_{ff}=0.01$ is imposed in the far-field boundaries, following the same
approach of other jet simulations\cite{Mendezetal2012, Bres2017}. A sponge 
layer model\cite{Fladetal2014} is imposed around the far-field boundary
condition, represented by the gray region in Fig.~\ref{fig:geo_domain}, to 
damp oscillations that could reach the boundary condition and be reflected to 
the jet flow.

\subsection{Simulation Settings}

Four simulations assess the influence of ``\textit{hp}'' mesh refinements 
and polynomial degrees on the representation of supersonic jet flows. The
simulations utilize the three numerical meshes presented in 
Tab.~\ref{tab:mesh}. Two polynomial degrees are used in the simulations to
represent the numerical solution, linear or quadratic polynomials, which 
yield second and third-order accurate spatial discretization, respectively. 
In conjunction with the two interpolating polynomial degrees, the ensemble of
numerical meshes produced simulations with a total number of degrees of 
freedom varying from $50 \times 10^6$ to $410 \times 10^6$ DOFs. The 
numerical simulations are named S-1 to S-4 simulations. The simulation 
ordering is associated with the increasing resolution of the numerical
simulations.

The four simulations utilize the same time marching method, a third-order 
accurate standard Runge-Kutta scheme with three stages. The time step 
applied to the simulations has a fixed value based on a fraction of the 
limit of stability for the Courant–Friedrichs–Lewy (CFL) number for the 
Runge-Kutta method. The CFL number stability limit varies with the 
polynomial degree used to represent the numerical solutions. High polynomial
degrees present larger restrictions on the maximum CFL than small ones. The
numerical simulations are performed with a fraction of the maximum CFL of
$0.71$, $0.46$, $0.39$, and $0.49$ from the S-1 to S-4 simulations. The S-3
simulation is performed with a conservative time step, while the other 
three are advanced in time with an overall CFL number closer to the 
corresponding stability limits. Table~\ref{table.sim} indicates the settings
used in the four numerical calculations.

\begin{table*}[htb!]
\centering
\caption{Summary of simulations settings.}
\begin{tabular}{ c c c c c } \hline
Simulation & Numerical & Polynomial &  Order of   & Degrees of \\ 
 &  & Degree & Accuracy & Freedom ($10^{6}$) \\ \hline
S-1 & M-1 & Linear &  2nd order & $\approx 50$ \\
S-2 & M-2 & Quadratic & 3rd order & $\approx 50$ \\
S-3 & M-3 & Linear & 2nd order & $\approx 120$ \\ 
S-4 & M-3 & Quadratic & 3rd order & $\approx 410$ \\
\hline
\end{tabular}
\label{table.sim}
\end{table*}

The development of the numerical simulations involved two steps. The jet flow
solution is let to develop from its initial condition to a statistically 
steady flow. Then, the data acquisition is performed. The initial condition 
of the simulation is a quiescent flow for simulations S-1 to S-3. The S-4
simulation used the solution from the S-3 simulation as the initial
condition, which reduced the simulation time to reach the statistically 
steady condition. A non-dimensional time scale is employed as a reference for 
the evolution of the simulation. The non-dimensional time, which is called
flow-through time (FTT), is defined with the jet inlet velocity ($U_j$) and
the jet inlet diameter ($D_j$). It was observed that starting the simulation
with a quiescent flow required approximately $240$ FTT to reach a 
statistically steady flow, and using a previous solution reduced the 
development time to approximately $100$ FTT.

{The simulations were conducted at the Jean-Zay supercomputer
from the IDRIS computing center \cite{jeanzay}. A total 
of $800$ to $1300$ CPUs were used to perform the simulations, depending of the 
test case addressed. The 
computational cost of the simulations was approximately $100 \times 10^3$ 
CPU hours for the S-1 and S-2 simulations and $300 \times 10^3$ for the S-3
simulation. The computational cost of the S-4 simulation was reduced due to
the restart from the solution of the S-3 simulation. After the restart, the
computational cost to conclude the S-4 simulation was approximately $500 
\times 10^3$ CPU hours.}

\subsection{Calculation of Statistical Properties}

The acquisition time is in agreement with the smallest Strouhal number ($St$) 
used in the experimental reference\cite{BridgesWernet2008} and from other
numerical work that evaluated a similar jet flow 
condition\cite{Mendezetal2012}. The acquisition frequency is determined to
provide a high value of the maximum Strouhal number to be investigated.
The acquisition frequency used is $200$ kHz. Estimating the maximum
Strouhal number through $St_{max}=fD_j/2U_j$ and the minimum Strouhal number
by $St_{min}=D_j/ t_s U_j$, where $f$ is the acquisition frequency and 
$t_s$ is the time sample, the values of $St_{max}=10.5$ and $St_{min}=0.01$
are obtained.

Flow properties' mean and root mean square (RMS) fluctuations are calculated
along the longitudinal distributions at the centerline and lipline and radial 
distributions at four streamwise positions. The centerline is defined as the
line in the center of the geometry $r/D_j=0.0$, whereas the lipline is a 
cylindrical surface at the nozzle diameter, $r/D_j=0.5$. The four streamwise
planes are at $x/D_j=2.5$, $x/D_j=5.0$, $x/D_j=10.0$, and $x/D_j=15.0$. 
%\textcolor{blue}{The results from the lipline and the four streamwise positions are obtained 
%by an azimuthal average from $120$ equally spaced positions. Such strategy is
%employed to increase the time sample, without an excessive data acquisition
%sample.} 
Spectral analysis
is conducted. The Power Spectral Density (PSD) of the longitudinal and radial
velocity fluctuations are performed at the lipline in the four streamwise
planes.

%-------------------------------------------------------------------------------
\section{Flow Field Result Analyses}
%-------------------------------------------------------------------------------
\label{sec.results}

The results of the four simulations are compared to assess the ``\textit{hp}''
mesh refinement and polynomial degree improvements obtained with the 
increase in the simulation resolution. Due to axisymmetry of the jet flow 
the results are presented in the polar coordinate system $\boldsymbol{U} = 
(U_x, U_r, U_{\theta})$. A qualitative analysis is performed with contours of
instantaneous and statistical properties. A quantitative analysis is performed
with the longitudinal and radial distributions at the centerline, lipline, and
the four streamwise planes. In the quantitative analysis, the results of the
four simulations are compared with experimental data. A signal analysis is
performed with the power spectral density of the velocity fluctuation in the
lipline.

\subsection{Instantaneous Flow Field}

\begin{figure*}[htb!]
\centering
\subfloat[S-1 simulation.]{
	\includegraphics[width=0.48\linewidth]{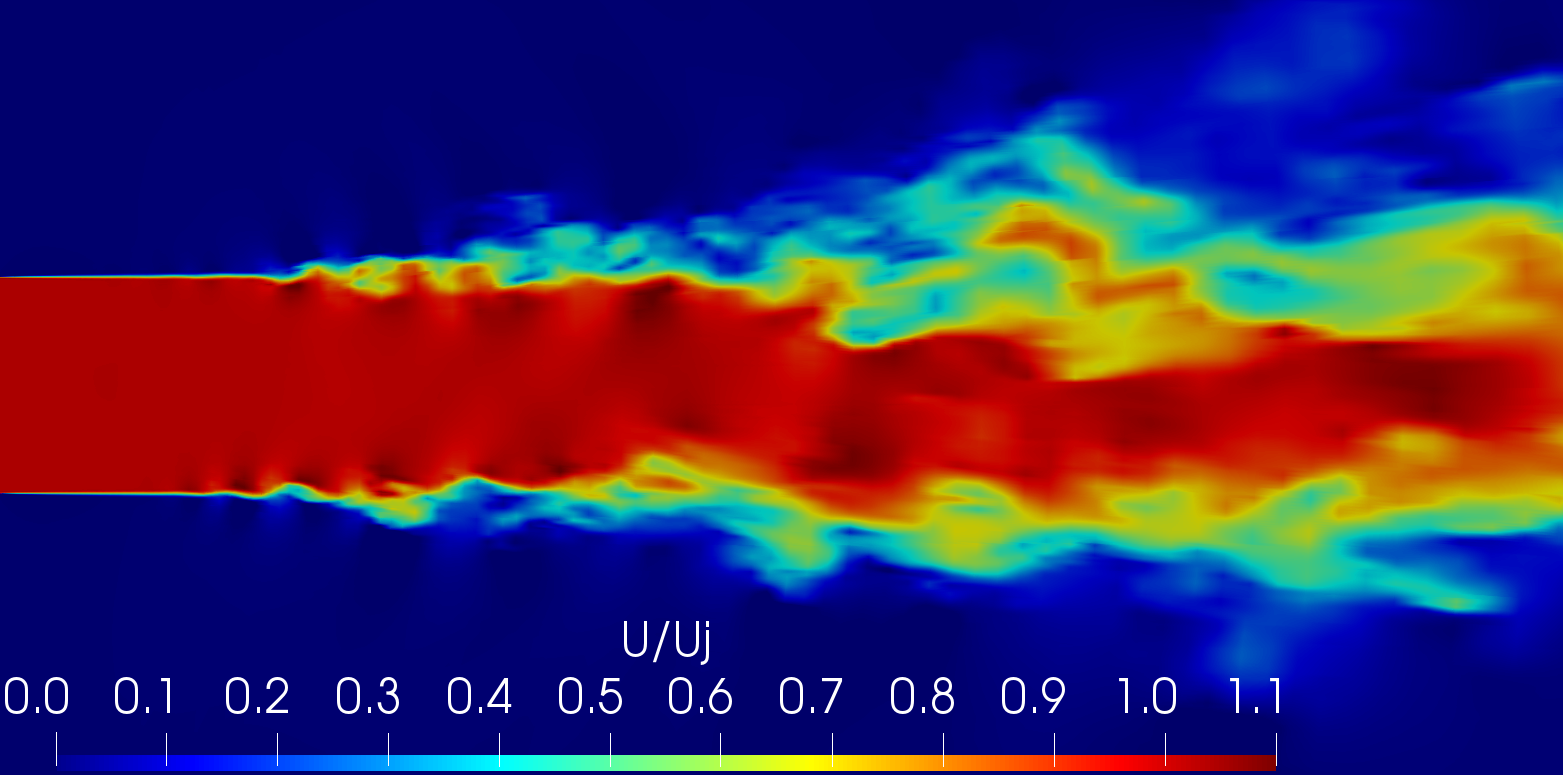}
	\label{res.vxsnap_s1}
	}
\subfloat[S-2 simulation.]{
	\includegraphics[width=0.48\linewidth]{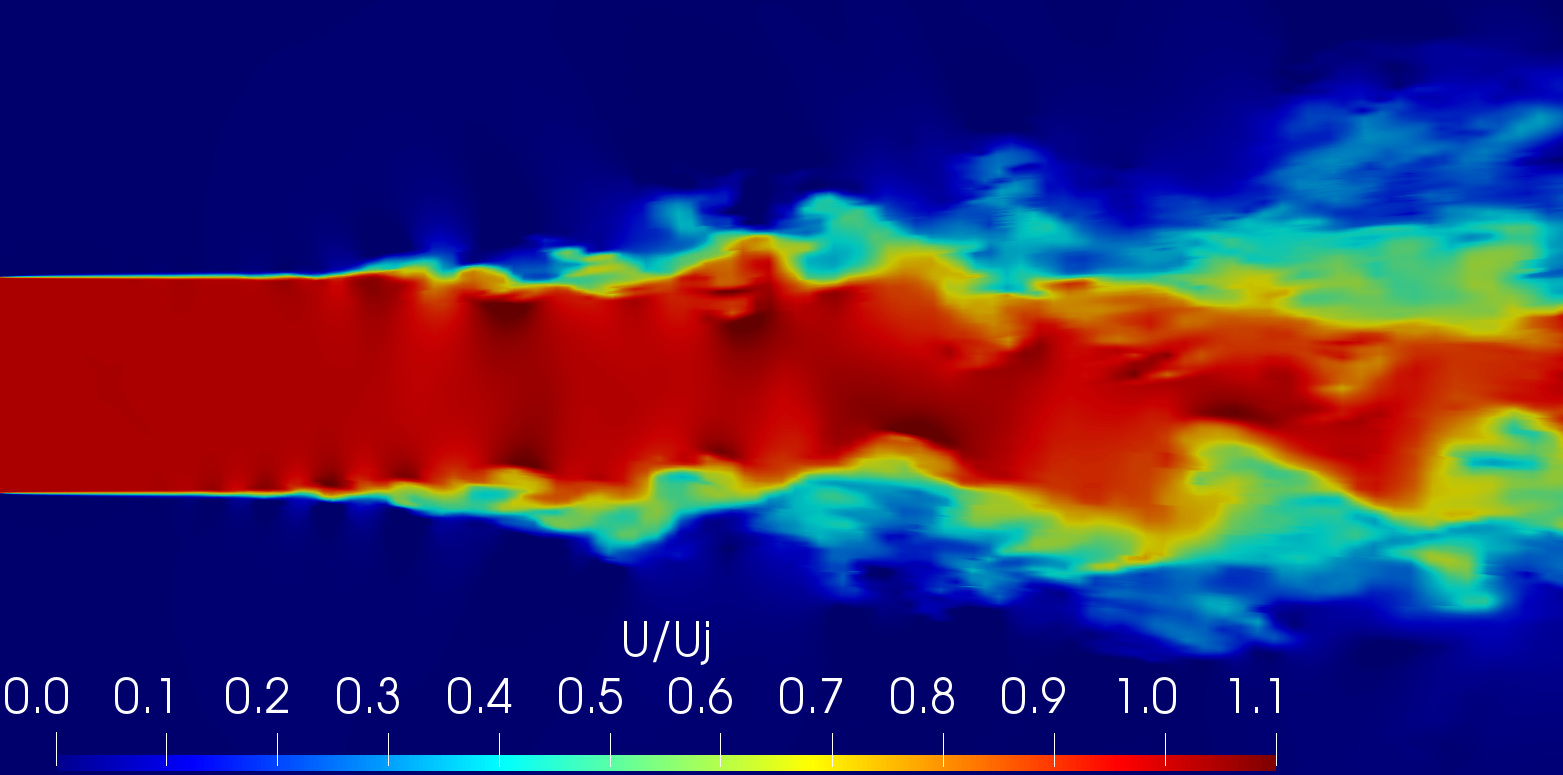}
	\label{res.vxsnap_s2}
	}
\\
\subfloat[S-3 simulation.]{
	\includegraphics[width=0.48\linewidth]{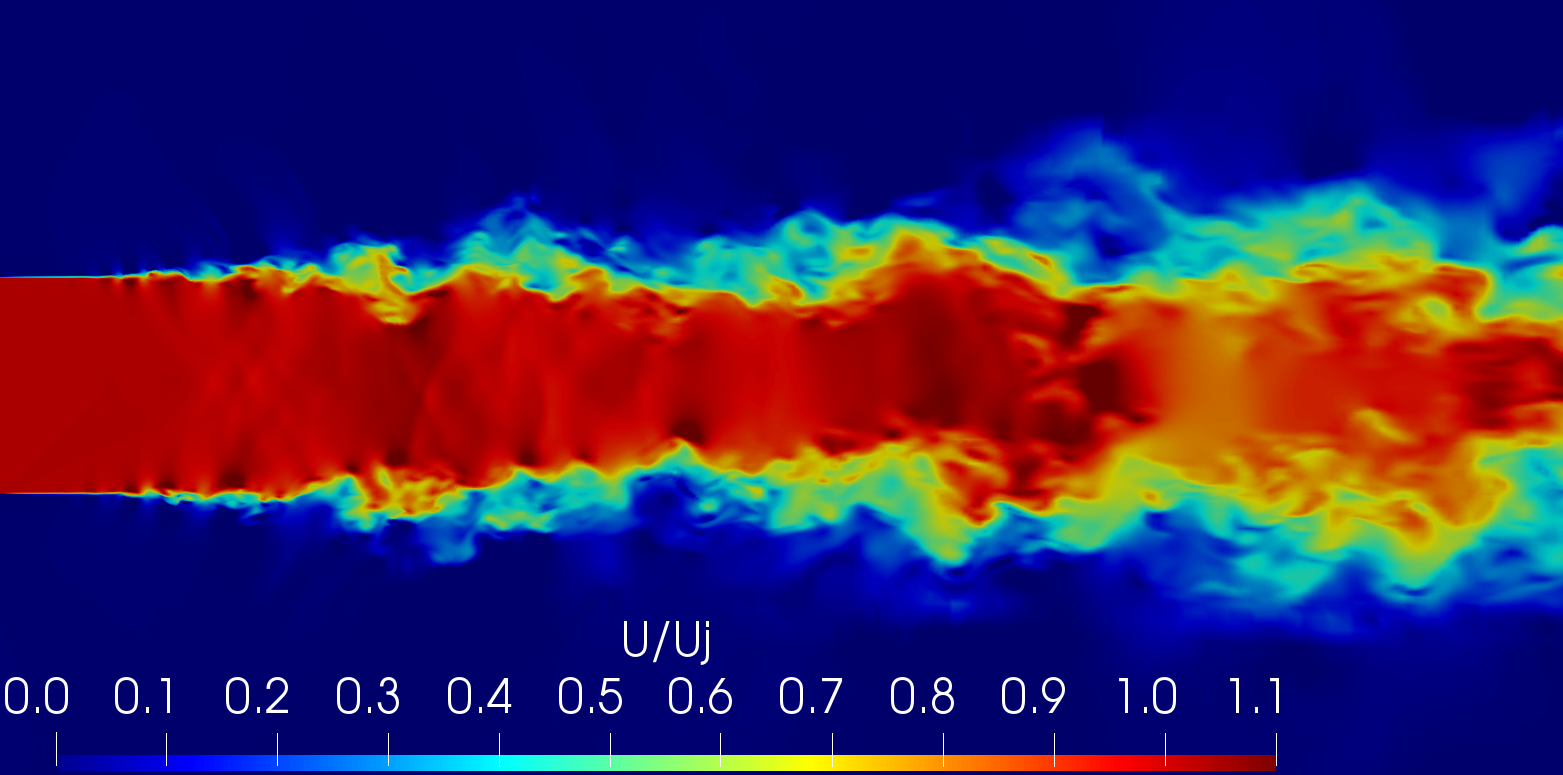}
	\label{res.vxsnap_s3}
	}
\subfloat[S-4 simulation.]{
	\includegraphics[width=0.48\linewidth]{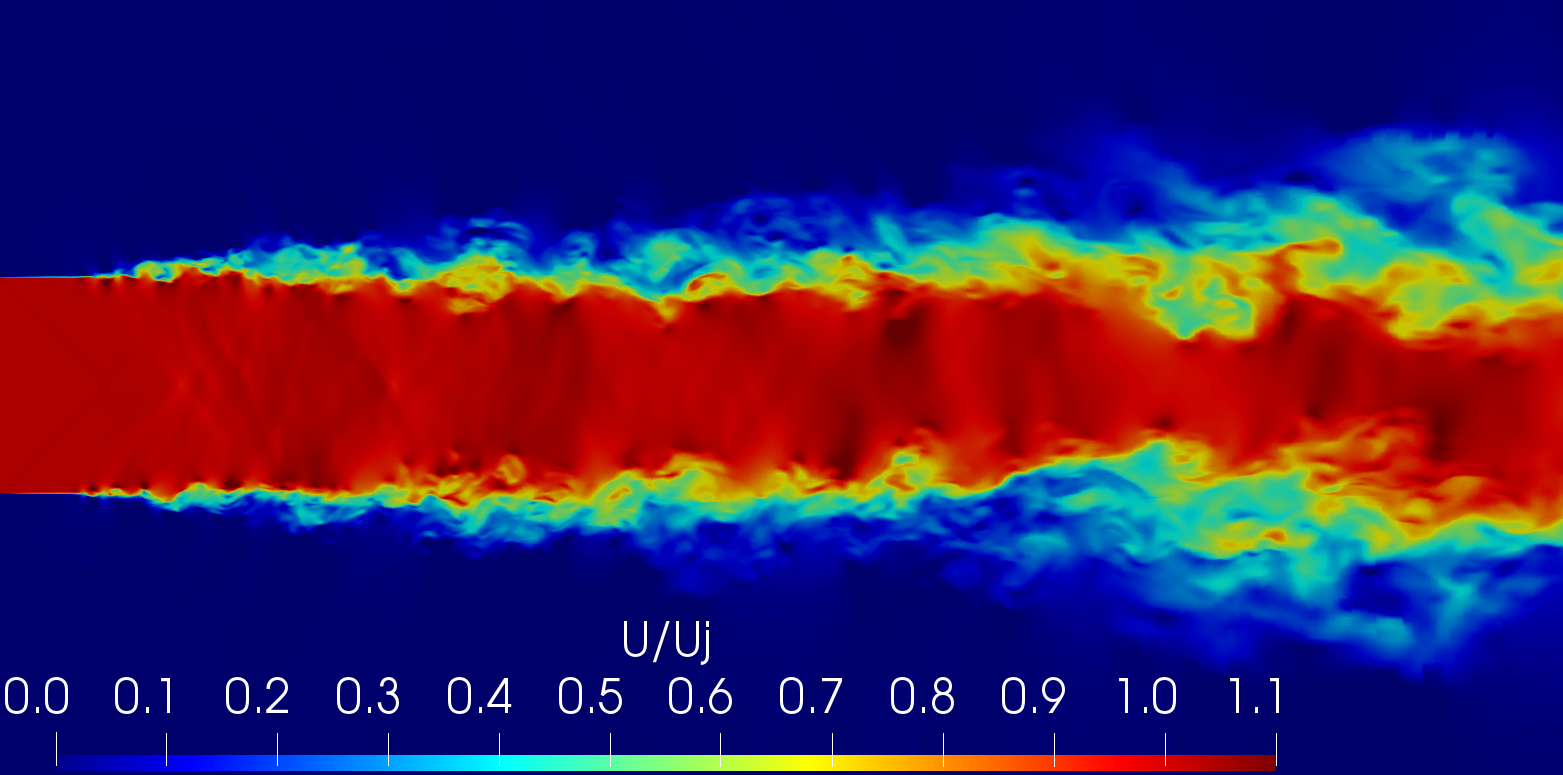}
	\label{res.vxsnap_s4}
	}
\caption{Instantaneous longitudinal velocity component contours on a plane
         through the centerline of the jet.} 
\label{res.vxsnap}
\end{figure*}

The instantaneous contours of velocity and pressure from the jet flow field 
are investigated in order to visualize the flow features presented by the
numerical simulations. The instantaneous longitudinal velocity component
contours presented in Fig.~\ref{res.vxsnap} show that in the mixing layer
from the S-3, Fig.~\ref{res.vxsnap_s3}, and S-4, Fig.~\ref{res.vxsnap_s4},
simulations present smaller eddies than those in the mixing layer
from the of the S-1, Fig.~\ref{res.vxsnap_s1}, and S-2, 
Fig.~\ref{res.vxsnap_s2}, simulations. The velocity contours in the lipline 
of the jet flow indicate a transition occurring closer to the jet inlet 
section for the S-3 and S-4 simulations than in the S-1 and S-2 simulations.
In the S-1 and S-2 simulations, the interaction of the shock waves is weak,
producing few changes to the velocity distribution in the interior of
the jet flow. The higher resolution from S-3 and S-4 simulations than the 
S-1 and S-2 simulations reproduce more intense shock interactions, which 
affects the flow in the interior of the jet flow, which can be visualized by
the many velocity changes in the interior of the jet flow.

\begin{figure*}[htb!]
\centering
\subfloat[S-1 simulation.]{
	\includegraphics[width=0.48\linewidth]{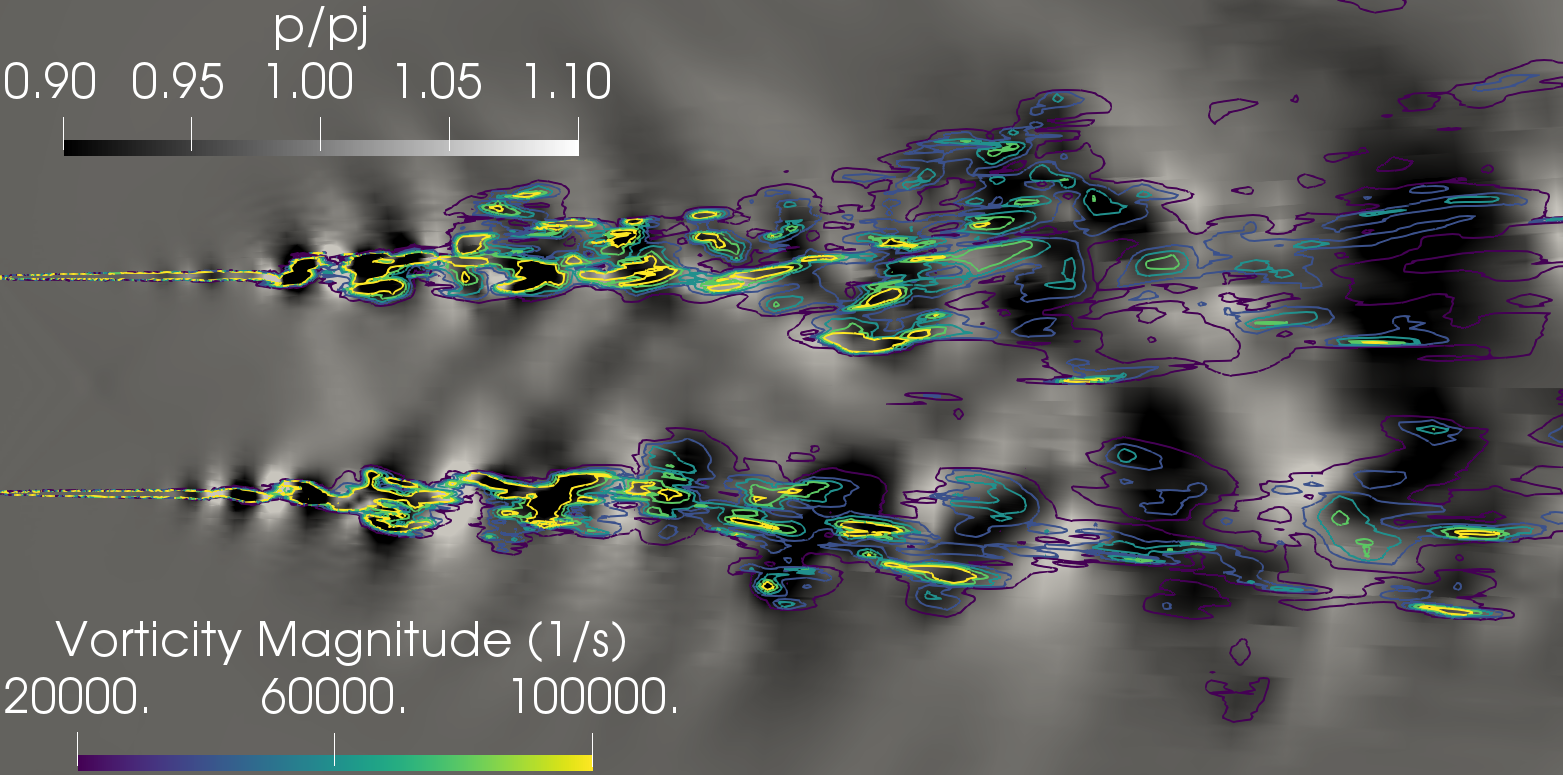}
	\label{res.psnap_s1}
	}
\subfloat[S-2 simulation.]{
	\includegraphics[width=0.48\linewidth]{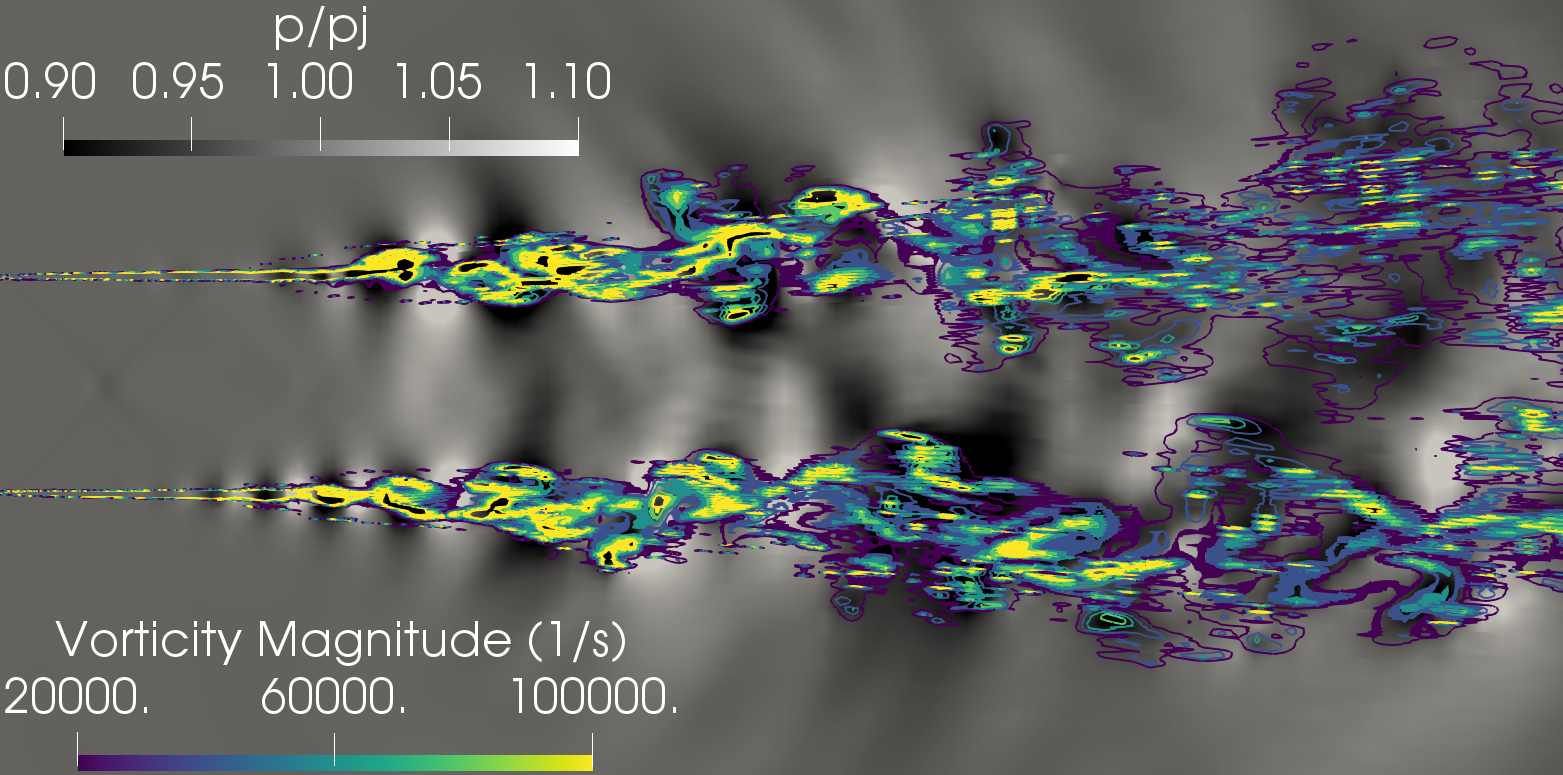}
	\label{res.psnap_s2}
	}
\\
\subfloat[S-3 simulation.]{
	\includegraphics[width=0.48\linewidth]{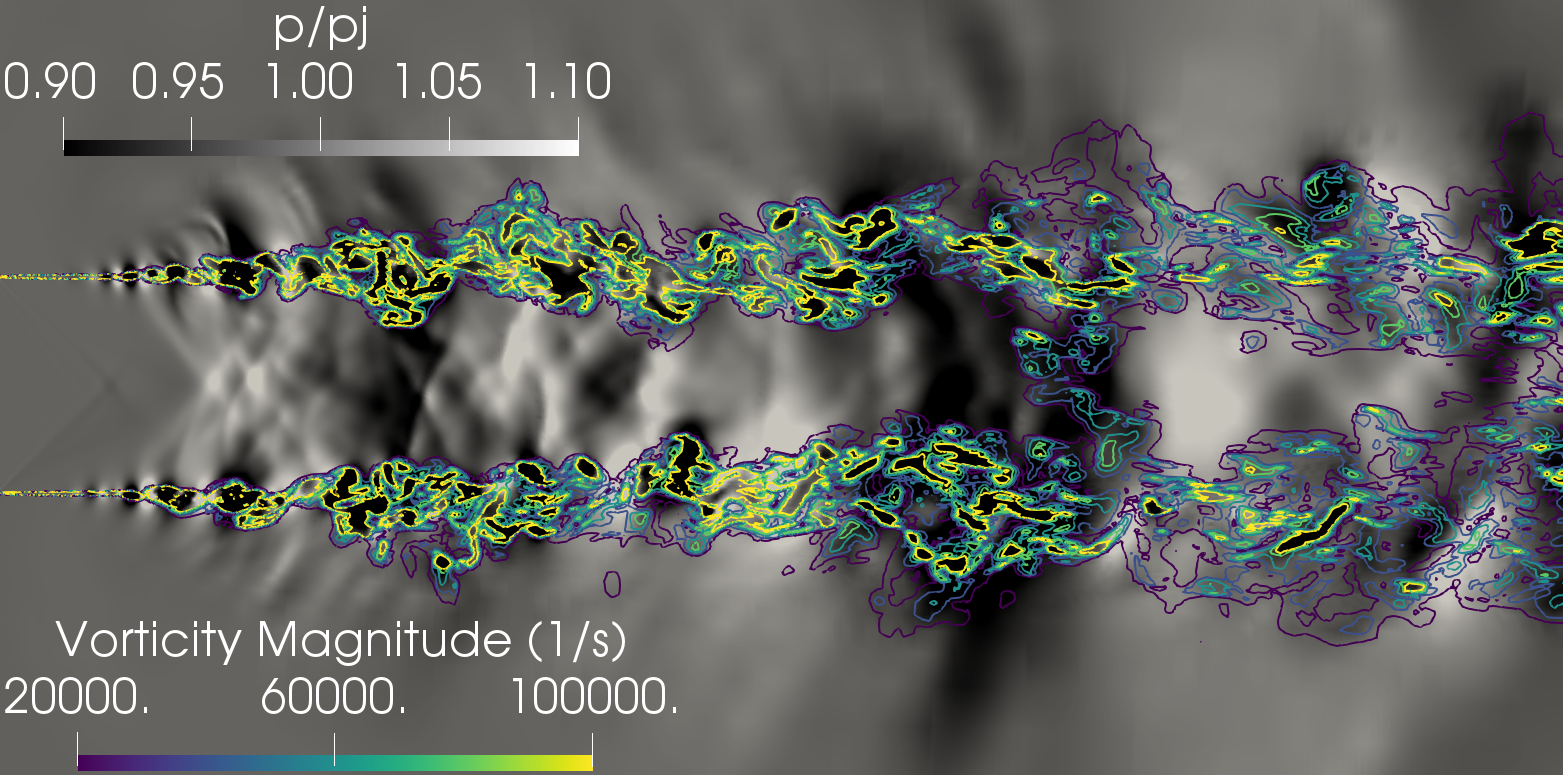}
	\label{res.psnap_s3}
	}
\subfloat[S-4 simulation.]{
	\includegraphics[width=0.48\linewidth]{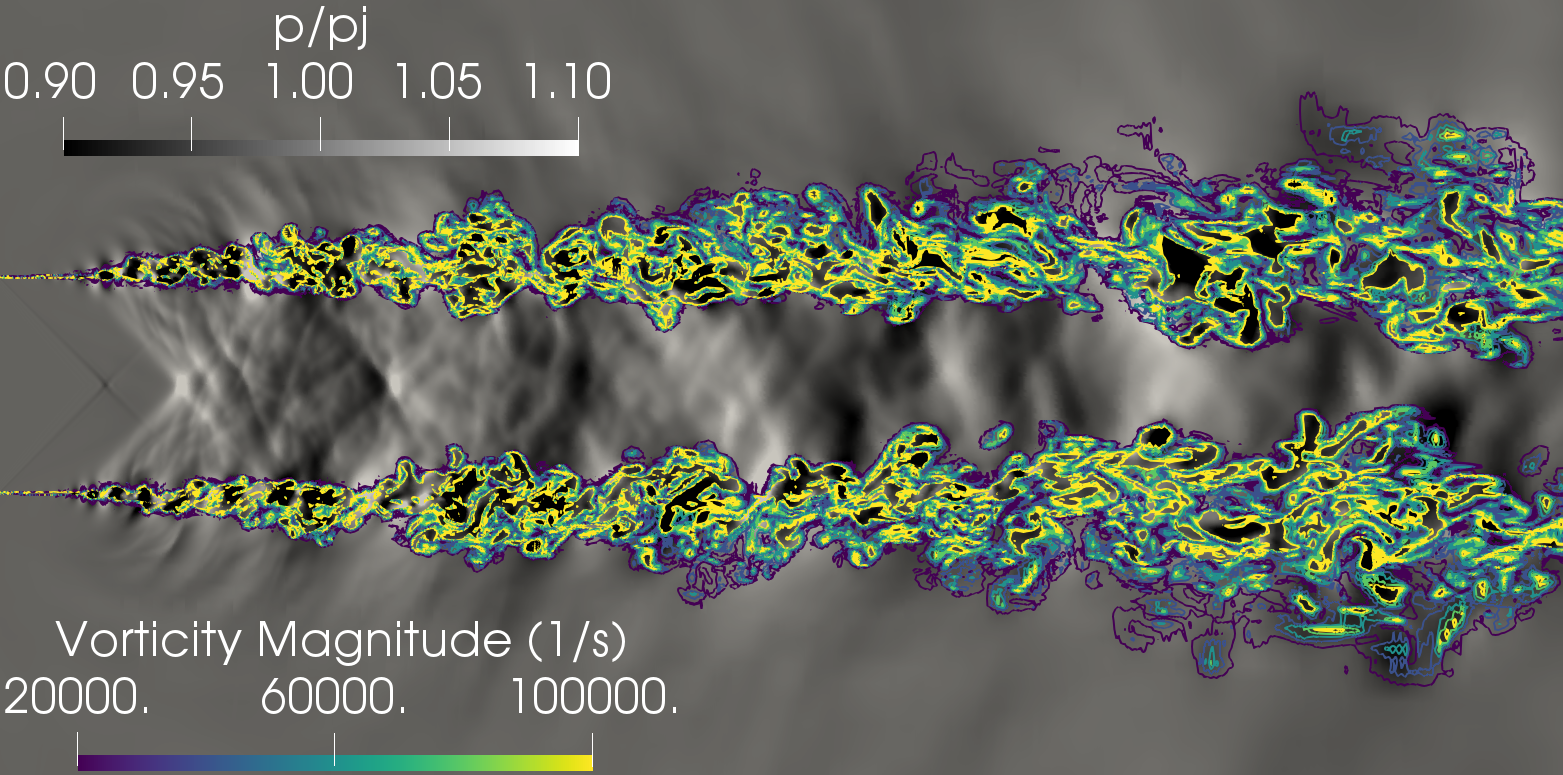}
	\label{res.psnap_s4}
	}
\caption{Snapshots of pressure with superimposed contours of vorticity 
         magnitude on a plane through the centerline of the jet flow.}
\label{res.psnap}
\end{figure*}

The vorticity contours, presented in Fig.~\ref{res.psnap} highlight the 
instabilities responsible for the transition from 
laminar to turbulent and the interaction between the shock waves and the
mixing layer. The pressure contours highlight the interaction between the
shock waves in the interior of the jet flow and the pressure wave propagation
external to the jet flow. The vorticity calculated by the S-1 simulation, 
Fig.~\ref{res.psnap_s1} indicate the presence of weak vortices close to 
the jet inlet section that almost dissipate when it advances in the 
domain. The increase in the order of accuracy from S-1 to S-2 simulation, 
Fig.~\ref{res.psnap_s2}, can produce better-defined vortices capable of 
sustaining the presence of turbulent eddies with the advance in the domain.
In the S-1 and S-2 simulations, the pressure contours indicate the presence
of large-scale pressure waves associated with the vortical structures
reproduced in the mixing layer. The simulations that used the M-3 mesh,
the S-3 and S-4 simulations, could represent additional features
in the pressure and vorticity contours. The vorticity contours from the S-3 
simulation, Fig.~\ref{res.psnap_s3}, shows that the simulation can reproduce
small-scale vortices responsible for the transition of the flow, and more
shock waves are present, which produce a series of pressure changes in the
interior of the jet flow. The S-4 simulation, Fig.~\ref{res.psnap_s4},
presents the best-defined eddies in the mixing layer and shock waves in the 
interior of the jet flow from the four simulations analyzed. 
{The comparison among the pressure contours and vorticity magnitude contours
from the four simulations show a reduction of the numerical dissipation, with
the capability of the high-resolved simulations to maintain the flow features 
correctly represented for longer distances and, also, to include the
representation of smaller features present in the flowfield}.

%In the
%supplemental material, videos of the longitudinal velocity component and
%pressure contours are available to comprehend further the flow
%features discussed.

\subsection{Mean Flow Field}

\begin{figure*}[htb!]
\centering
\subfloat[S-1 simulation.]{
	\includegraphics[width=0.48\linewidth]
    {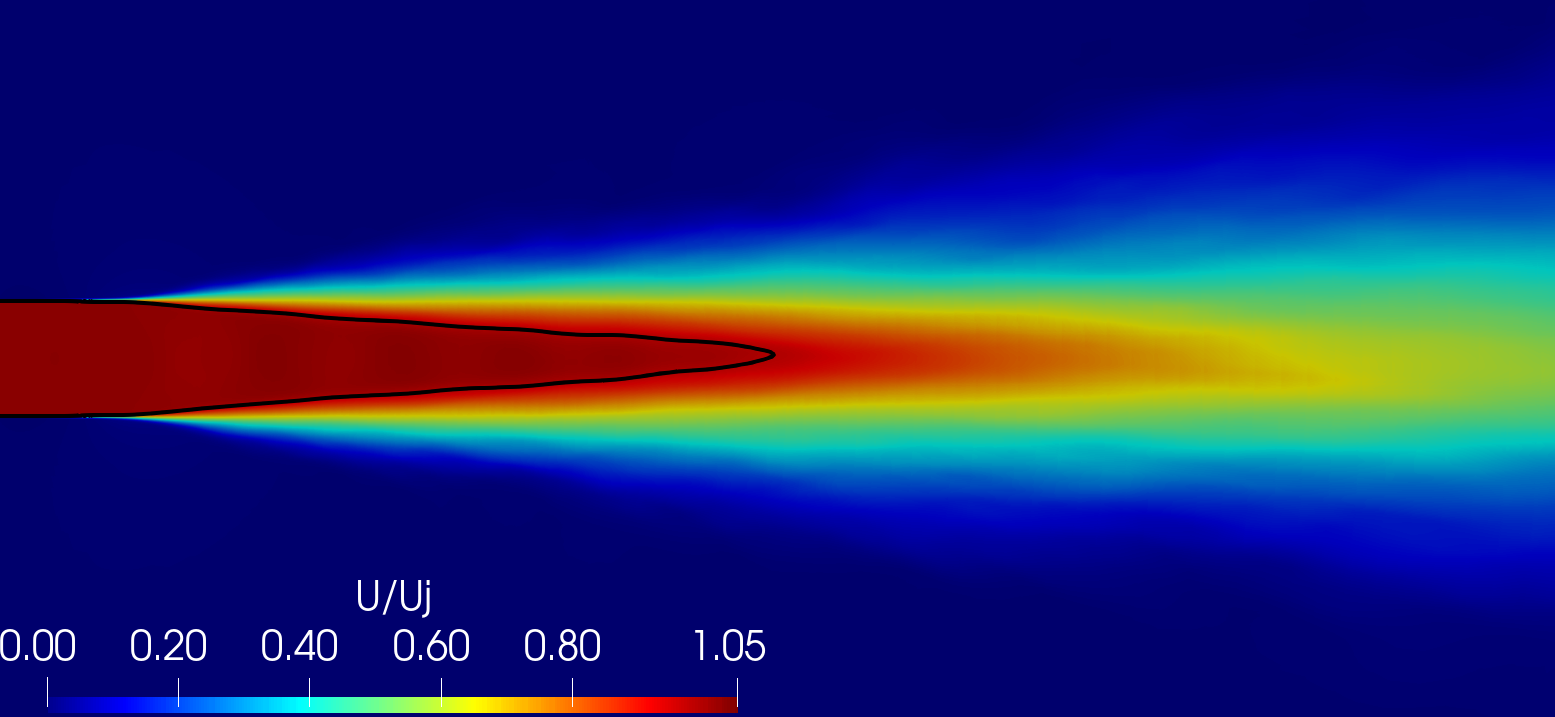}
	\label{res.vxmean_s1}
	}
\subfloat[S-2 simulation.]{
	\includegraphics[width=0.48\linewidth]
    {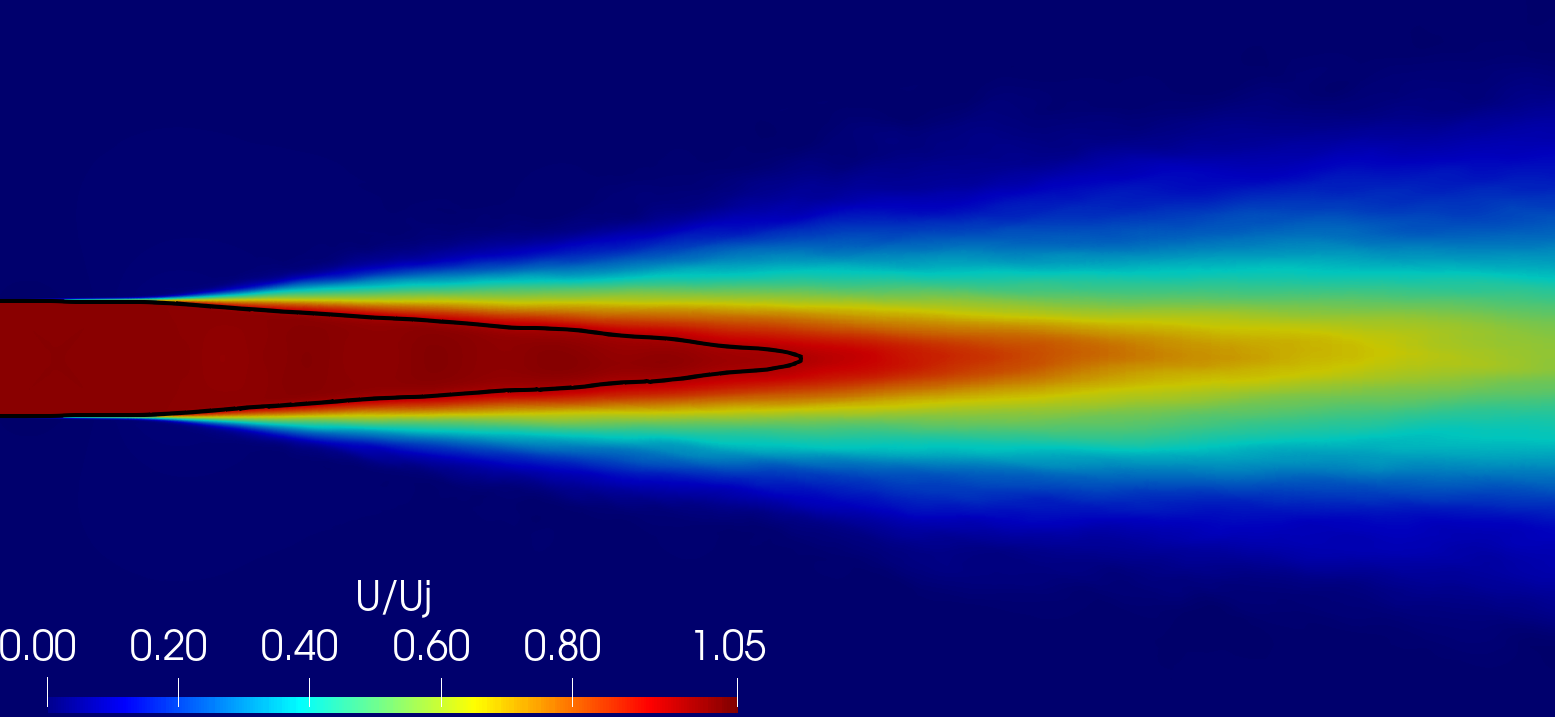}
	\label{res.vxmean_s2}
	}
\\
\subfloat[S-3 simulation.]{
	\includegraphics[width=0.48\linewidth]
    {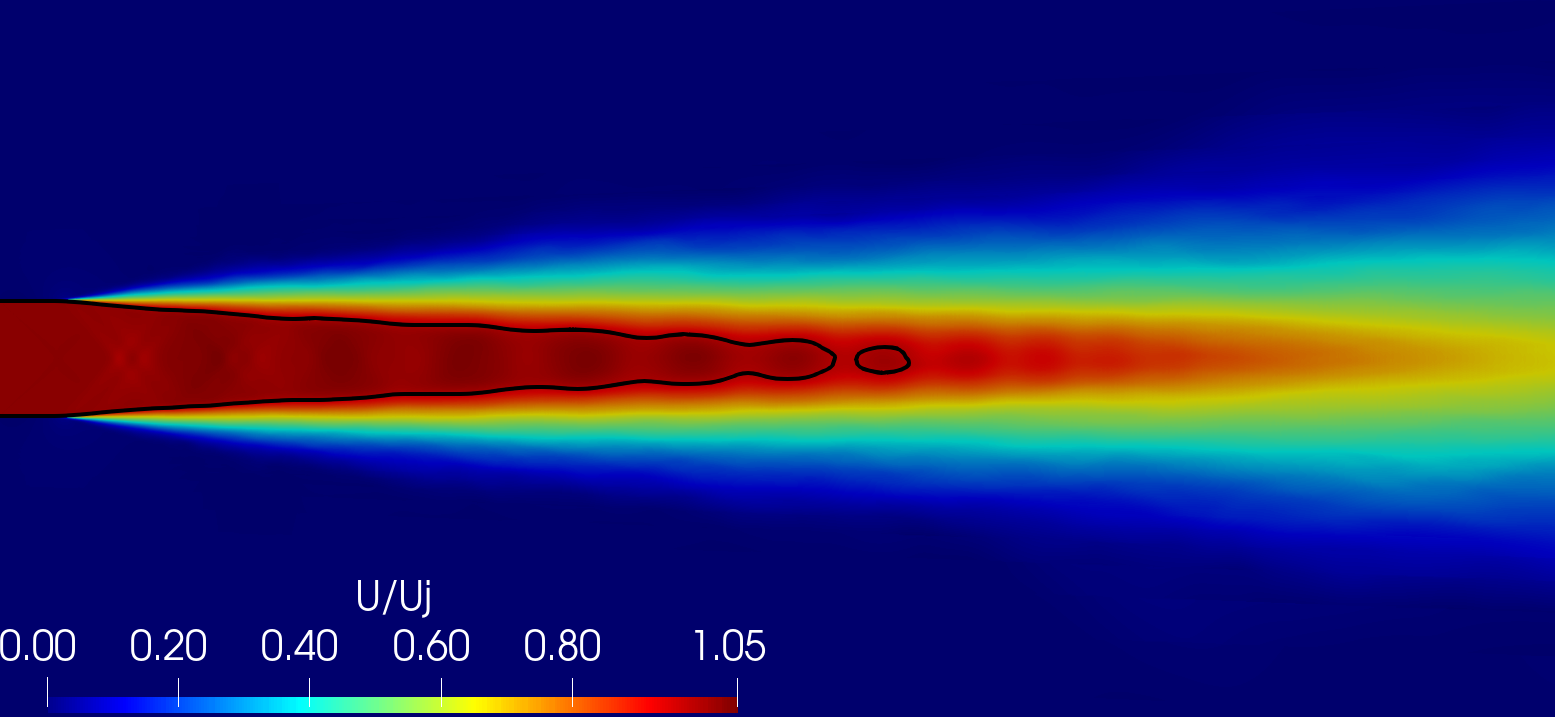}
	\label{res.vxmean_s3}
	}
\subfloat[S-4 simulation.]{
	\includegraphics[width=0.48\linewidth]
    {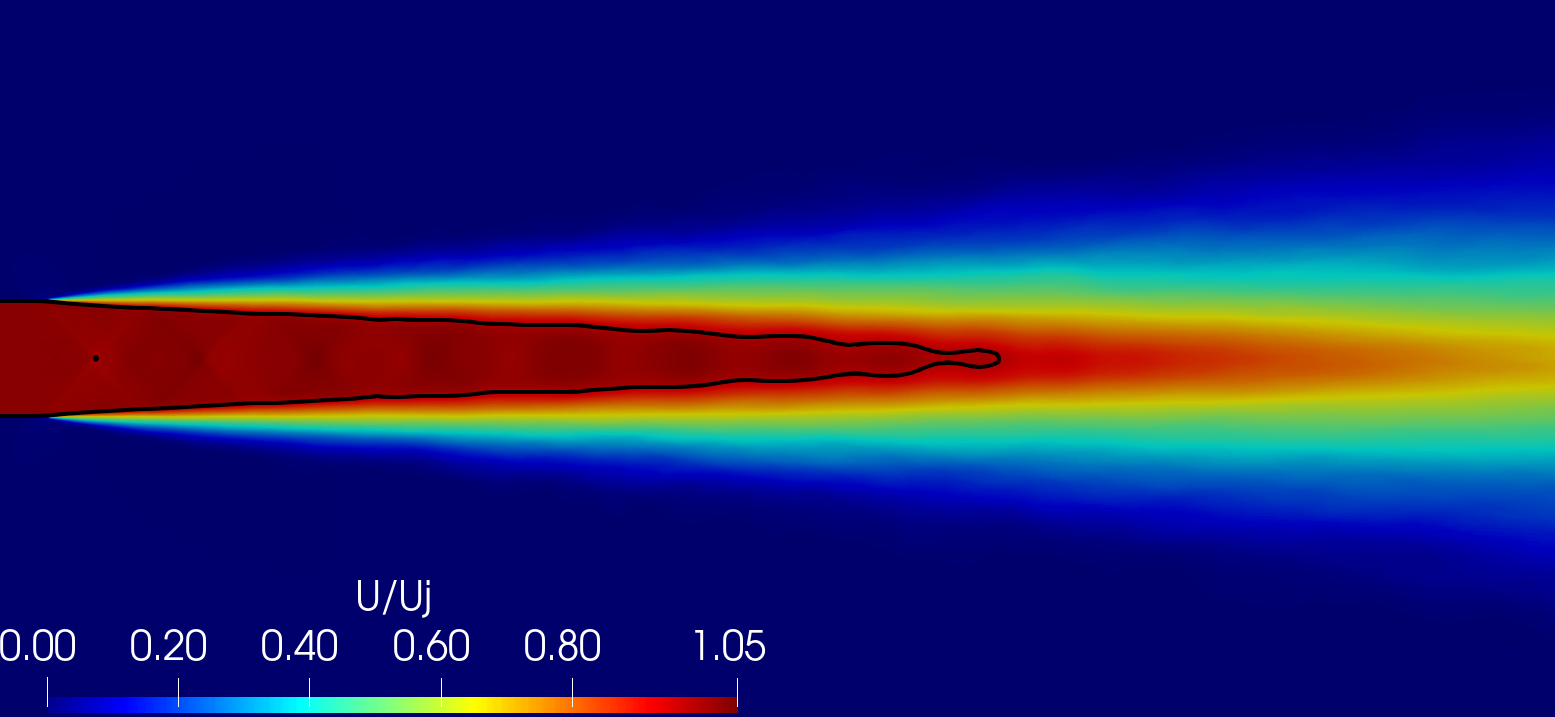}
	\label{res.vxmean_s4}
	}
\caption{Contours of mean non-dimensional longitudinal velocity component 
         for the simulations performed. The black contour indicate the
         boundaries of the jet potential core, $U=0.95U_j$.}
\label{res.vxmean}
\end{figure*}

The contours of the mean longitudinal velocity component, presented 
in Fig.~\ref{res.vxmean}, provide some qualitative information on the distinct
calculations of the potential core from each simulation. The jet potential core 
is the region of the jet with a longitudinal velocity higher than $0.95 U_j$, 
where $U_j$ is the jet inlet velocity. The length of the potential core is the 
distance in the centerline of the jet from the inlet section to the position 
where the velocity reaches the boundary of the potential core, $U=0.95 U_j$. 
Comparing the velocity contours in Fig.~\ref{res.vxmean_s1} and 
Fig.~\ref{res.vxmean_s2}, from S-1 and S-2 simulations, one can observe that 
the S-2 simulation presents a slightly longer core. Both simulations present 
a subtle velocity change associated with the shock waves. The velocity 
contours from the simulations performed with the more refined mesh, 
Figs.~\ref{res.vxmean_s3} and \ref{res.vxmean_s4}, present longer jet cores,
larger shock waves, and the mixing layer development commences closer to the
jet inlet condition than the contours from the S-1 and S-2 simulations. The
comparison of the velocity contours from simulations S-3 and S-4 shows a 
longer potential core with a narrow mixing layer when increasing the 
resolution from the S-3 to S-4 simulations.

\begin{figure*}[htb!]
\centering
\subfloat[S-1 simulation.]{
	\includegraphics[width=0.48\linewidth]
    {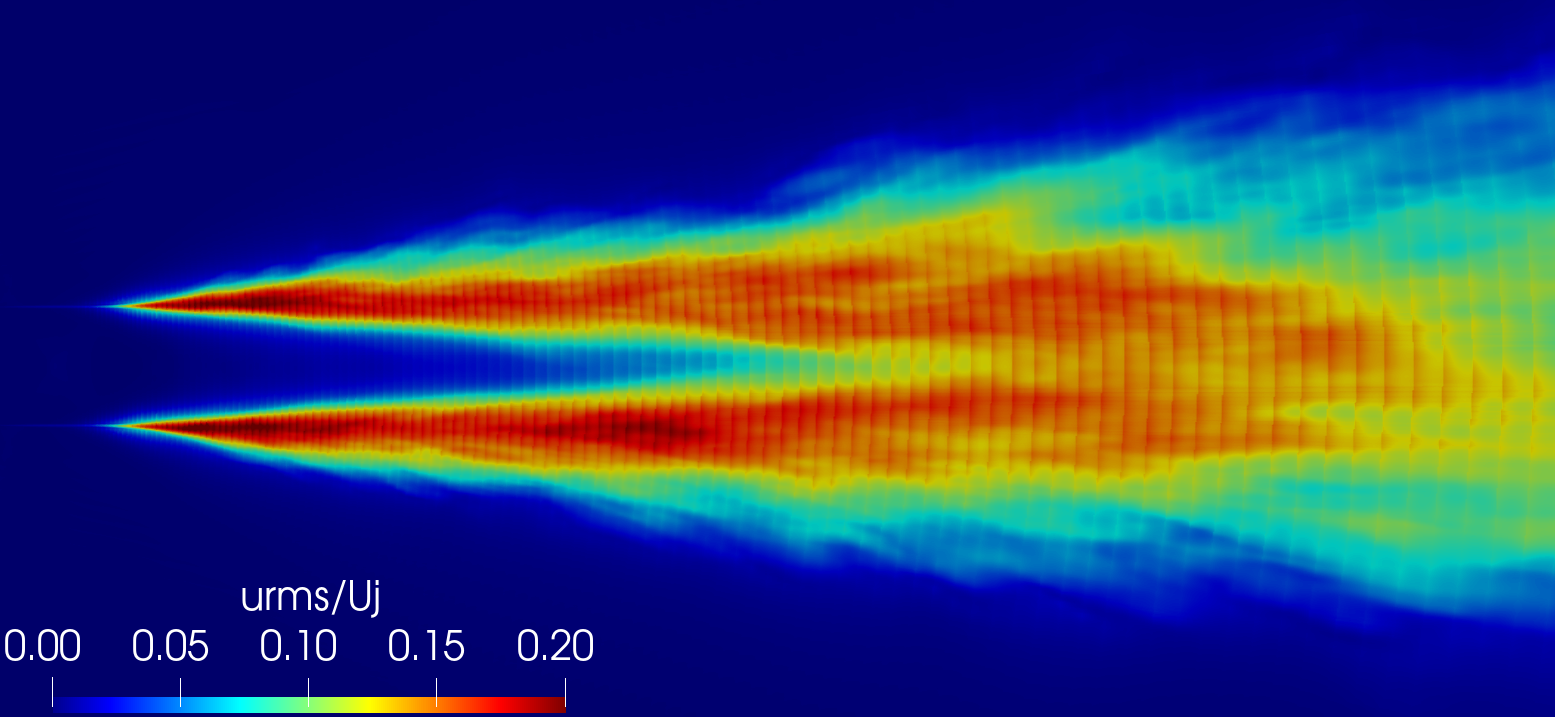}
	\label{res.vxrms_s1}
	}
\subfloat[S-2 simulation.]{
	\includegraphics[width=0.48\linewidth]
    {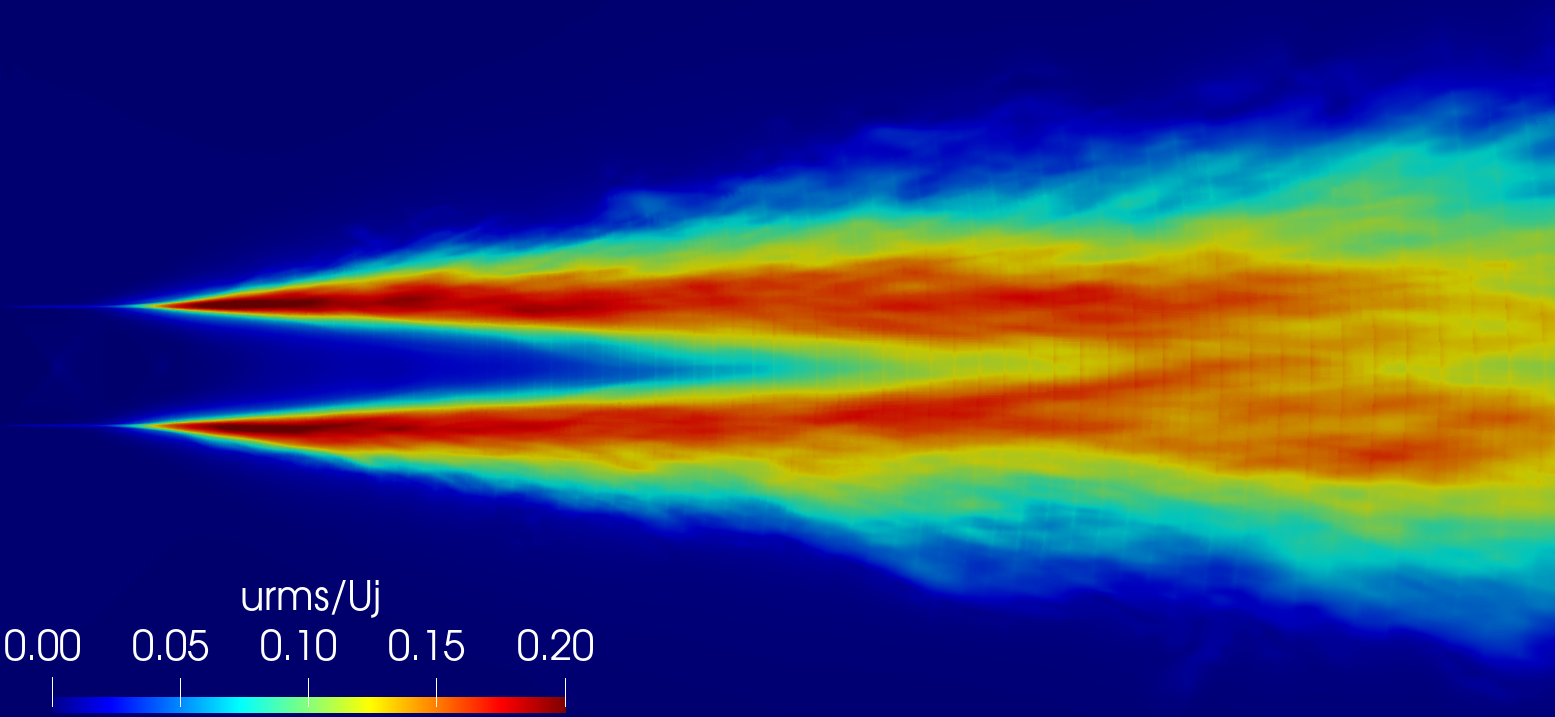}
	\label{res.vxrms_s2}
	}
\\
\subfloat[S-3 simulation.]{
	\includegraphics[width=0.48\linewidth]
    {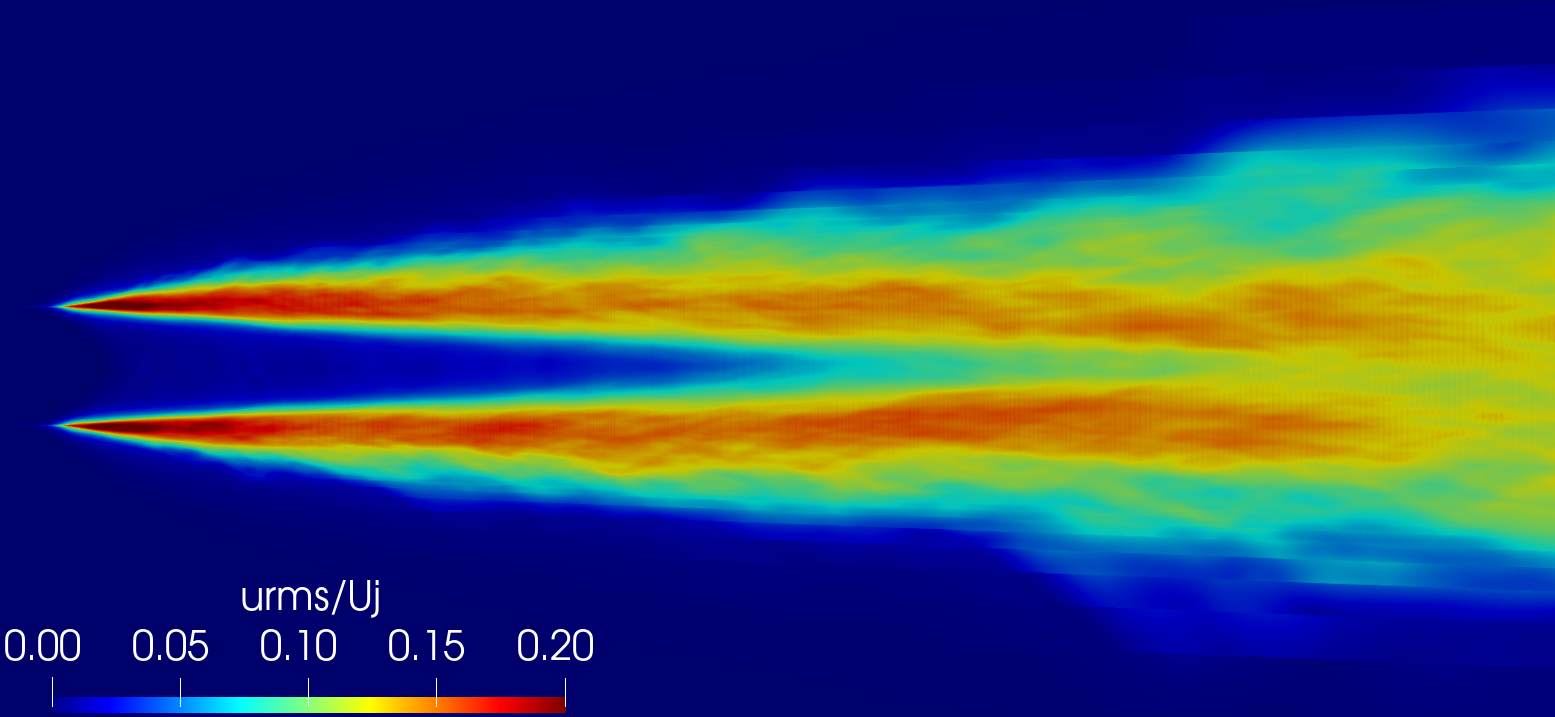}
	\label{res.vxrms_s3}
	}
\subfloat[S-4 simulation.]{
	\includegraphics[width=0.48\linewidth]
    {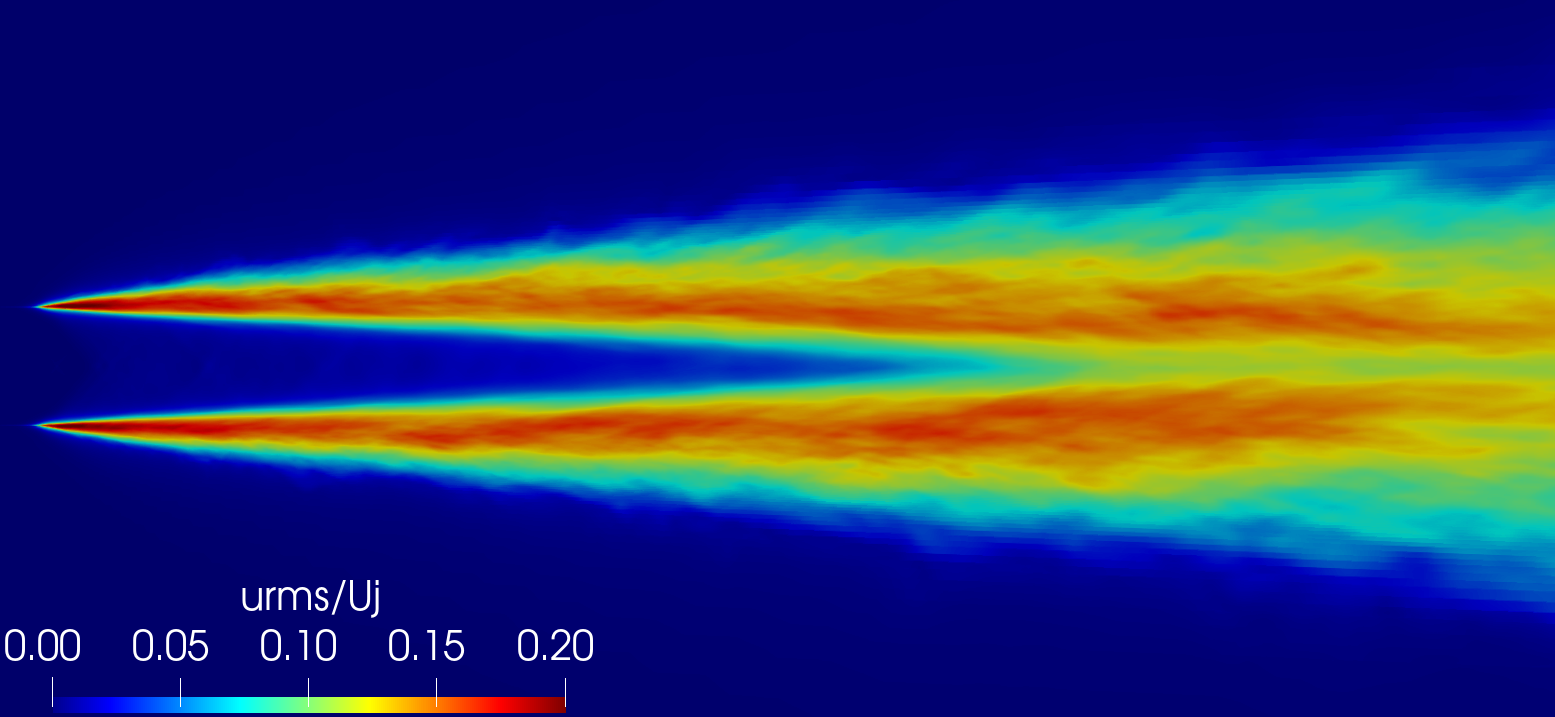}
	\label{res.vxrms_s4}
	}
\caption{Contours of root mean square non-dimensional longitudinal velocity 
         component fluctuations.}
\label{res.vxmrms}
\end{figure*}

A deeper investigation of the mixing layer can be performed by visualizing 
the contours of the root mean square (RMS) of the longitudinal 
velocity fluctuation, presented in Fig.~\ref{res.vxmrms}. Due to the
interaction between the jet flow and the ambient air, the turbulence levels
in the mixing layer are higher than in other parts of the flow, and the 
longitudinal velocity fluctuation presents information on the turbulence
levels. In the contours presented in Figs.~\ref{res.vxrms_s1} and
\ref{res.vxrms_s2}, from the S-1 and S-2 simulations, the values of the peak
of RMS of the longitudinal velocity component and the spreading of velocity
fluctuation in the longitudinal and transversal directions are similar. 
The velocity contours from the S-3 and S-4 simulations, Figs.~\ref{res.vxrms_s3}
and \ref{res.vxrms_s4}, show some differences in the mixing layer development.
The commencement of the development of the shear layer and the peak values 
occurs close to the inlet section. The contours present minor transversal 
spreading of velocity fluctuation, which can be observed by the more extended
region unaffected by the mixing layer in the centerline and the narrow region
with high fluctuation.

\begin{figure*}[htb!]
\centering
\subfloat[S-1 simulation.]{
	\includegraphics[width=0.48\linewidth]
    {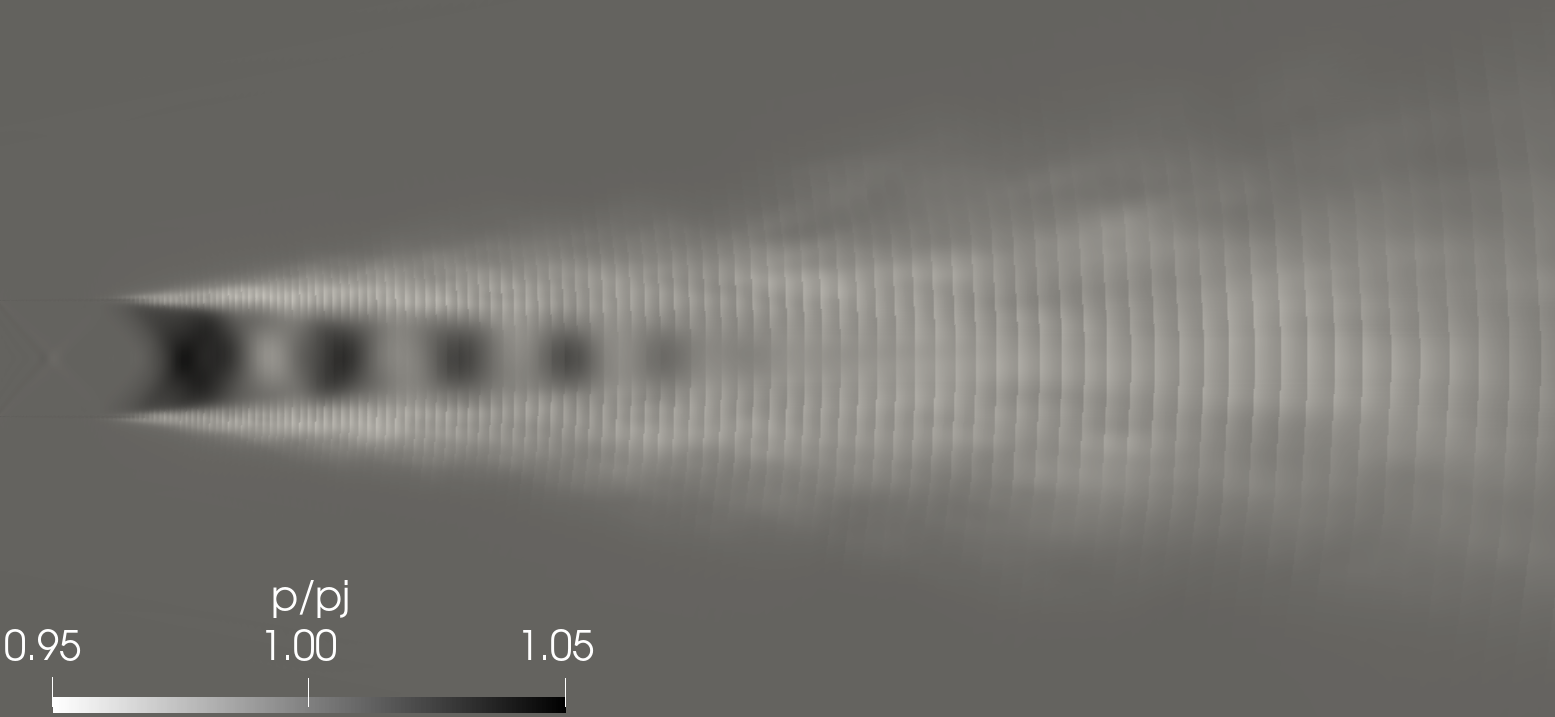}
	\label{res.presmean_s1}
	}
\subfloat[S-2 simulation.]{
	\includegraphics[width=0.48\linewidth]
    {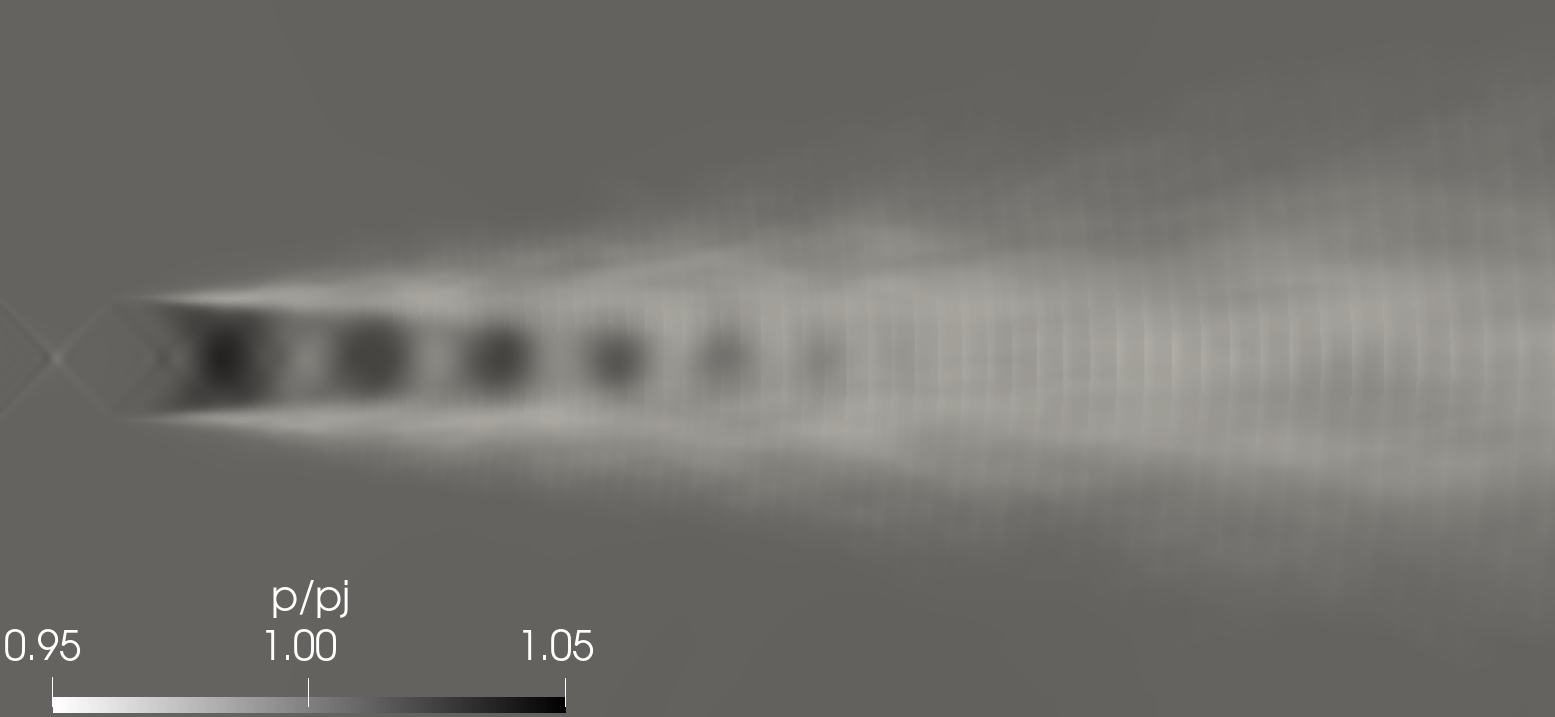}
	\label{res.presmean_s2}
	}
\\
\subfloat[S-3 simulation.]{
	\includegraphics[width=0.48\linewidth]
    {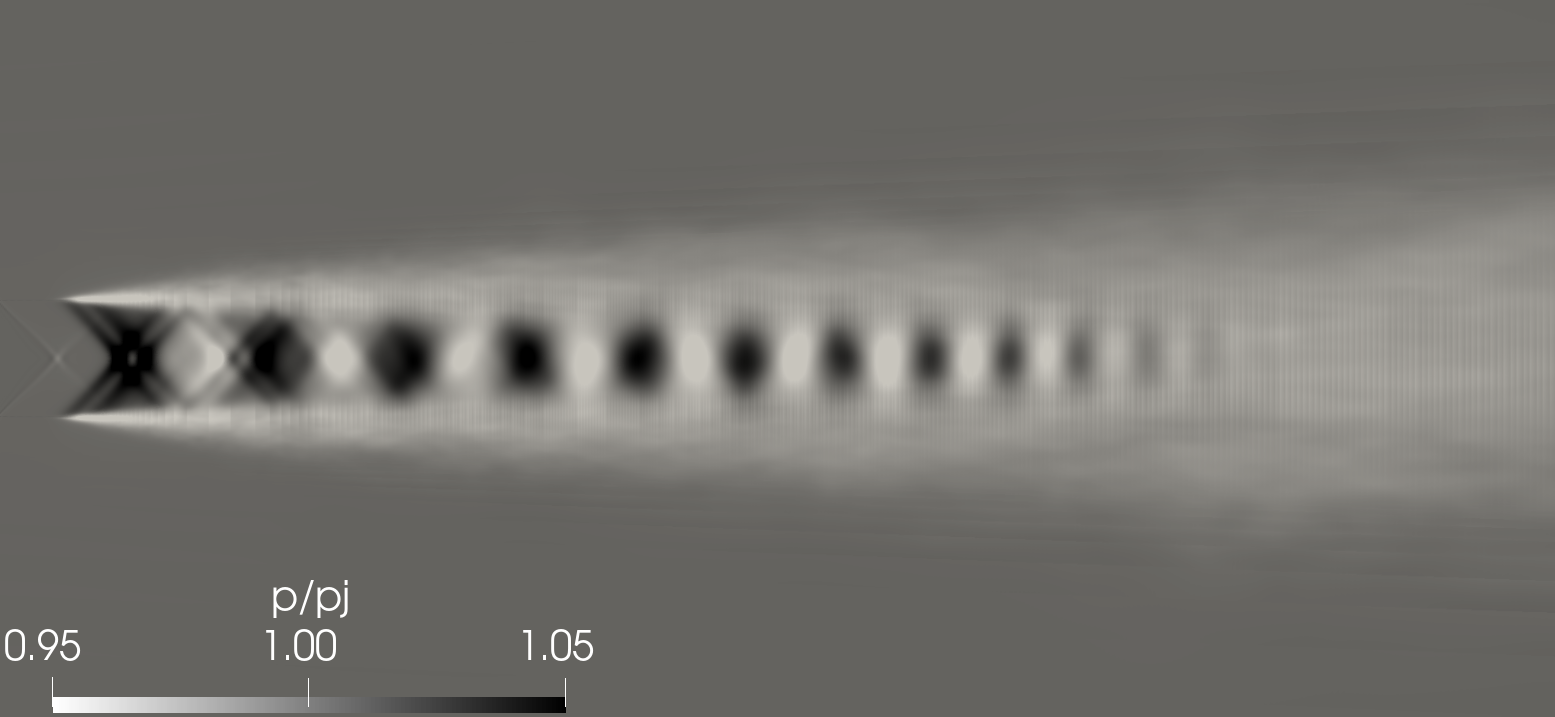}
	\label{res.presmean_s3}
	}
\subfloat[S-4 simulation.]{
	\includegraphics[width=0.48\linewidth]
    {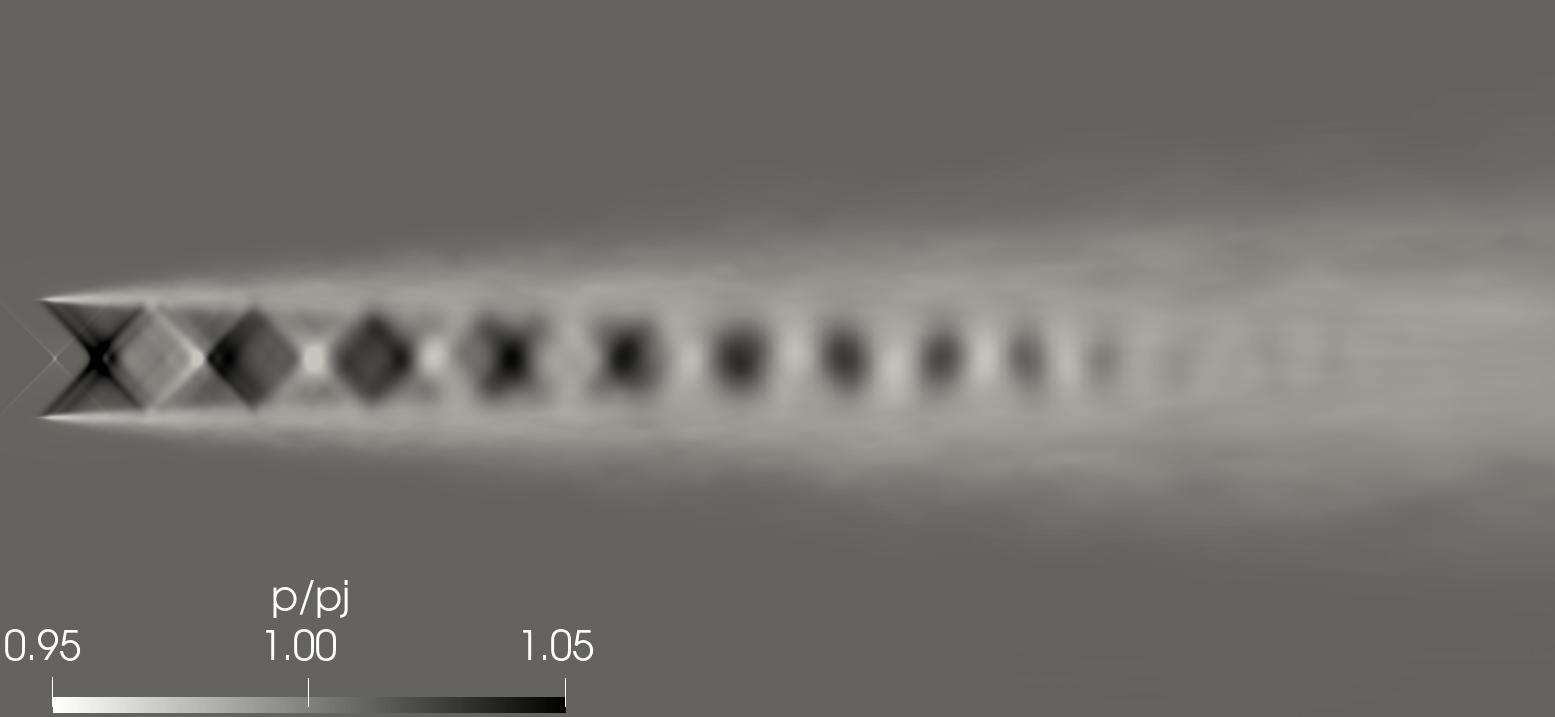}
	\label{res.presmean_s4}
	}
\caption{Mean non-dimensional pressure contours from the simulations performed.}
\label{res.presmean}
\end{figure*}

The contours of mean pressure highlight the differences in the simulations 
concerning the shock wave formation present in the jet core, 
Fig.~\ref{res.presmean}. The contours from the S-1 simulation, 
Fig.~\ref{res.presmean_s1}, indicate a non-uniform pressure distribution, 
which is associated with the lack of \textit{hp}-resolution of the simulation 
{in conjunction with the interpolation of the simulation data
to the set of probes used for the statistical analyses}.
The pressure contours from the S-2 simulation, Fig.~\ref{res.presmean_s2},
presents a smoother transition when compared to the contours from the S-1
simulation, and additional shock waves are present. Significant differences
appear when comparing the pressure contours from the S-3 and S-4 simulations,
Figs.~\ref{res.presmean_s3} and \ref{res.presmean_s4}, with the contours from
the S-1 and S-2 simulations, The shock waves present a larger number of
repetitions distributed over a larger region than the reproduced by the S-1
and S-2 simulations. The first shock wave is closer to the inlet section in
the S-3 simulation than the S-2 simulation contours. The first shock waves
from the S-3 and S-4 simulations present a smaller thickness when compared
to the pressure contours from the S-1 and S-2 simulations. The better-defined
shock waves are associated with a better resolution from the S-3 and S-4
simulations when compared to the S-1 and S-2 simulations. The pressure
contours from the S-4 simulation indicate thin shock waves compared to the
S-3 simulation. However, the shock waves from the S-3 simulation present
larger pressure differences between the repetitions than those of the S-4
simulation.

\subsection{Longitudinal Profiles of Velocity and Reynolds Stress Tensor}

\begin{figure*}[htb!]
\centering
%\subfloat[Mean longitudinal velocity component at centerline]{
\subfloat[Mean velocity distribution at the centerline.]{
	\includegraphics[trim = 0mm 0mm 10mm 5mm, clip, width=0.48\linewidth]
    {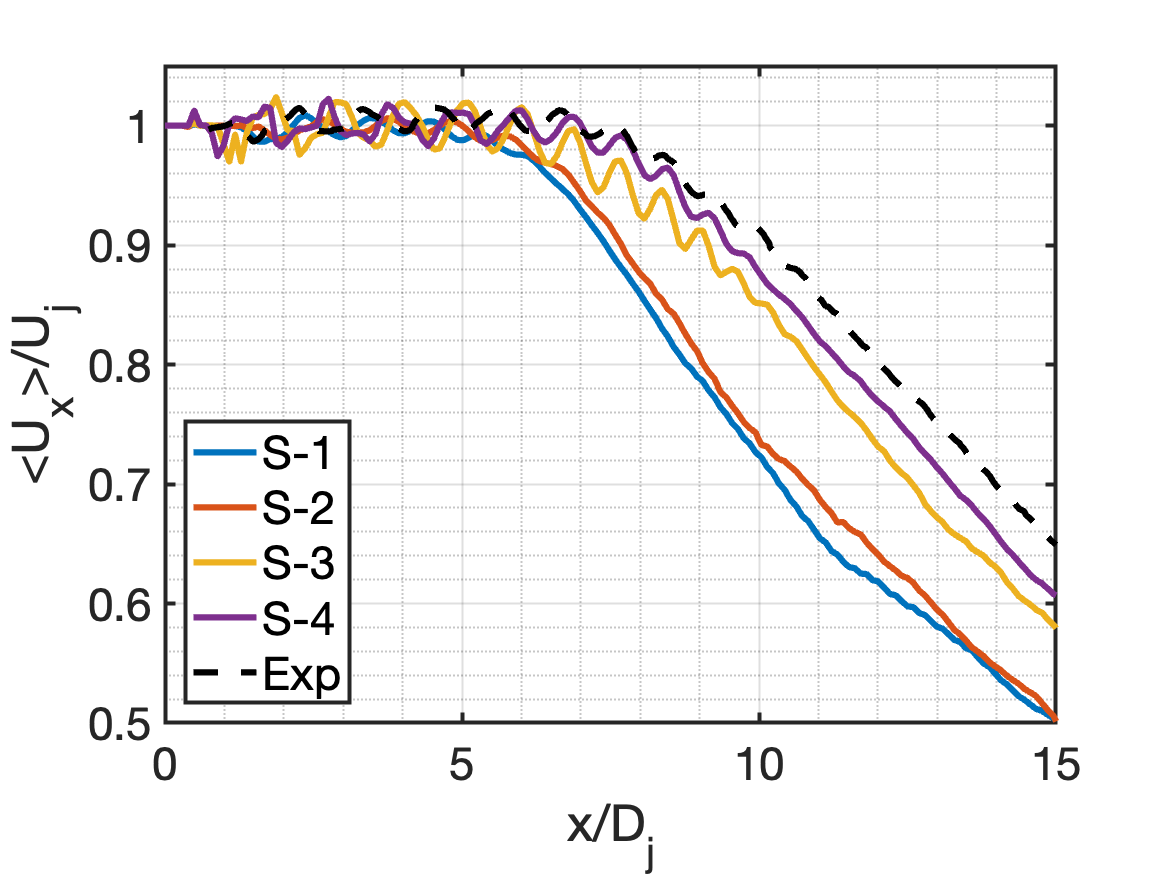}
	\label{res:velx_mean_cl}	
	}%
%\subfloat[RMS of longitudinal velocity fluctuation at centerline]{
\subfloat[RMS of velocity fluctuation distribution at the centerline.]{
    \includegraphics[trim = 0mm 0mm 10mm 5mm, clip, width=0.48\linewidth]
    {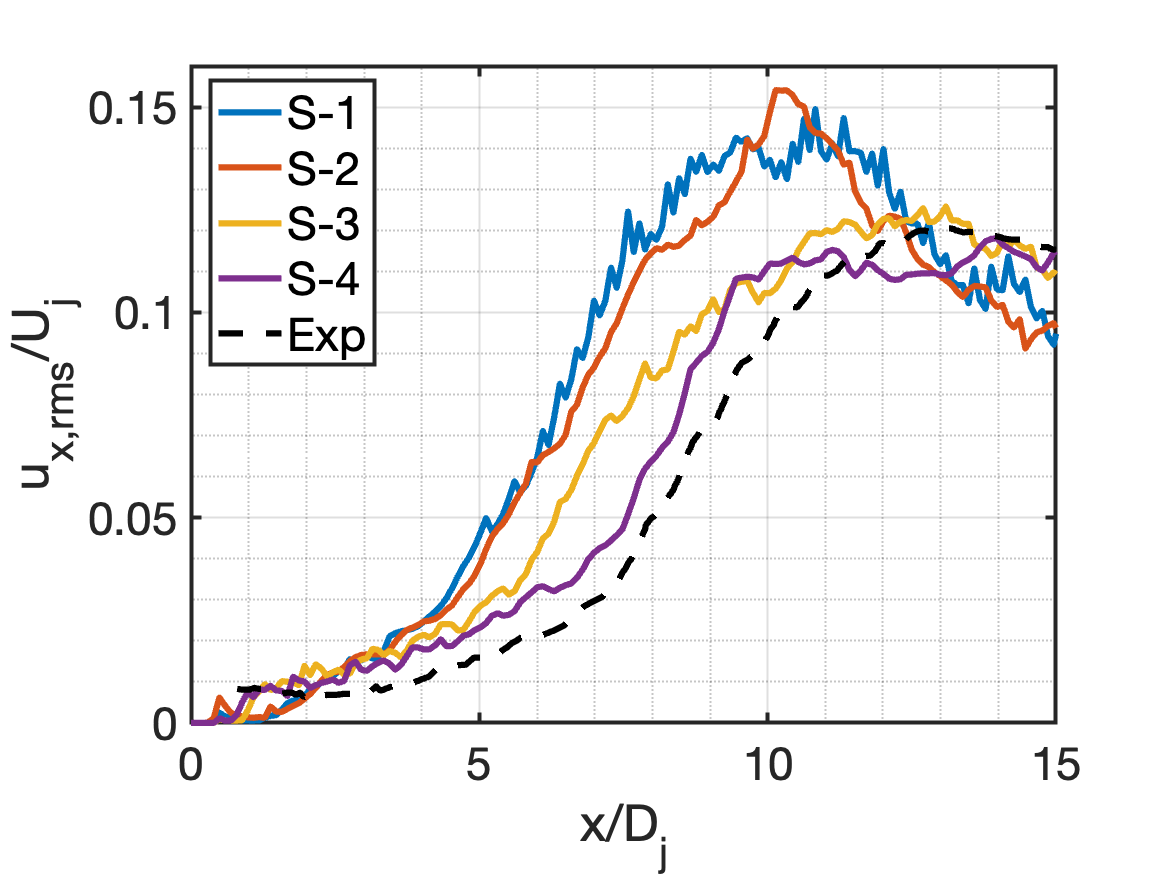}
	\label{res:velx_rms_cl}	
	}
%\newline
\\
%\subfloat[Mean longitudinal velocity component at lipline]{
\subfloat[Mean velocity distribution at the lipline.]{
	\includegraphics[trim = 0mm 0mm 10mm 5mm, clip, width=0.48\linewidth]
    {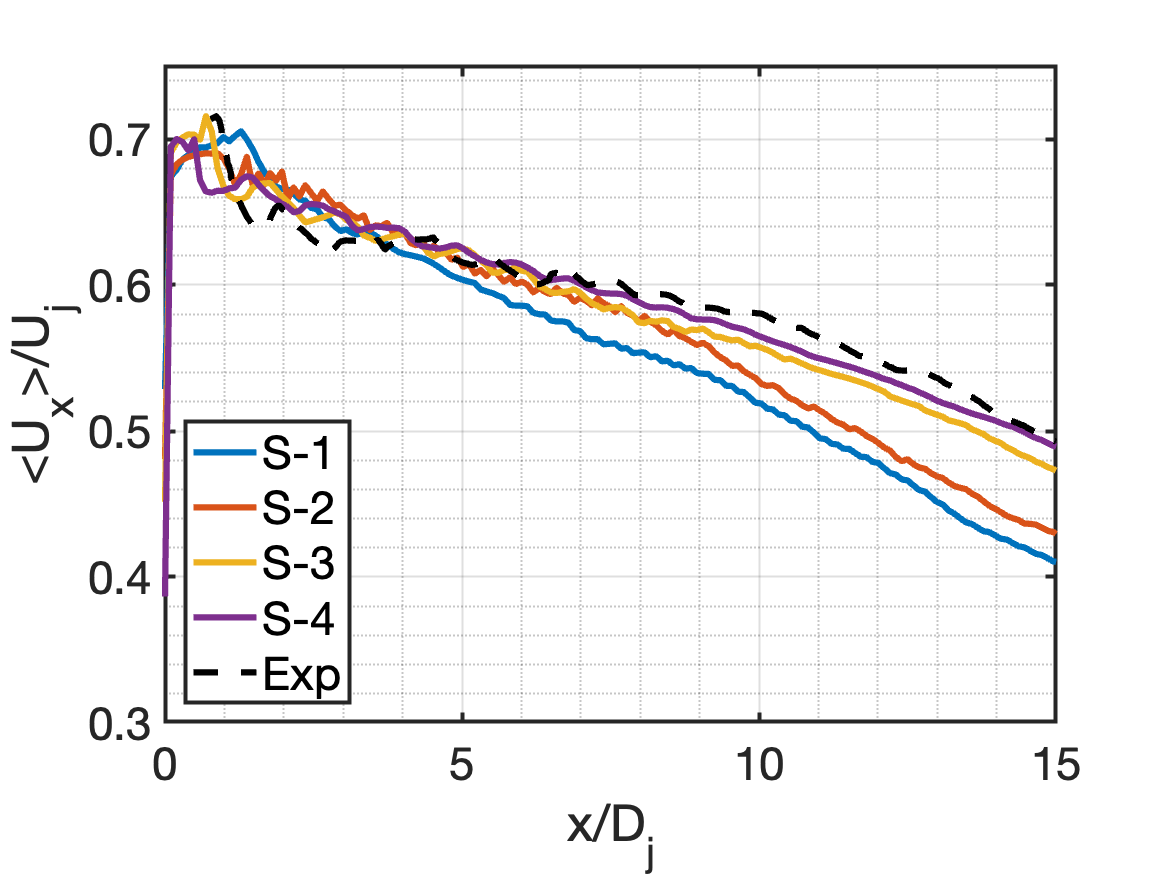}
	\label{res:velx_mean_ll}
	}
%\subfloat[RMS of longitudinal velocity fluctuation at lipline]{
\subfloat[RMS of velocity fluctuation distribution at the lipline.]{
	\includegraphics[trim = 0mm 0mm 10mm 5mm, clip, width=0.48\linewidth]
    {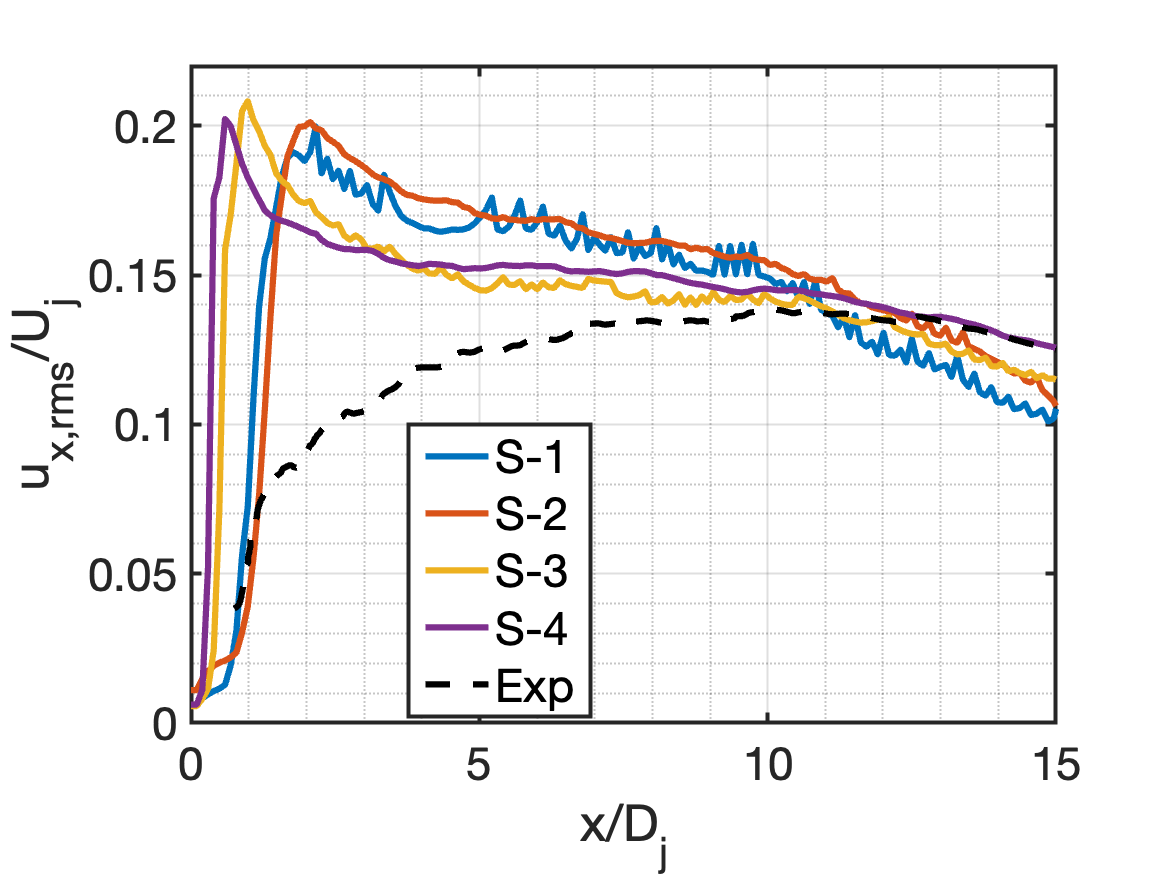}
	\label{res:velx_rms_ll}	
	}
\caption{Longitudinal distributions of the mean longitudinal velocity component
         and the rms of the longitudinal velocity fluctuation at the jet
         centerline, $r/D_j=0.0$, and lipline, $r/D_j=0.5$.}
\label{res:velx_cl_ll}
\end{figure*}

The velocity distribution of the four simulations is quantitatively 
compared with experimental data\cite{BridgesWernet2008} at the centerline 
and lipline, Fig.~\ref{res:velx_cl_ll}. Figures~\ref{res:velx_mean_cl} and 
\ref{res:velx_rms_cl} present the mean longitudinal velocity component 
distribution and the RMS of the longitudinal velocity fluctuation distribution 
at the centerline of the jet flow. The mean longitudinal velocity component 
distribution presents a general behavior between the simulations. Close to
the jet inlet section, inside the potential core of the jet flow, the velocity
values are only affected by the shock waves. Close to the potential core
length, the velocity distribution presents a negative slope 
associated with the mixing layer of the jet flow spreading the velocity from
the jet core. The S-1 and S-2 simulations present a shorter potential core
length when compared to the S-3 and S-4 simulations. The short potential
core length, which is close to the change of the velocity slope, occurs
close to $x/D_j=7.0$. The change in the slope from the experimental data 
is only observed downstream of the jet flow. The velocity distribution from
the S-2 simulation indicates a slightly longer potential core when compared 
with the S-1 simulation. The velocity distribution from the S-3 
and S-4 simulations are closer to the experimental reference when compared 
to the S-1 and S-2 simulations. The potential core length is close to 
$x/D_j=8.0$, and there is a visible influence of the set of shock waves 
through the jagged distribution in the potential core. In the S-1 and S-2
simulations, a series of shock waves with small velocity peak to peak
velocity amplitude can be observed up to the section $x/D_j \approx 
6$, while in the S-3 simulation the set of shock waves extend up to $x/D_j 
\approx 10$. The S-4 simulation presents a better agreement with experimental
data when compared to the other simulations. The potential core from the S-4
simulation is longer than the potential core from the S-3 simulation. The
peak-to-peak velocity amplitude in the potential core from the S-4 simulation
is minor than the S-3 simulation. The potential core length is used
as a parameter to assess the error of the numerical simulations. The results
of the error assessment are presented in %Fig.~\ref{fig:error_plot}.
Tab.~\ref{tab:potcore}.

The RMS of the longitudinal velocity fluctuation distribution in the 
centerline, Fig.~\ref{res:velx_rms_cl}, indicates an improvement from the 
simulation results with increased resolution compared to the experimental 
reference. The longitudinal velocity fluctuation distributions from the S-1 
and S-2 simulations present an increase in the velocity slope occurring
closer to the jet inlet section than the S-3 and S-4 simulations and the
experimental data. The velocity fluctuation distribution from the S-1
simulation presents oscillation in the velocity distribution that may
be associated with employing a relatively coarse mesh and linear polynomial
degree with the probe processing employed to interpolate data in the flow 
field. The results with large mesh refinement or third-order accurate scheme
does not present a similar oscillation. The velocity fluctuation of the S-3
simulation presents a reduced maximum RMS value when compared to the maximum
values from the S-1 and S-2 simulations. The reduced maximum values from the
velocity fluctuation distribution and the change in the velocity slope 
occurring far from the inlet section led the velocity fluctuation distribution
to approach the experimental reference. In the present property evaluated, the
S-4 simulation provided results that better match the experimental reference
than the other numerical simulations.

Figures~\ref{res:velx_mean_ll} and \ref{res:velx_rms_ll} present the 
distribution of the mean longitudinal velocity component and the RMS of the 
longitudinal velocity fluctuation at the lipline of the jet flow. The 
mean longitudinal velocity component distribution at the lipline,
Fig.~\ref{res:velx_mean_ll}, indicate that in the region $0 < x/D_j <5.0$
the four simulations and the experimental data present similar values. The
velocity distributions from the S-1 and S-2 simulations present a change
in the velocity slope, which resulted in smaller mean velocity velocities.
The change in the velocity slope occurs after the position $x/D_j \approx 5.0$
for the S-1 simulation and for $x/D_j \approx 7.0$ for the S-2 simulations. 
The mean velocity distributions from the S-3 and S-4 simulations indicate 
similar velocity values along the longitudinal region evaluated.

The distributions presented for the RMS of the longitudinal velocity
fluctuation at the lipline, Fig.~\ref{res:velx_rms_ll}, have a different
behavior than the previous analyses from Figs. \ref{res:velx_mean_cl}, 
\ref{res:velx_rms_cl}, and \ref{res:velx_mean_ll}. The four simulations 
present an abrupt change in the slope of the velocity fluctuation distribution
in the region $0.0 < x/D_j < 2.0$. After the abrupt increase in the velocity
fluctuation levels, the distributions reach maximum velocity fluctuation 
values significantly larger than those from the experimental
reference. After reaching the maximum value, an inversion in the velocity
distribution slope is observed. The velocity slope inversion led to the S-1,
S-2 and S-3 simulations to present velocity fluctuation levels above the
experimental reference for $x/D_j>12.0$. The S-3 and S-4 simulations 
present a similar behavior of velocity fluctuation in the lipline. In these
two simulations, the abrupt change in the velocity slope occurs in a station
even closer to the jet inlet section, $x/D_j \approx 0.5$, when compared to 
the change in the slope from the S-1 and S-2 simulations, $x/D_j \approx 
1.0$, and reaches its peak close to $x/D_j=1.0$. The velocity fluctuation
distribution from the S-4 simulation never reaches velocity values below 
those of experimental data. The velocity fluctuation distribution shows that
the growth in the velocity fluctuation occurs closer to the jet inlet section
with increased resolution, except for the S-1 and S-2 simulations. Even 
though the S-1 simulation is simulated with a second-order accurate 
spatial discretization, it has a more refined mesh than the S-2 simulation, 
which may be responsible for allowing the simulation to capture smaller 
eddies than those of the S-2 simulation. Once the same mesh is employed with 
different spatial discretization accuracy, the third-order accurate method 
could better reproduce the turbulent eddies than the second-order accurate,
allowing the simulation to anticipate the transition mechanism. The results
from the velocity fluctuation in the lipline indicates important differences
to the experimental data, which are associated with the inviscid profile 
imposed in the jet inlet section and not related to a lack of accuracy of the 
numerical method. 

% \begin{figure}
%     \centering
%     \includegraphics[width=0.48\linewidth]{Figures/error_plot.png}
%     \caption{Absolute error from the numerical simulations compared to the
%              experimental reference based on the potential core length.}
%     \label{fig:error_plot}
% \end{figure}

\begin{table}[htb!]
\caption{Summary of potential core length information for the four numerical
         simulations performed.}
\centering
\begin{tabular}{c c c} \hline 
Simulation     & Potential core & Error to  \\
  & length ($x/D_j$) & experimental data ($\%$) \\ \hline
 S-1  & 6.6 & 25.0 \\
 S-2  & 6.9 & 21.6 \\
 S-3  & 7.7 & 12.5 \\
 S-4  & 8.5 & 3.4 \\ \hline 
\end{tabular}
\label{tab:potcore}
\end{table}

\begin{figure*}[htb!]
\centering
%\subfloat[Mean longitudinal velocity component at centerline]{
\subfloat[S1 simulation along the centerline.]{
	\includegraphics[trim = 0mm 0mm 10mm 5mm, clip, width=0.49\linewidth]
    {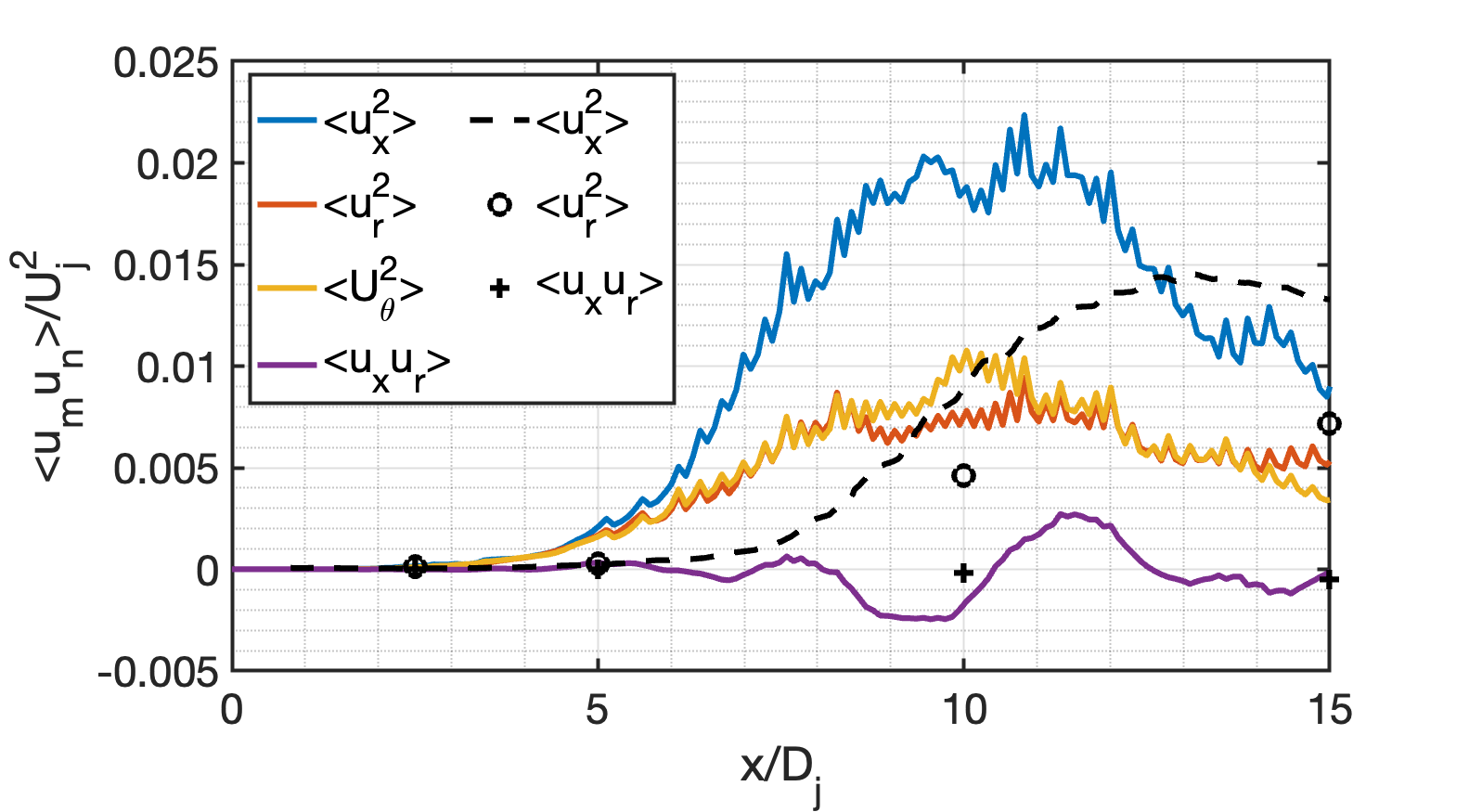}
	\label{res:Re_stress_cl_s1}	
	}%
%\subfloat[RMS of longitudinal velocity fluctuation at centerline]{
\subfloat[S1 simulation along the lipline.]{
	\includegraphics[trim = 0mm 0mm 10mm 5mm, clip, width=0.49\linewidth]
    {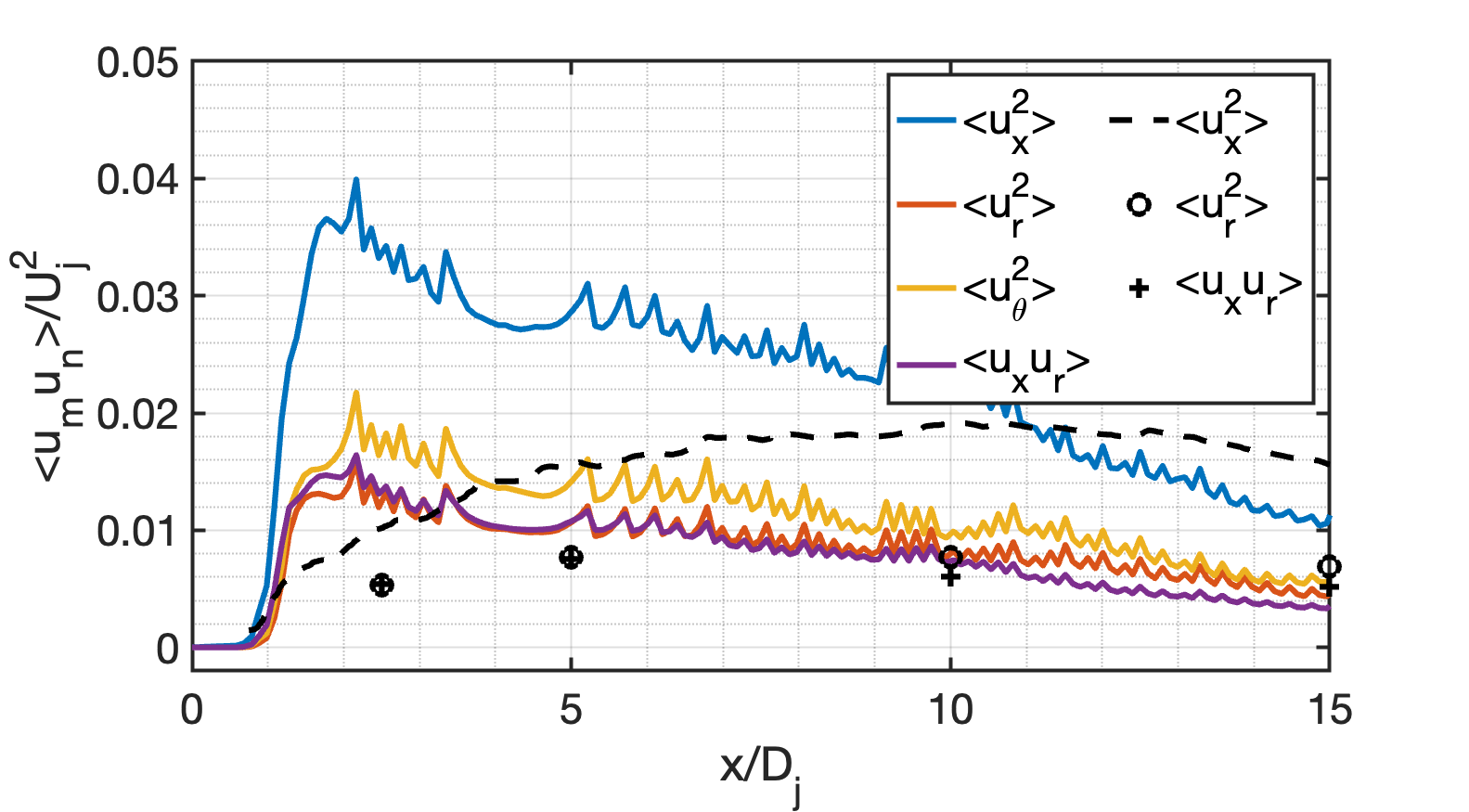}
	\label{res:Re_stress_ll_s1}	
	}
\\
\subfloat[S2 simulation along the centerline.]{
	\includegraphics[trim = 0mm 0mm 10mm 5mm, clip, width=0.49\linewidth]
    {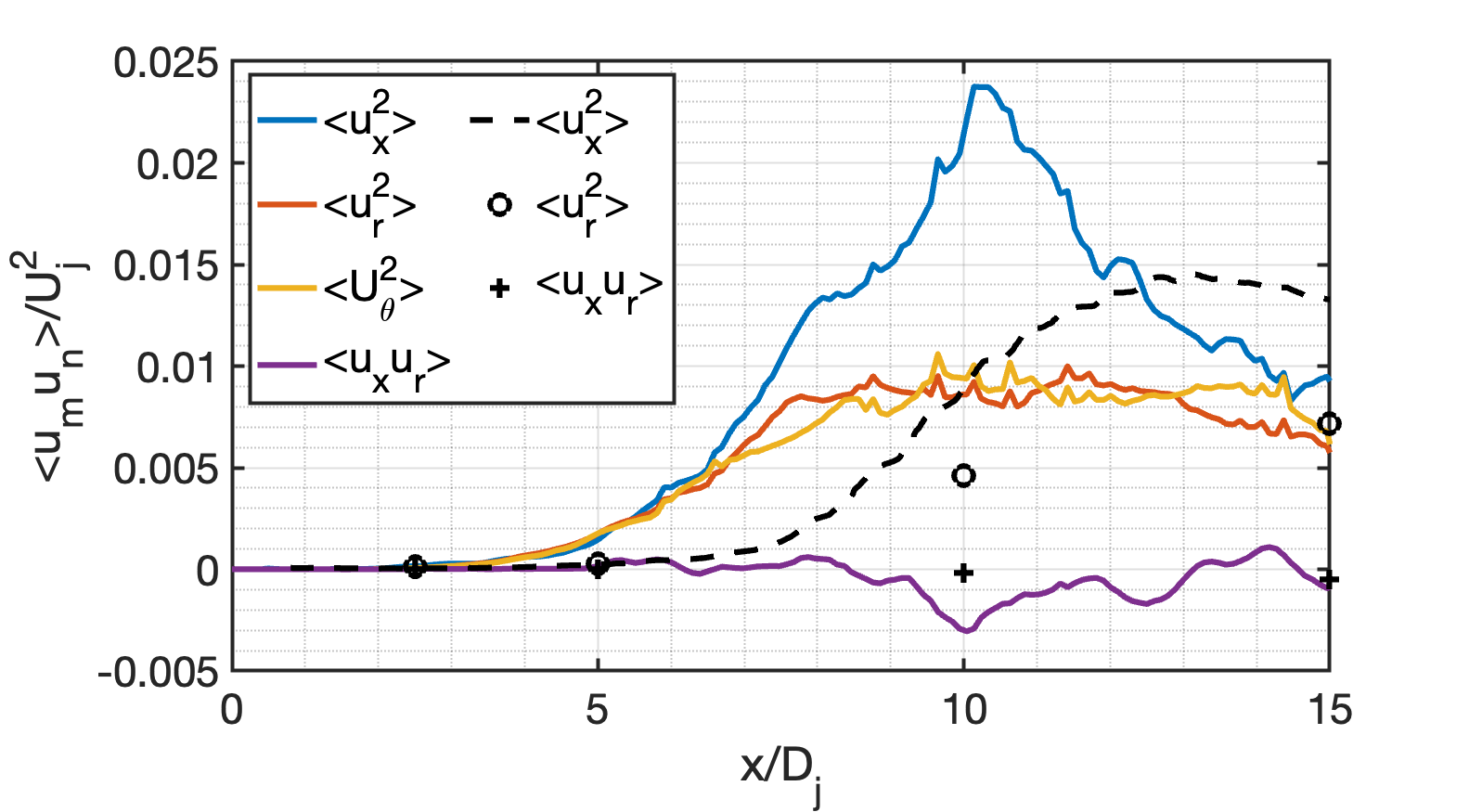}
	\label{res:Re_stress_cl_s2}	
	}%
%\subfloat[RMS of longitudinal velocity fluctuation at centerline]{
\subfloat[S2 simulation along the lipline.]{
	\includegraphics[trim = 0mm 0mm 10mm 5mm, clip, width=0.49\linewidth]
    {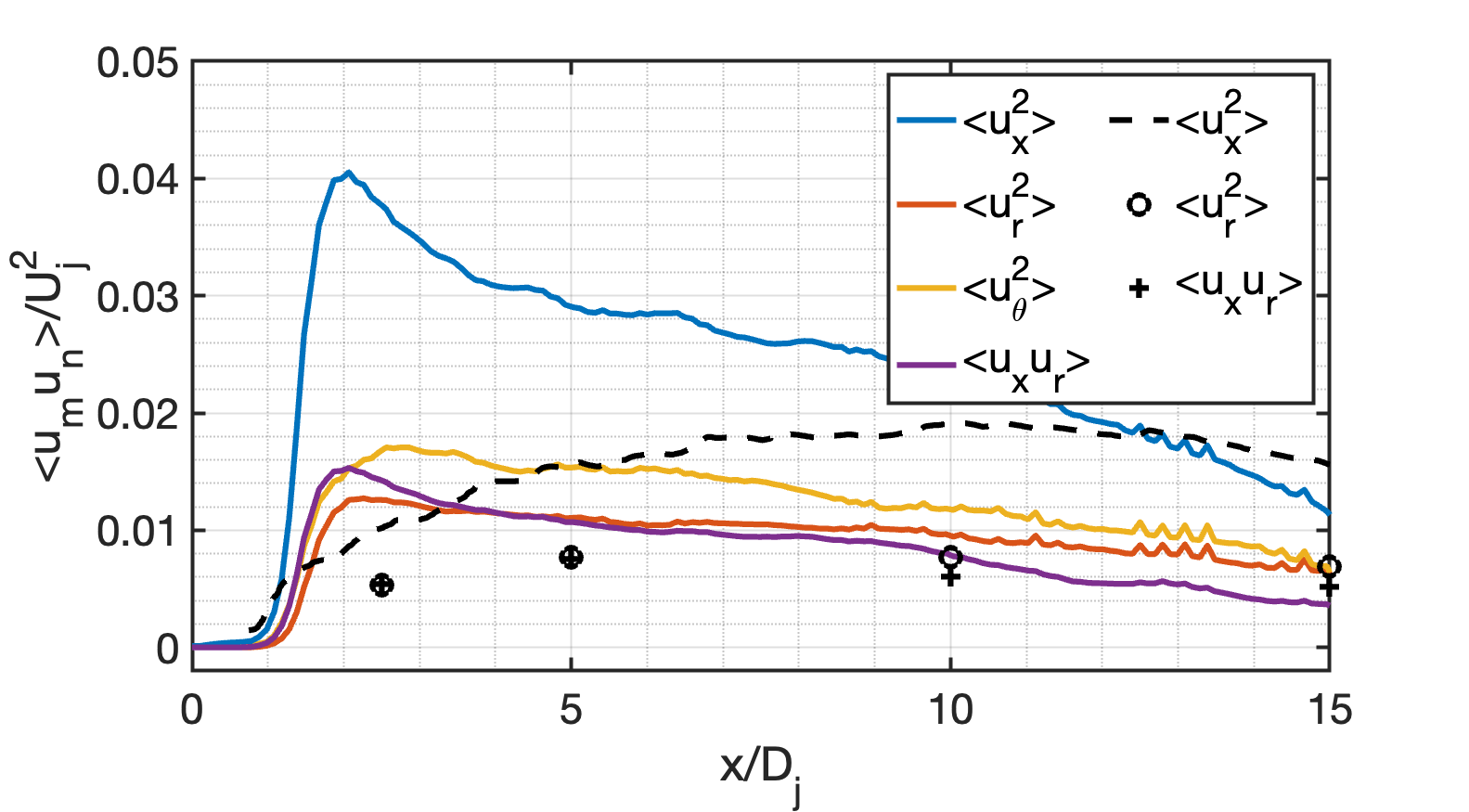}
	\label{res:Re_stress_ll_s2}	
	}
\\
\subfloat[S3 simulation along the  centerline.]{
	\includegraphics[trim = 0mm 0mm 10mm 5mm, clip, width=0.49\linewidth]
    {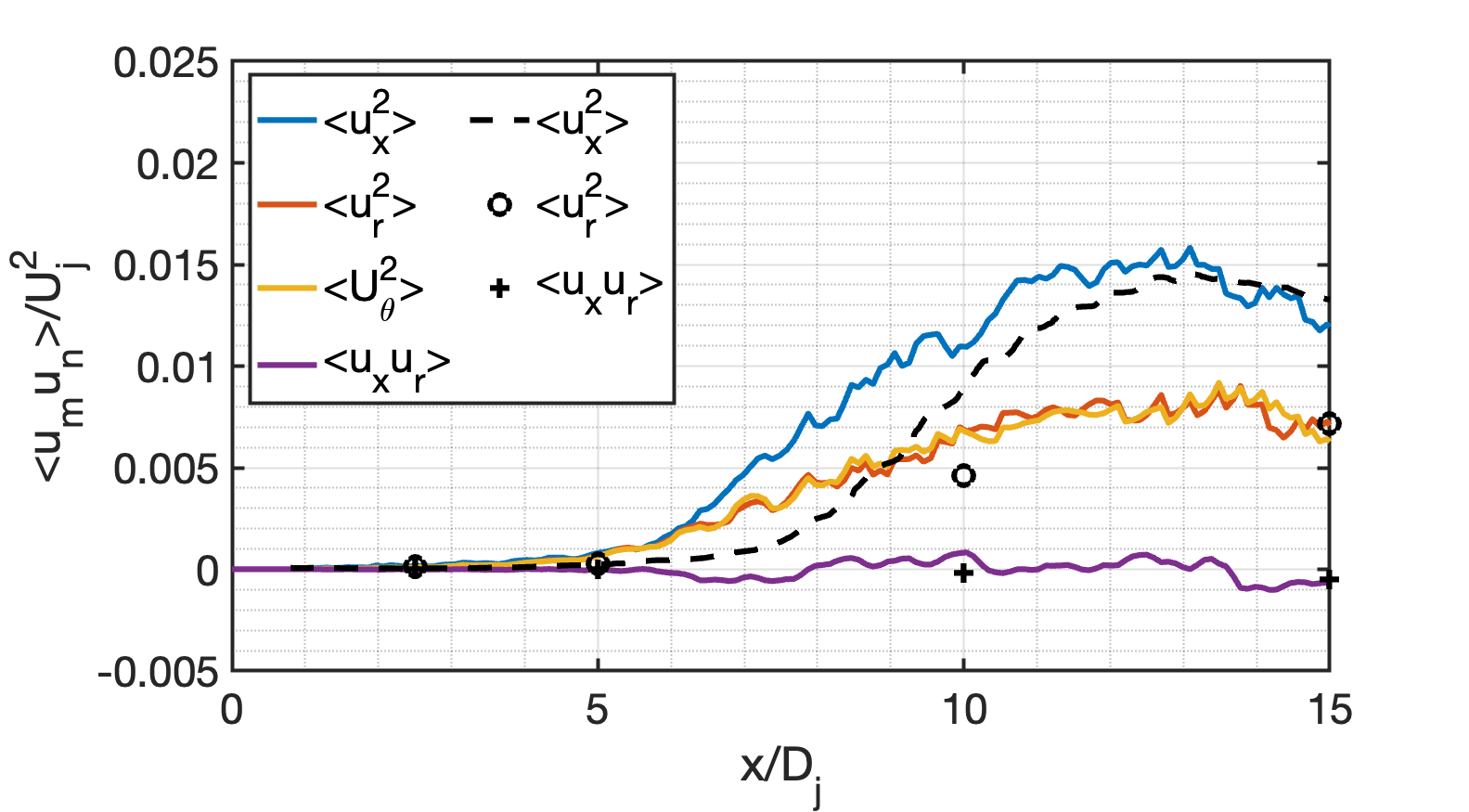}
	\label{res:Re_stress_cl_s3}	
	}%
%\subfloat[RMS of longitudinal velocity fluctuation at centerline]{
\subfloat[S3 simulation along the lipline.]{
	\includegraphics[trim = 0mm 0mm 10mm 5mm, clip, width=0.49\linewidth]
    {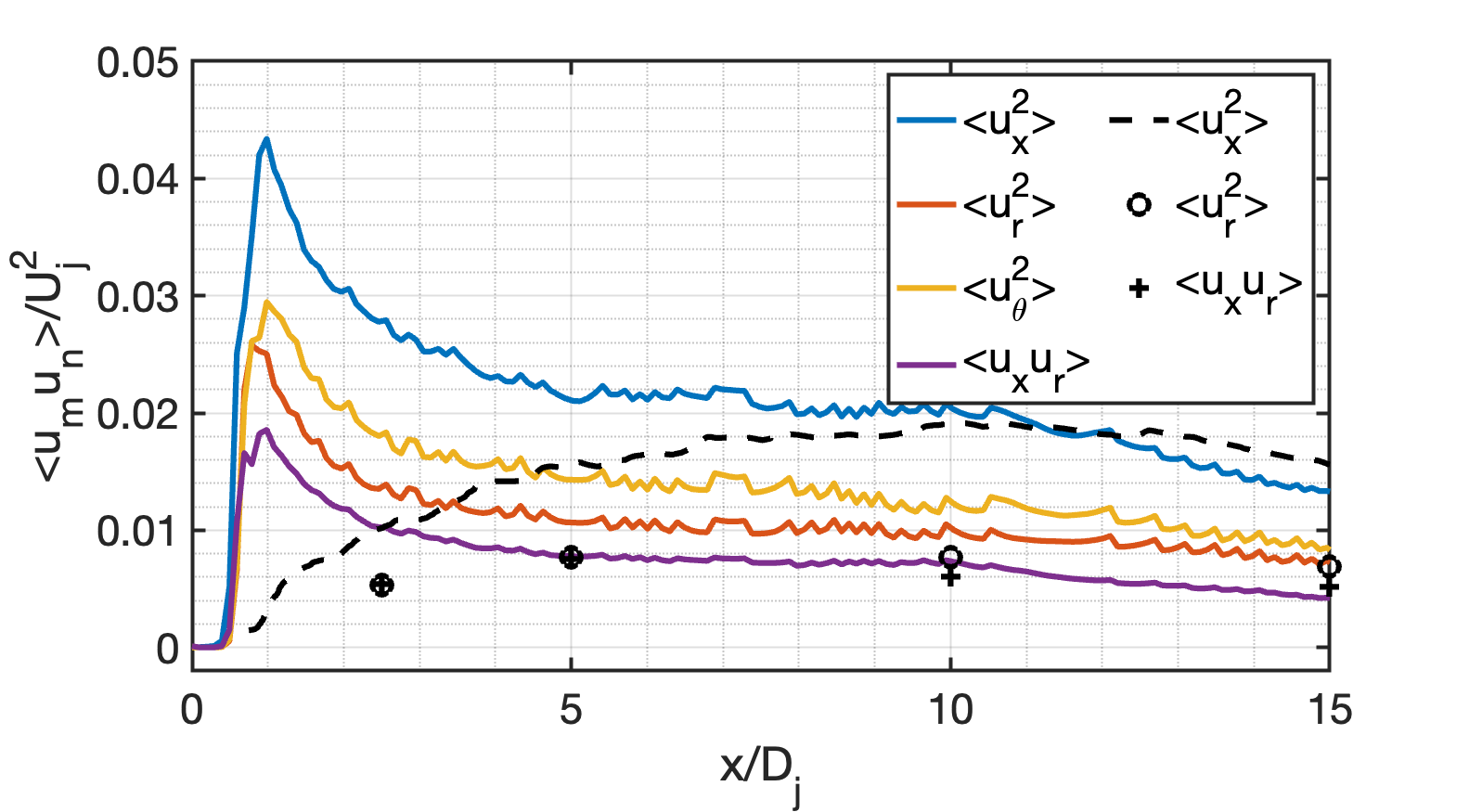}
	\label{res:Re_stress_ll_s3}	
	}
\\
\subfloat[S4 simulation along the centerline.]{
	\includegraphics[trim = 0mm 0mm 10mm 5mm, clip, width=0.49\linewidth]
    {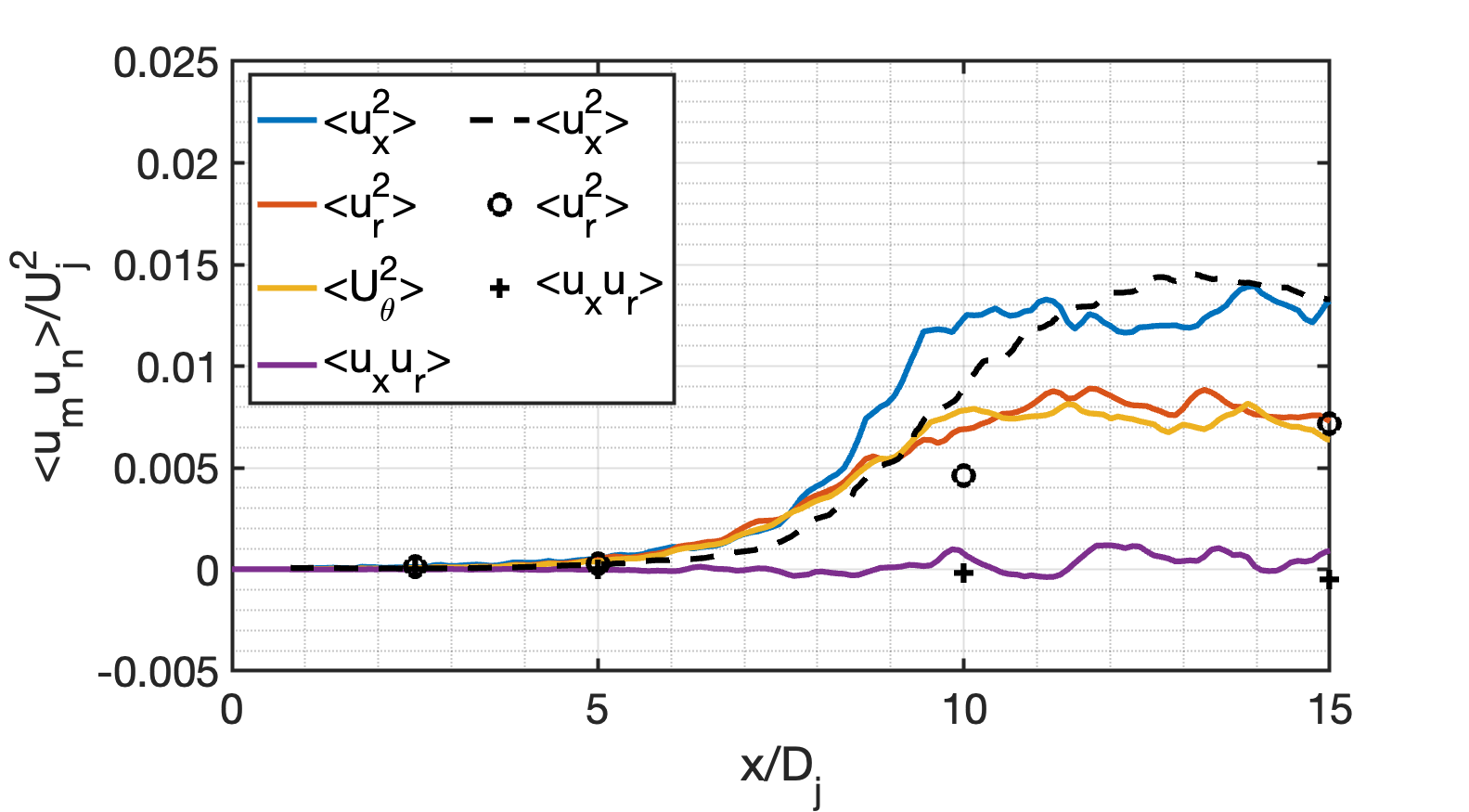}
	\label{res:Re_stress_cl_s4}	
	}%
%\subfloat[RMS of longitudinal velocity fluctuation at centerline]{
\subfloat[S4 simulation along the  lipline.]{
	\includegraphics[trim = 0mm 0mm 10mm 5mm, clip, width=0.49\linewidth]
    {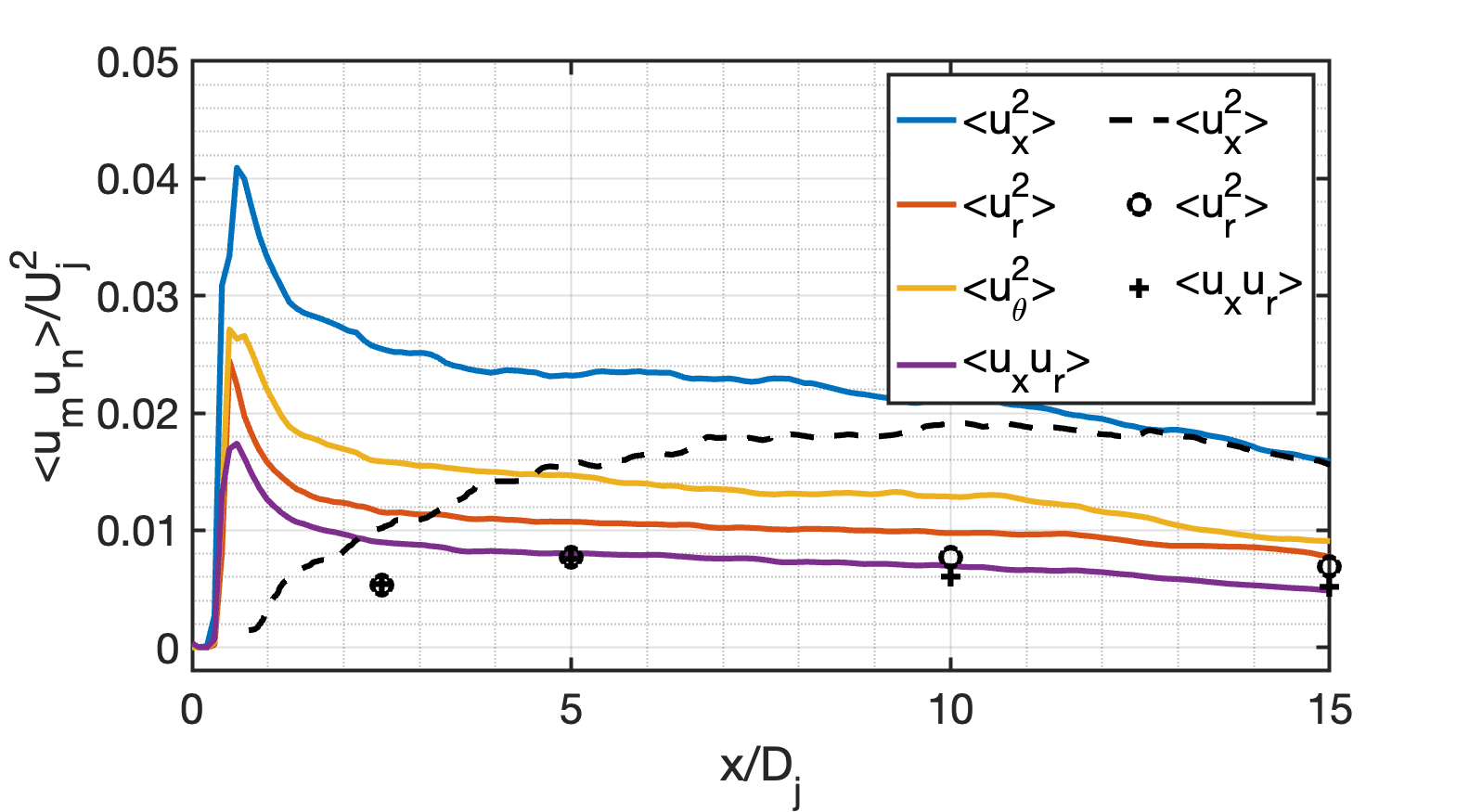}
	\label{res:Re_stress_ll_s4}	
	}
\caption{Reynolds stress tensor components longitudinal distributions from the
         four simulations analyzed at the jet centerline, $r/D_j=0.0$, and the jet lipline, $r/D_j=0.5$.}
\label{res:Re_stress}
\end{figure*}

Figure~\ref{res:Re_stress} presents the Reynolds stress tensor components 
at the centerline and lipline of the jet flow. Differently from the 
previously presented profiles, in Fig.\ \ref{res:Re_stress} each chart 
presents the numerical results from one simulation. In 
Figs.~\ref{res:Re_stress_cl_s1}, \ref{res:Re_stress_cl_s2},
\ref{res:Re_stress_cl_s3}, and \ref{res:Re_stress_cl_s4} the Reynolds stress
tensor components from the S-1 to S-4 simulations at the jet centerline are
presented, and Figs.~\ref{res:Re_stress_ll_s1}, \ref{res:Re_stress_ll_s2},
\ref{res:Re_stress_ll_s3}, and \ref{res:Re_stress_ll_s4} present the Reynolds
stress tensor components from the S-1 to S-4 simulations at the jet lipline. 
The longitudinal velocity fluctuation profiles are the same as presented in 
Figs.~\ref{res:velx_rms_cl} and \ref{res:velx_rms_ll} squared, and
they are left to reappear in the charts presented in Fig.~\ref{res:Re_stress}
to simplify the comparison. The Reynolds stress tensor components distribution
from the S-1 simulations present the same oscillations observed in 
Fig.~\ref{res:velx_rms_cl} and Fig.~\ref{res:velx_rms_ll}, which may be
associated probe processing used in the data extraction from the numerical
simulations. In the Reynolds stress tensor components distributions at the jet
centerline from the S-1 and S-2 simulations, the increase in the three velocity
fluctuations occurs closer to the jet inlet section than observed in the
experimental reference. The values of the longitudinal fluctuation component
present a larger increase than the other components, which, after the maximum
values, present an inversion in the Reynolds stress tensor slope until reach 
values closer to the radial and azimuthal components at $x/D_j=15.0$. The
Reynolds stress tensor components distribution from the S-3 and S-4 simulations
present a better agreement with experimental data than the distributions from 
the S-1 and S-2 simulations. The increase in the values of the Reynolds stress
tensor components occur farther from the inlet section than in the S-1 and 
S-2 simulations. It can be observed that, except for the longitudinal 
velocity fluctuation component the from S-1 and S-2 simulations close to 
station $x/D_j=15.0$, the components of the Reynolds stress tensor agrees
with experimental data.

The profiles of Reynolds stress tensor components at the lipline indicate
that its peak occurs in a station closer to the jet inlet section for the 
S-3 and S-4 simulations than the S-1 and S-2 simulations due to the higher
mesh refinement. The higher mesh refinement allows the S-3 and S-4 
simulations to capture small eddies and anticipate the flow transition. 
The shear-stress tensor component distribution from the S-1 and S-2 
simulations present a similar level of the radial component, while the S-3
and S-4 simulations present different velocity levels between all the
components. The comparison with experimental data indicates that close to 
the inlet section, the shear-stress and radial components should present
similar results that split with the progress in the flow field. The 
simulations did not capture the behavior. The S-4 calculation is 
the one that could get closer to the experimental data when compared to the
other calculations. However, it is of most importance to mention that the 
inlet velocity profile is imposed as an inviscid one, which plays an 
essential role in the turbulent transition and, consequently, also in the
Reynolds stress tensor components. One should notice that the article's goal
is to study the effects of resolution on the calculation, and a boundary 
condition comparison is out of the scope of the current study.

\subsection{Radial Profiles of Velocity and Reynolds Stress Tensor}

\begin{figure}[htb!]
\centering
%\subfloat[Mean longitudinal velocity component at $x/D_j=2.5$]{
\subfloat[]{
	\includegraphics[trim = 20mm 0mm 15mm 0mm, clip, height=0.26\linewidth]
    {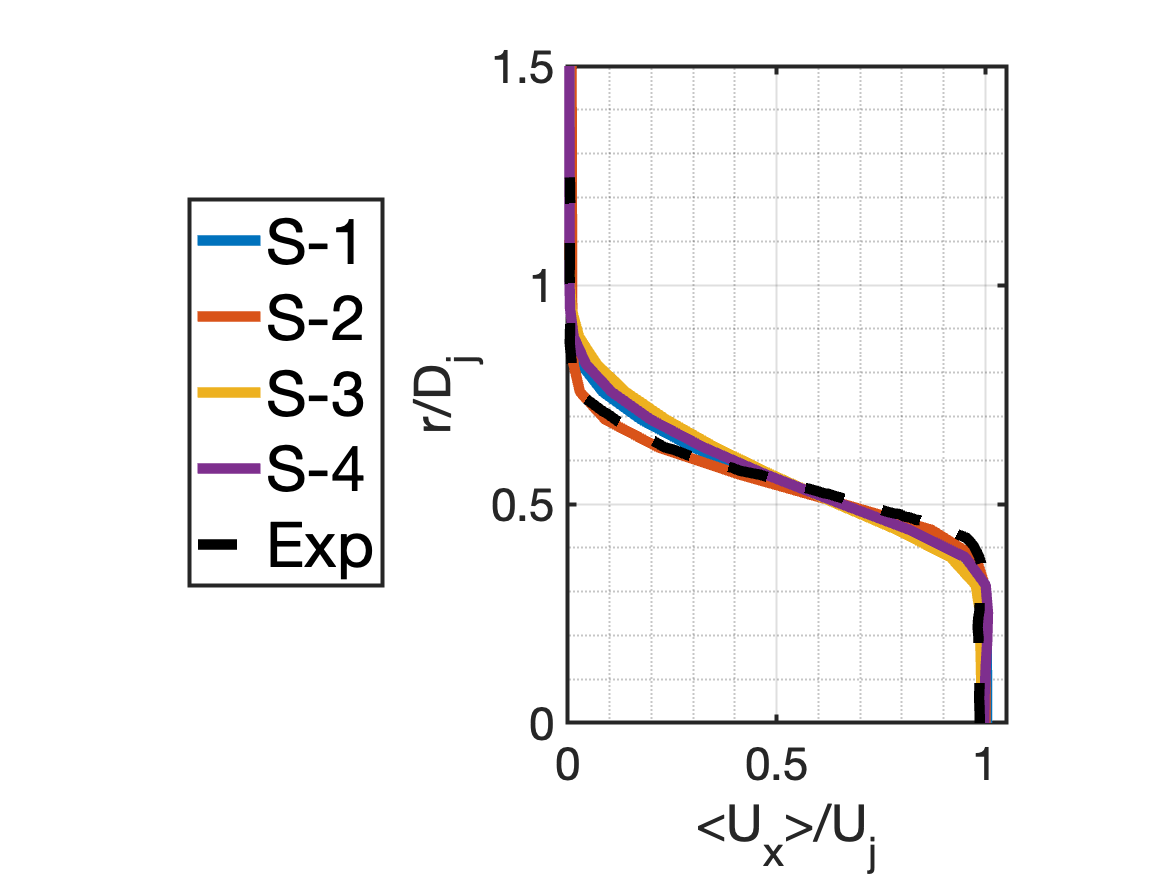}
	\label{res:radial_vxmean1}	
	}
%\subfloat[Mean longitudinal velocity component at $x/D_j=5$]{
\subfloat[]{
    \includegraphics[trim = 30mm 0mm 40mm 0mm, clip, height=0.26\linewidth]
    {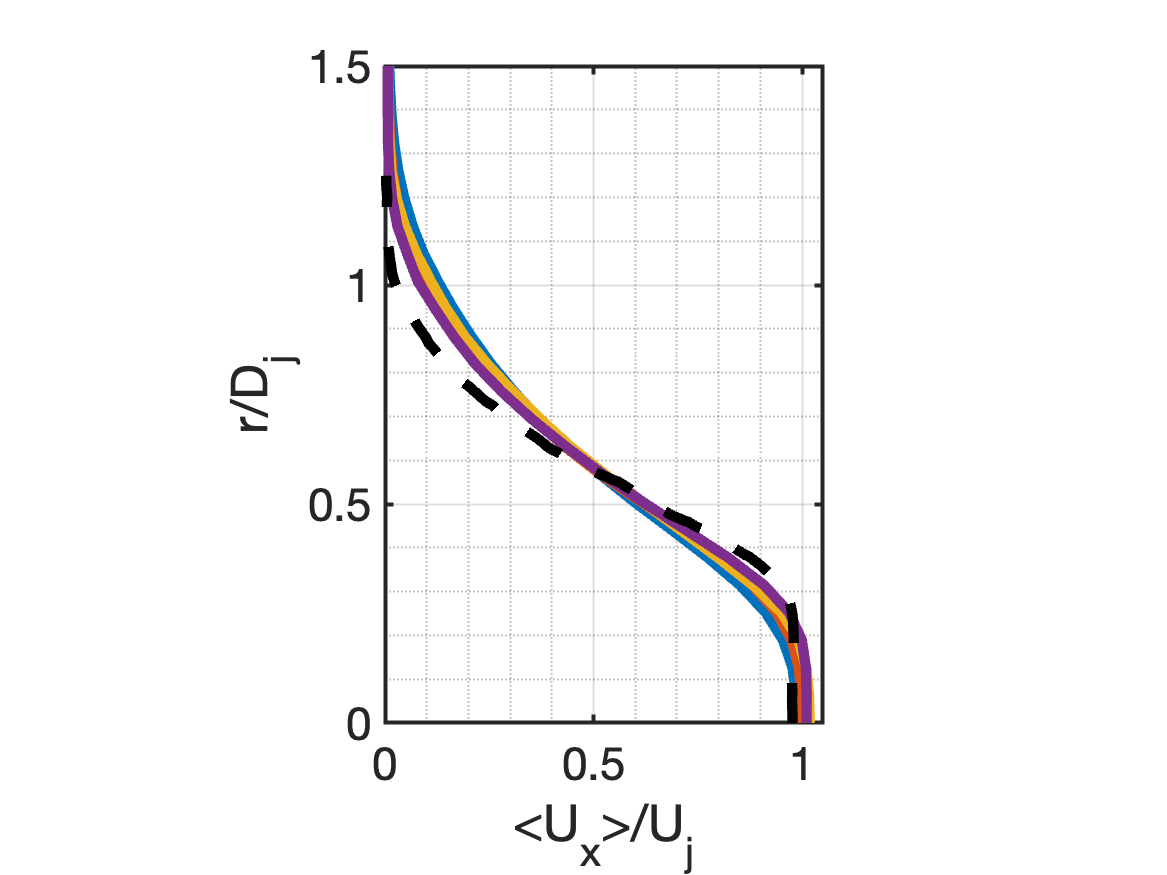}
	\label{res:radial_vxmean2}	
	}
%\subfloat[Mean longitudinal velocity component at $x/D_j=10$]{
\subfloat[]{
	\includegraphics[trim = 30mm 0mm 40mm 0mm, clip, height=0.26\linewidth]
    {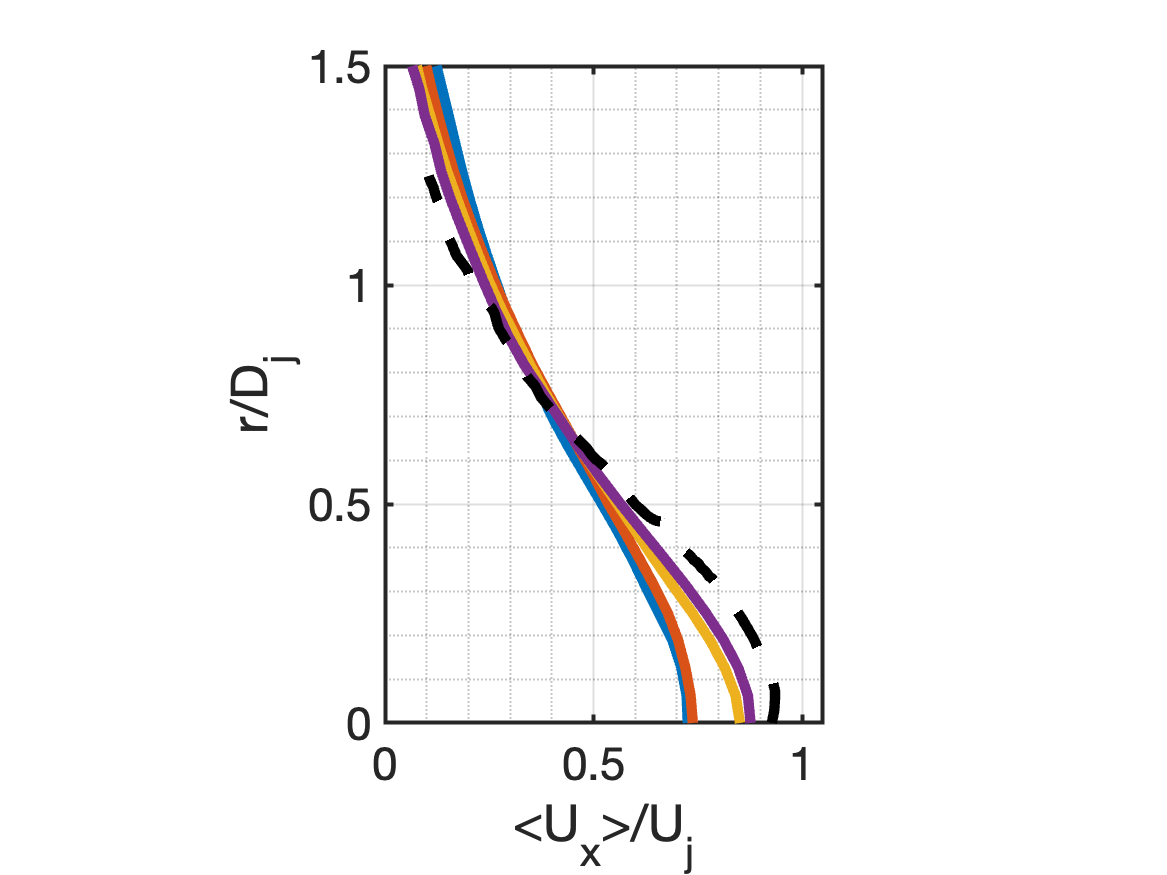}
	\label{res:radial_vxmean3}
	}
%\subfloat[Mean longitudinal velocity component at $x/D_j=15$]{
\subfloat[]{
	\includegraphics[trim = 30mm 0mm 40mm 0mm, clip, height=0.26\linewidth]
    {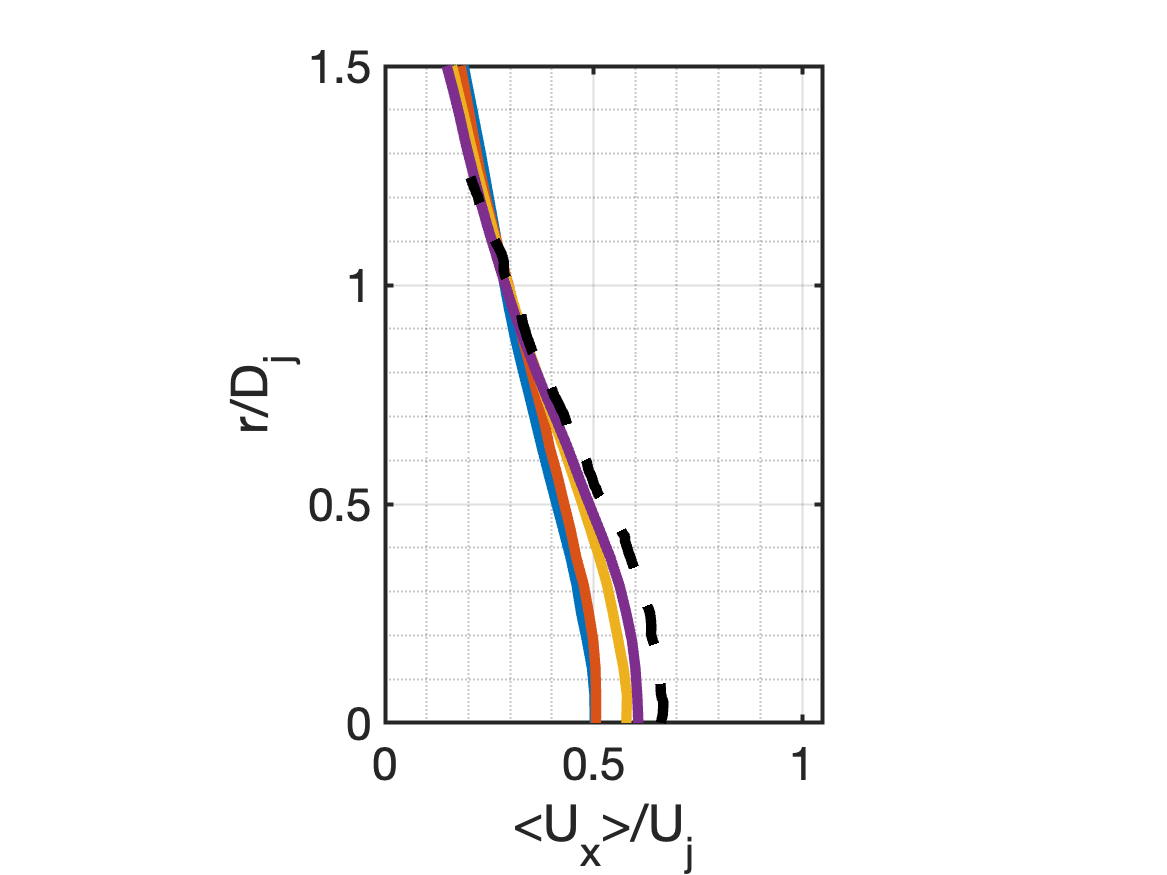}
	\label{res:radial_vxmean4}	
	}
\\
%\newline
%\subfloat[RMS of longitudinal velocity fluctuation at $x/D=2.5$]{
\subfloat[]{
	\includegraphics[trim = 20mm 0mm 15mm 0mm, clip, height=0.26\linewidth]
    {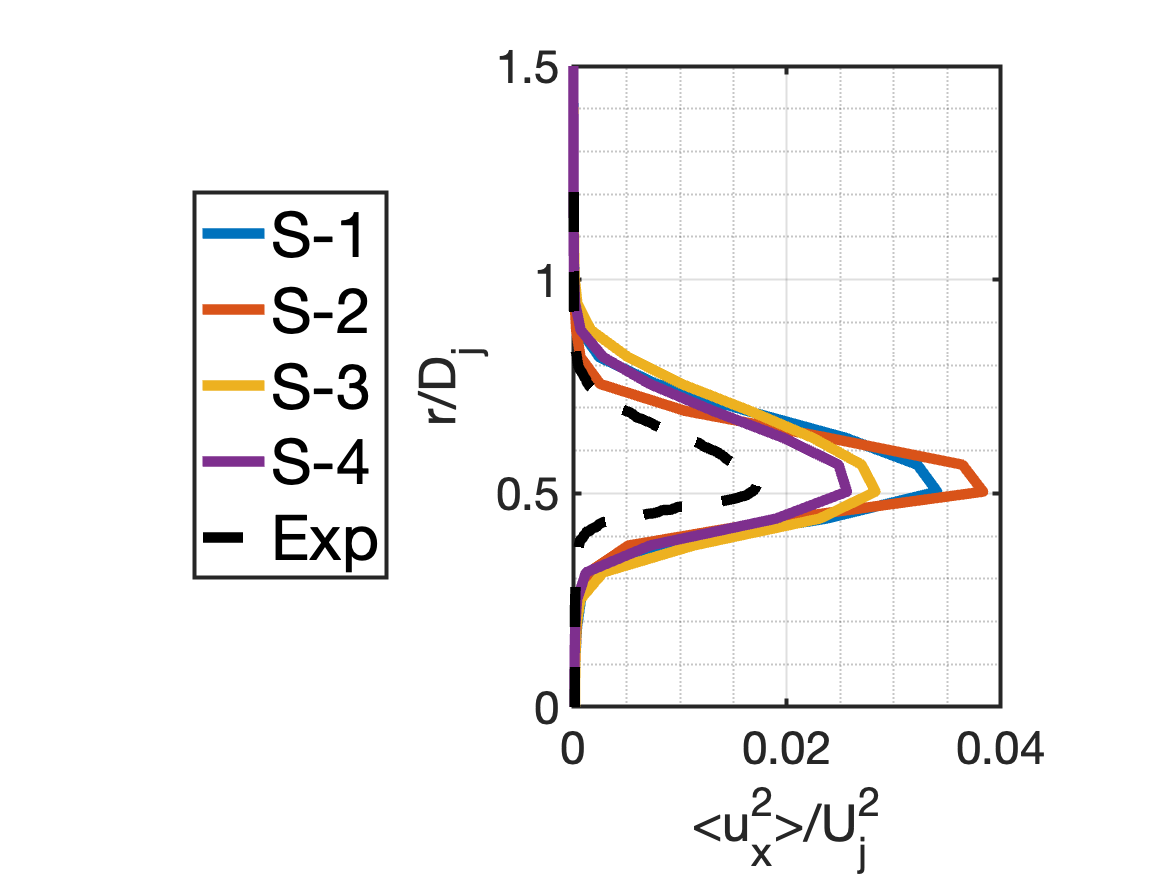}
	\label{res:radial_vxrms1}	
	}
%\subfloat[RMS of longitudinal velocity fluctuation at $x/D=5$]{
\subfloat[]{
	\includegraphics[trim = 30mm 0mm 40mm 0mm, clip, height=0.26\linewidth]
    {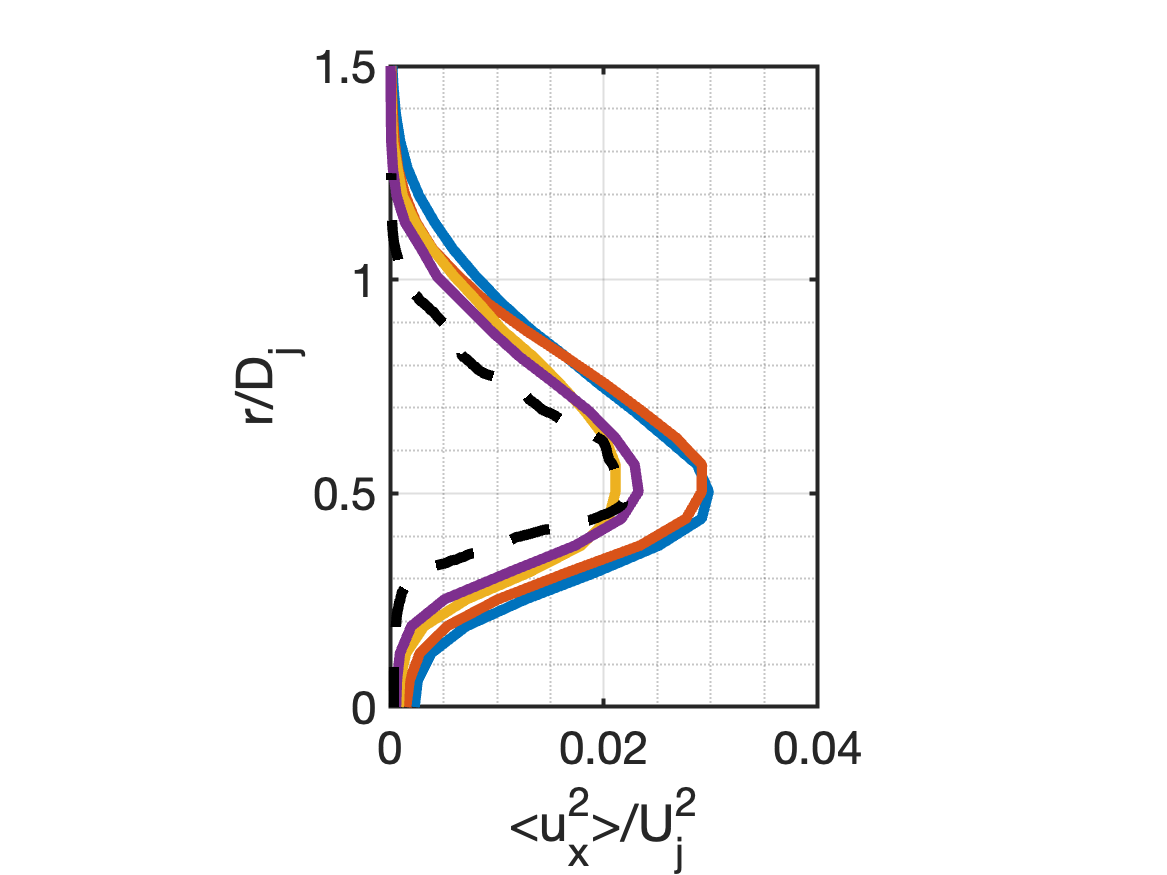}
	\label{res:radial_vxrms2}	
	}
%\subfloat[RMS of longitudinal velocity fluctuation at $x/D=10$]{
\subfloat[]{
	\includegraphics[trim = 30mm 0mm 40mm 0mm, clip, height=0.26\linewidth]
    {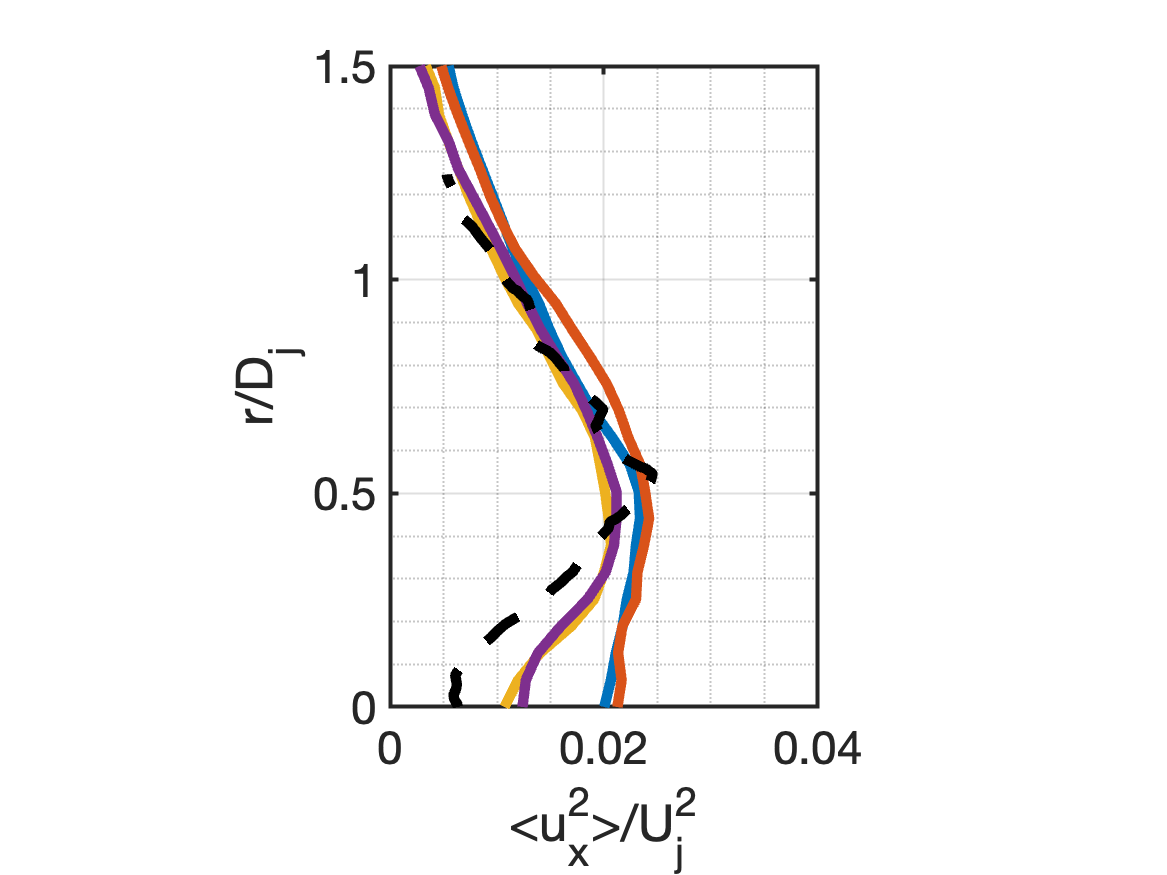}
	\label{res:radial_vxrms3}
	}
%\subfloat[RMS of longitudinal velocity fluctuation at $x/D=15$]{
\subfloat[]{
	\includegraphics[trim = 30mm 0mm 40mm 0mm, clip, height=0.26\linewidth]
    {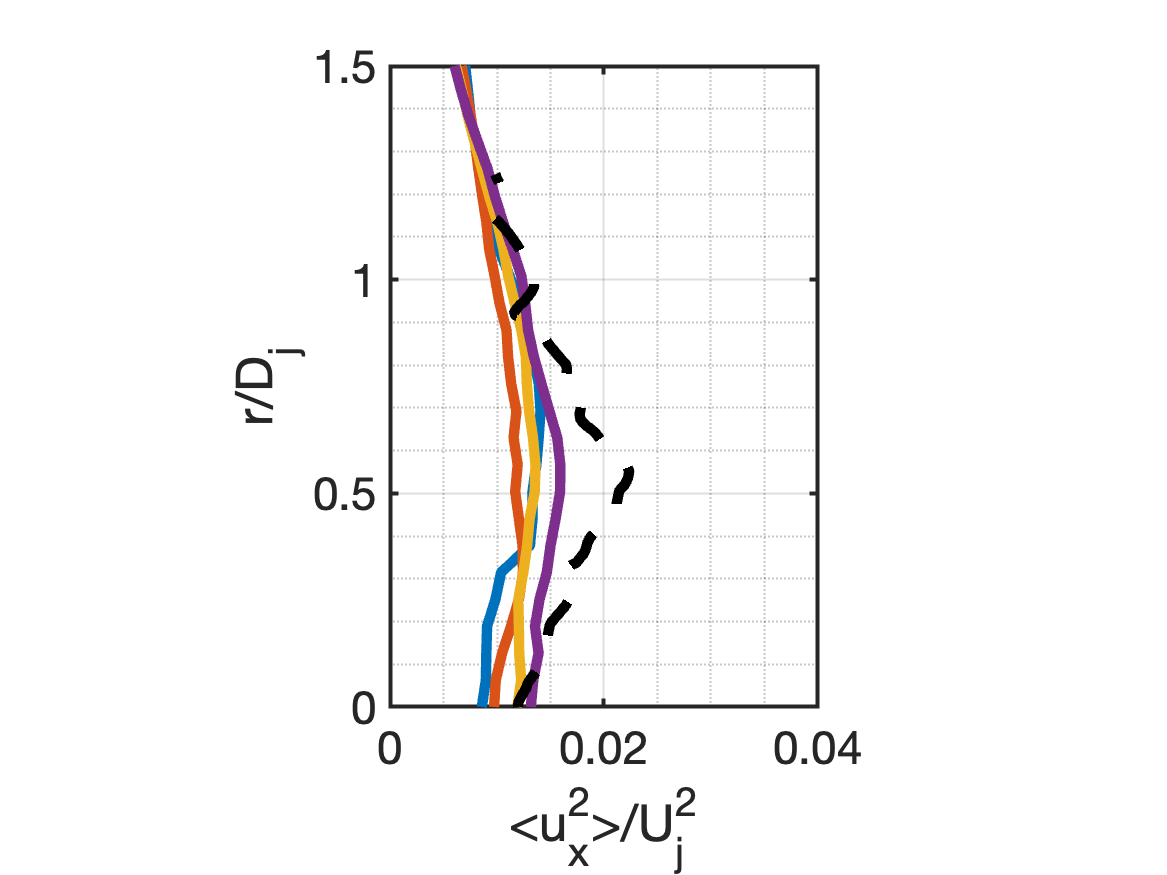}
	\label{res:radial_vxrms4}	
	}
\\
%\newline
%\subfloat[RMS of radial velocity fluctuation at $x/D=2.5$]{
\subfloat[]{
	\includegraphics[trim = 20mm 0mm 15mm 0mm, clip, height=0.26\linewidth]
    {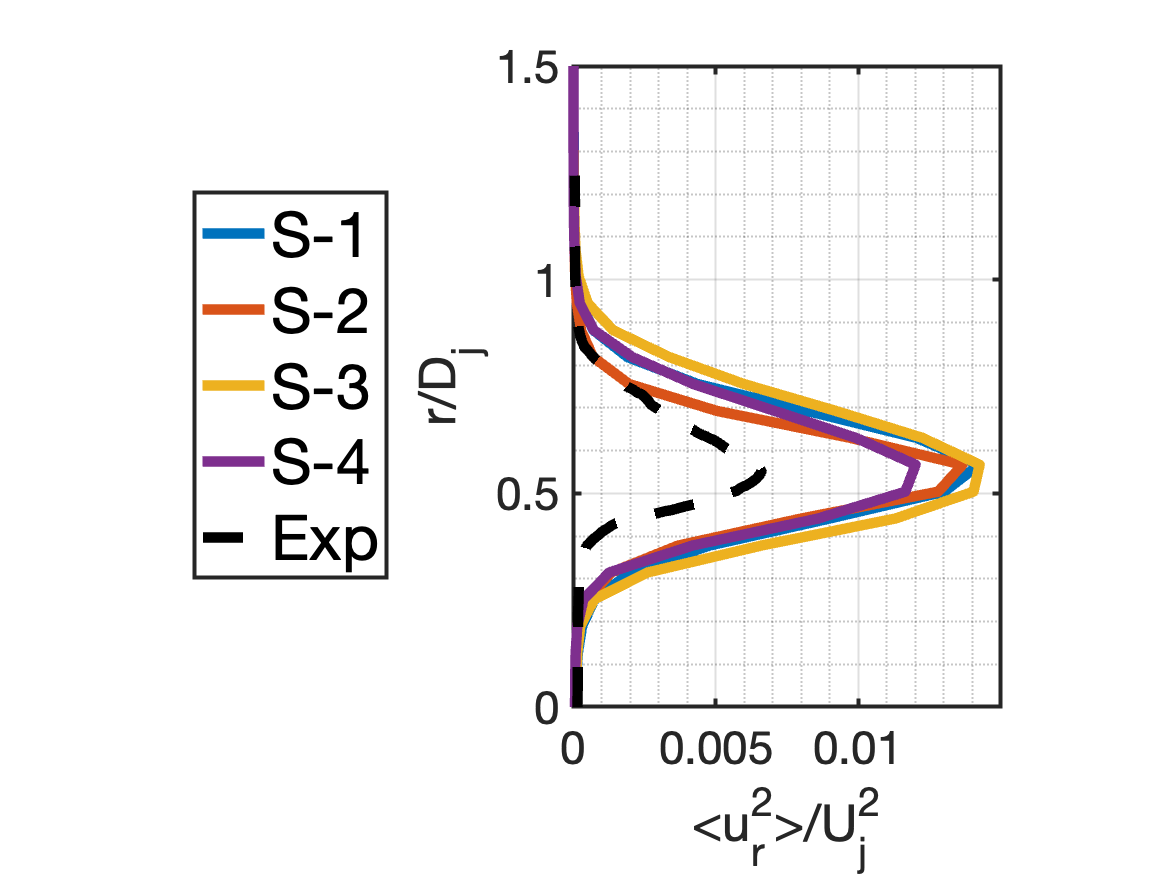}
	\label{res:radial_vyrms1}	
	}
%\subfloat[RMS of radial velocity fluctuation at $x/D=5$]{
\subfloat[]{
	\includegraphics[trim = 30mm 0mm 40mm 0mm, clip, height=0.26\linewidth]
    {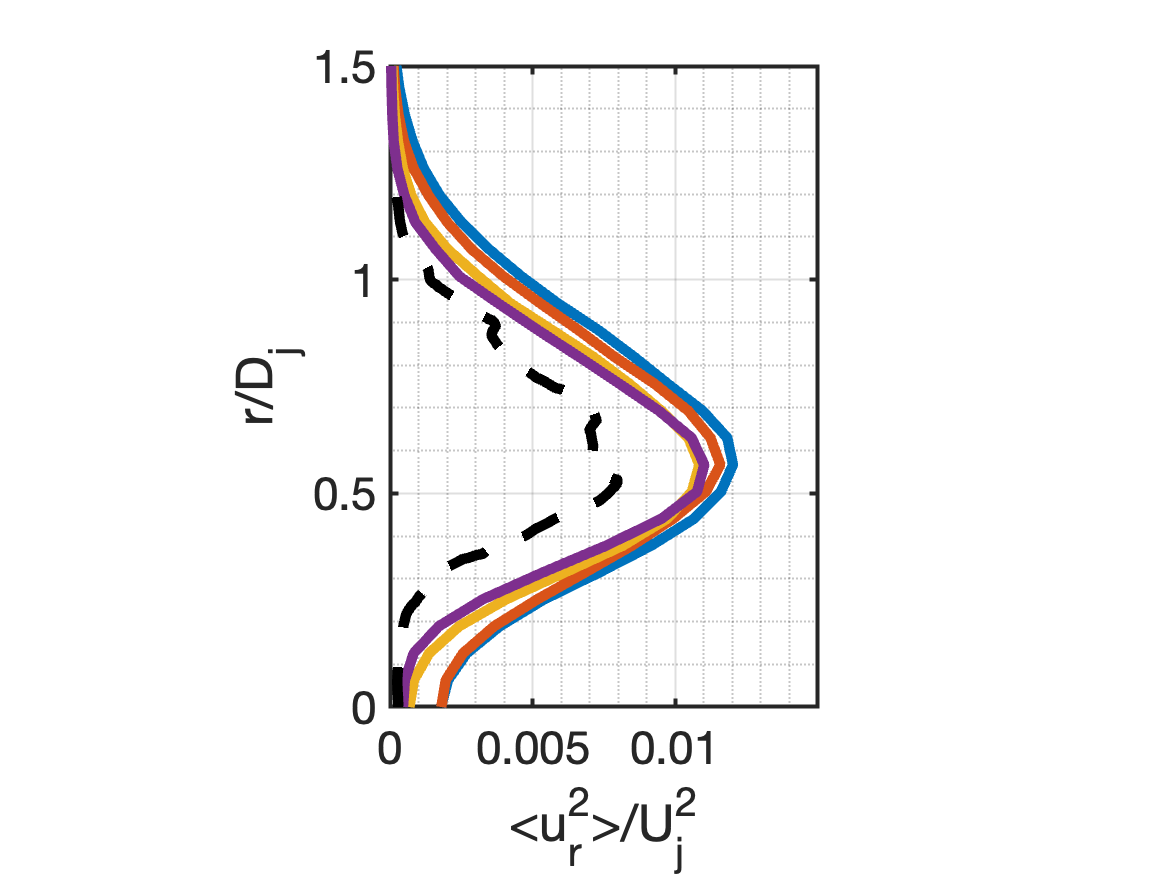}
	\label{res:radial_vyrms2}	
	}
%\subfloat[RMS of radial velocity fluctuation at $x/D=10$]{
\subfloat[]{
	\includegraphics[trim = 30mm 0mm 40mm 0mm, clip, height=0.26\linewidth]
    {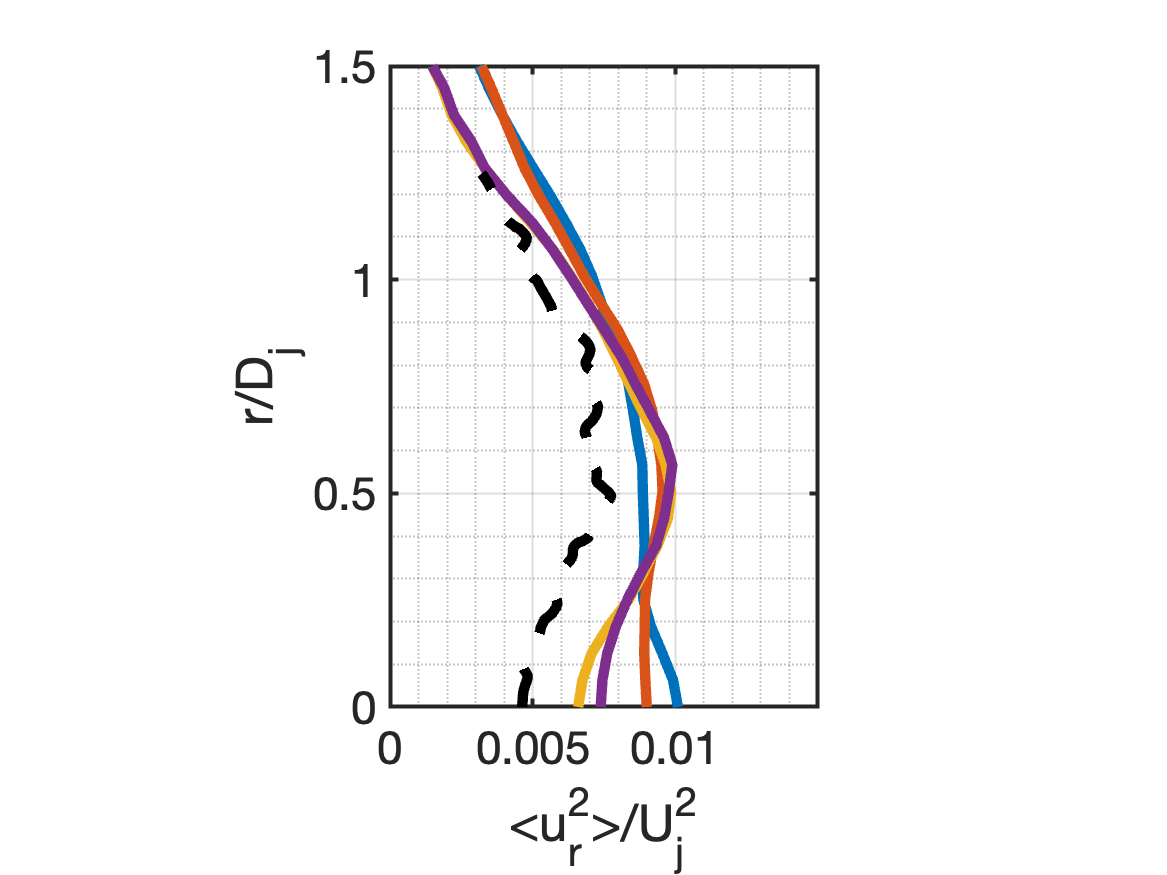}
	\label{res:radial_vyrms3}
	}
%\subfloat[RMS of radial velocity fluctuation at $x/D=15$]{
\subfloat[]{
	\includegraphics[trim = 30mm 0mm 40mm 0mm, clip, height=0.26\linewidth]
    {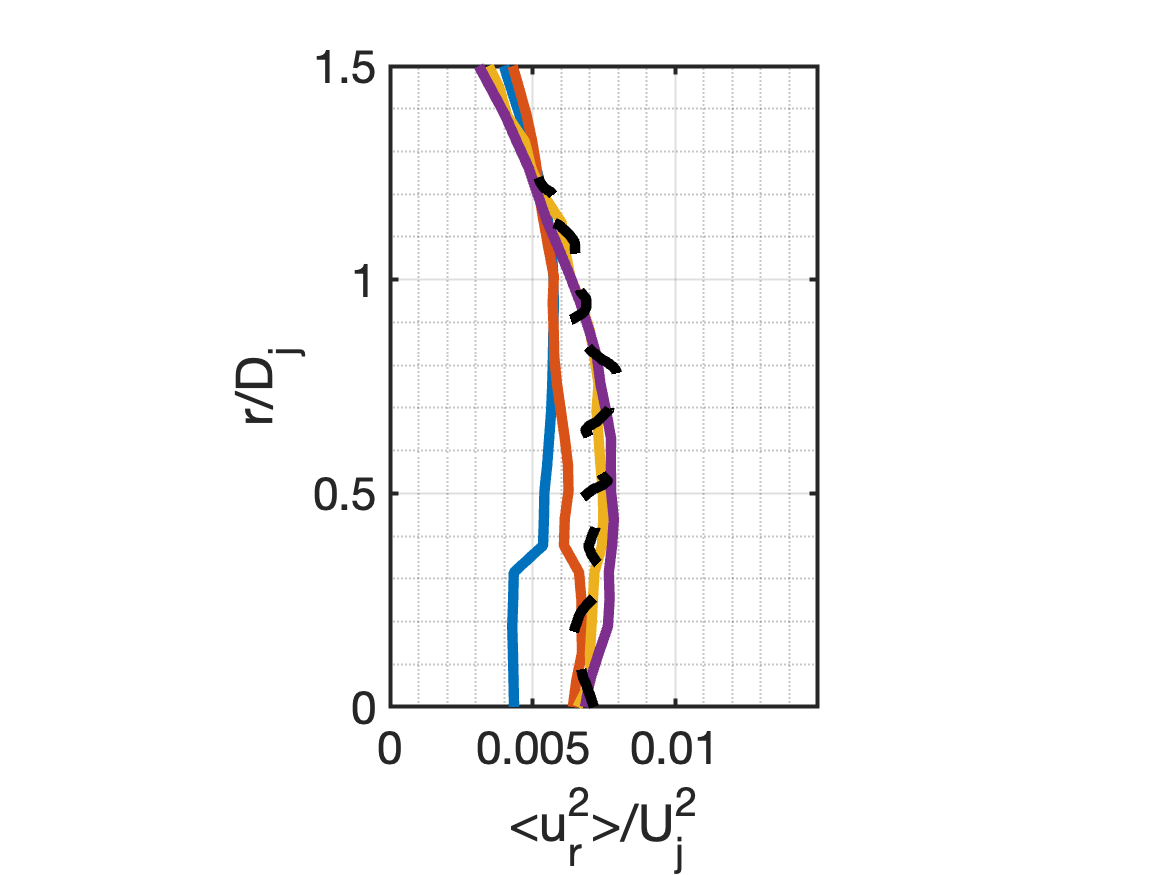}
	\label{res:radial_vyrms4}	
	}
\\
%\newline
%\subfloat[Mean shear stress tensor component at $x/D=2.5$]{
\subfloat[]{
	\includegraphics[trim = 20mm 0mm 15mm 0mm, clip, height=0.26\linewidth]
    {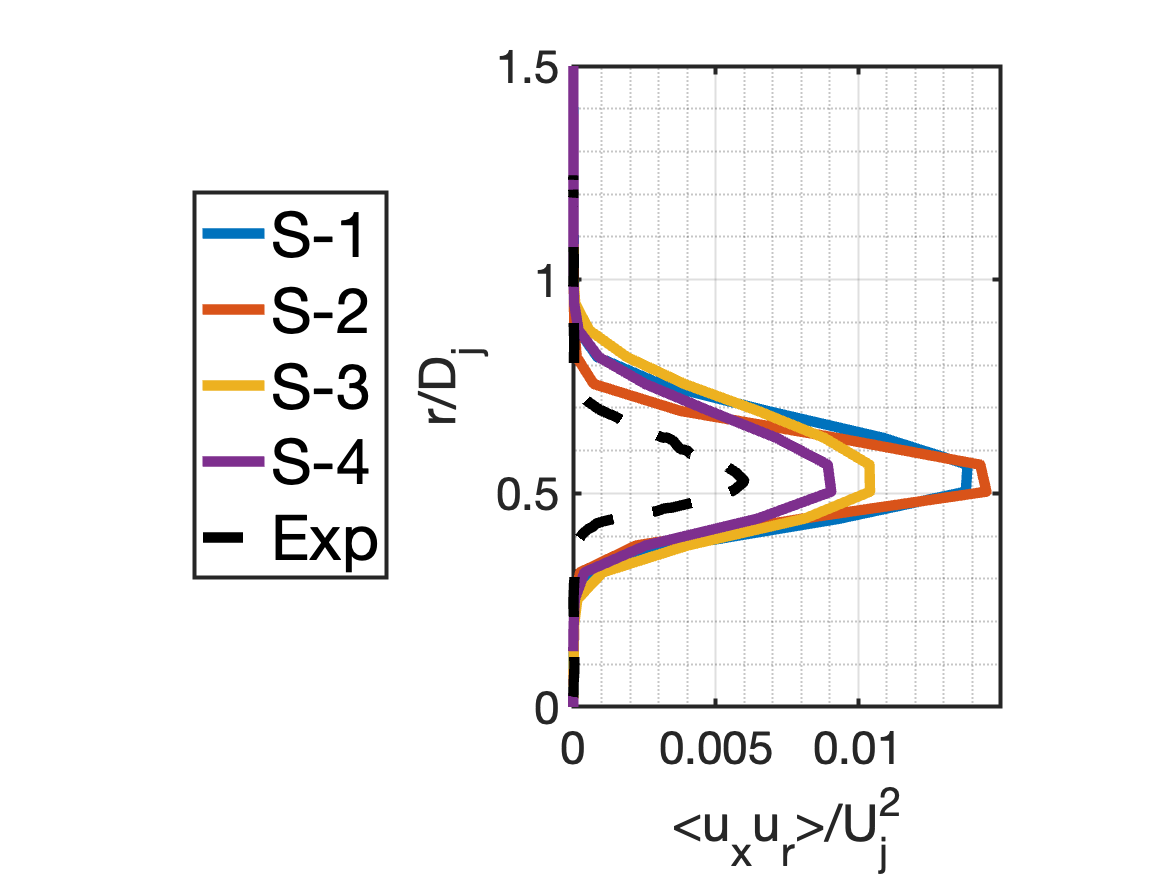}
	\label{res:radial_uvmean1}	
	}
%\subfloat[Mean shear stress tensor component at $x/D=5$]{
\subfloat[]{
	\includegraphics[trim = 30mm 0mm 40mm 0mm, clip, height=0.26\linewidth]
    {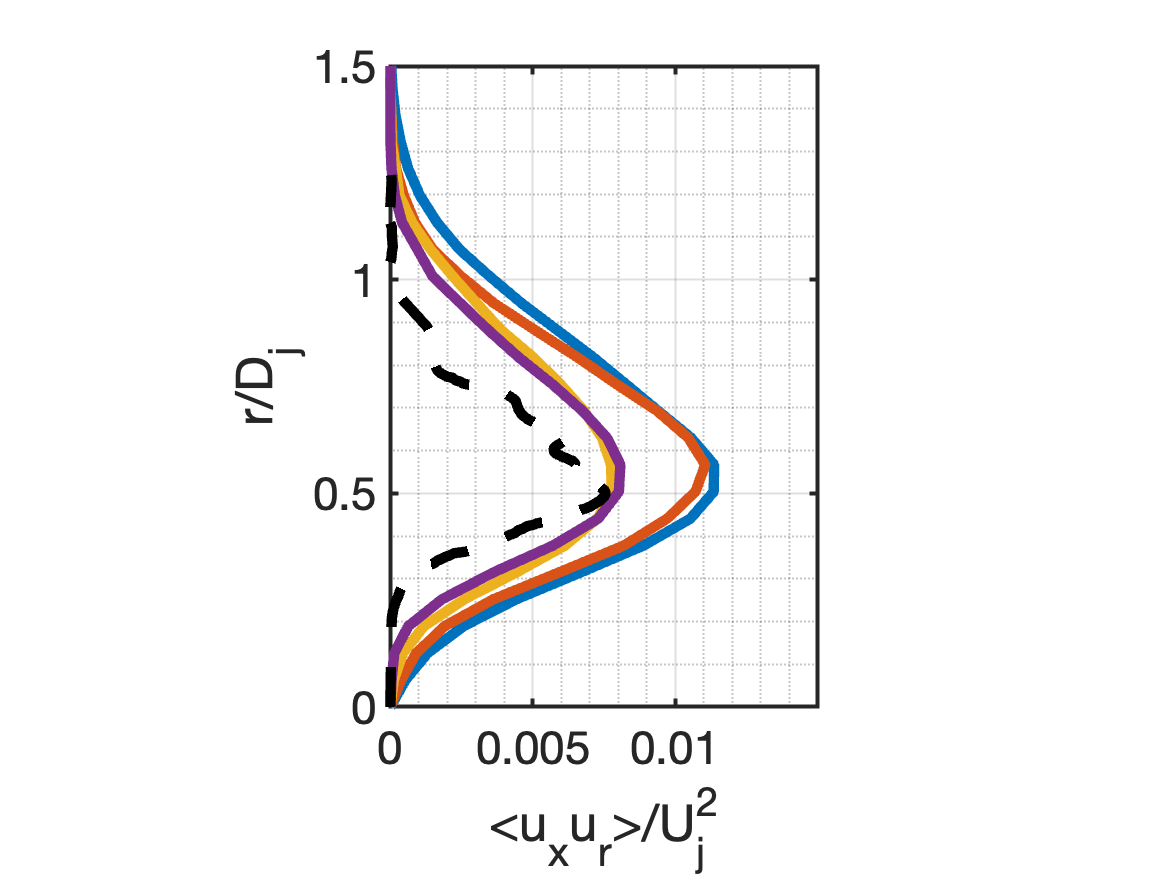}
	\label{res:radial_uvmean2}	
	}
%\subfloat[Mean shear stress tensor component at $x/D=10$]{
\subfloat[]{
	\includegraphics[trim = 30mm 0mm 40mm 0mm, clip, height=0.26\linewidth]
    {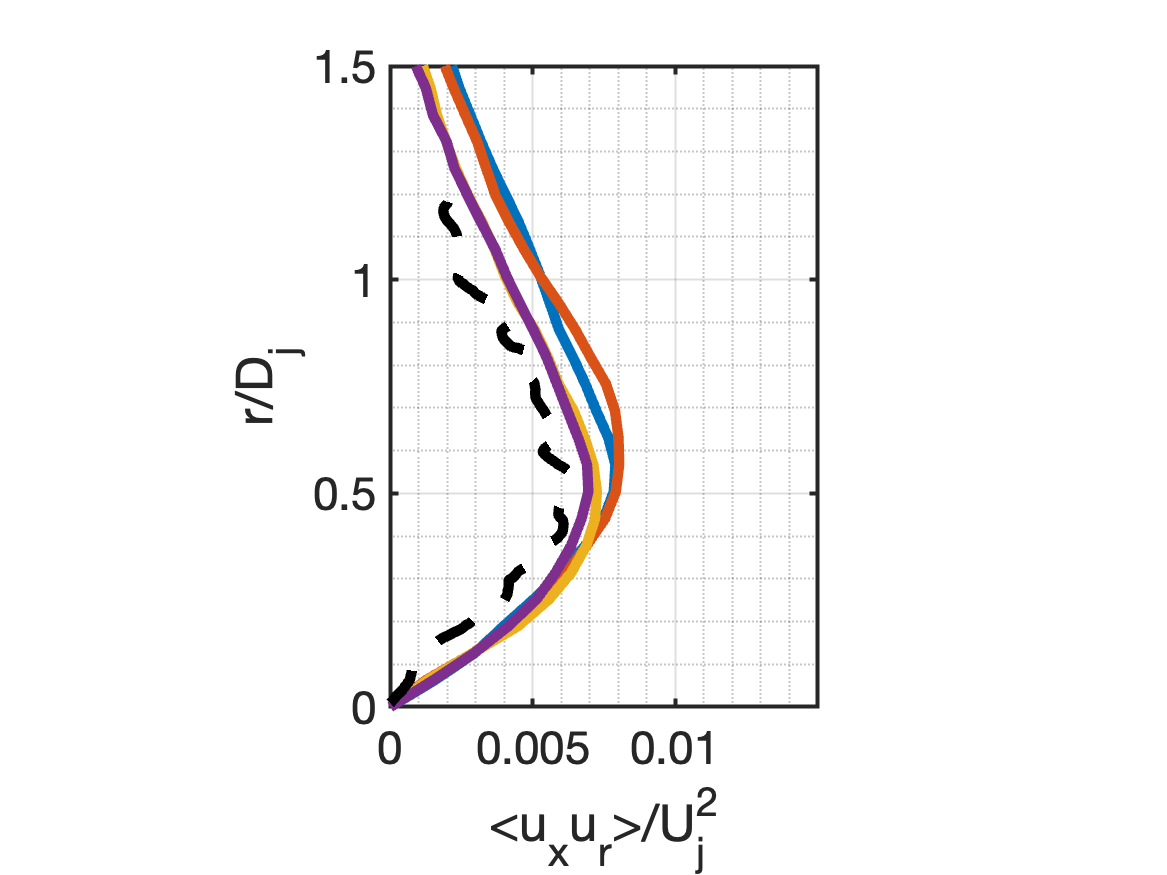}
	\label{res:radial_uvmean3}
	}
%\subfloat[Mean shear stress tensor component at $x/D=15$]{
\subfloat[]{
	\includegraphics[trim = 30mm 0mm 40mm 0mm, clip, height=0.26\linewidth]
    {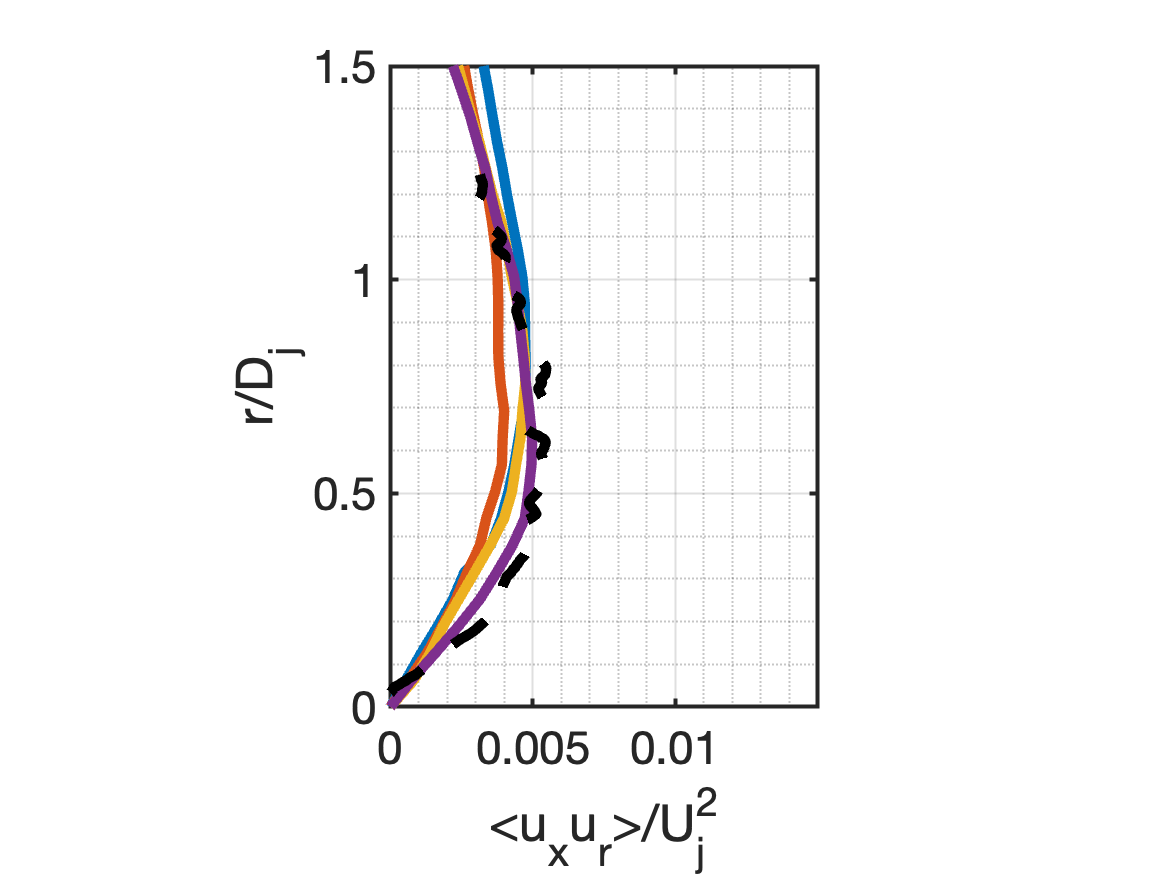}
	\label{res:radial_uvmean4}	
	}
\caption{Profiles of mean longitudinal velocity component, RMS of 
         longitudinal velocity fluctuation, RMS of radial velocity 
         fluctuation, and mean shear stress tensor component (from top to
         bottom) at four streamwise positions $x/D_j=2.5$, $x/D_j=5.0$,
         $x/D_j=10.0$, and $x/D_j=15.0$ (from left to right).}
\label{res:radial_profiles}
\end{figure}

Figure~\ref{res:radial_profiles} presents a set of radial profiles from the
four simulations compared with experimental data. The radial profiles were
obtained in four longitudinal positions: $x/D_j=2.5$, $x/D_j=5.0$, 
$x/D_j=10.0$, and $x/D_j=15.0$. The variables investigated are the mean 
longitudinal velocity component and three components from the Reynolds stress
tensor: the mean squared longitudinal velocity fluctuation component, the 
mean squared radial velocity fluctuation component, and the mean shear-stress
tensor component. The mean longitudinal velocity component distribution, 
Figs.~\ref{res:radial_vxmean1} to \ref{res:radial_vxmean4} present 
information on the development of the mean flow field of the jet flow. In 
Fig.~\ref{res:radial_vxmean1}, which is the closest station to the jet inlet 
section, the profiles from the four simulations present a similar shape when
compared with the experimental data, with a larger slope in the region 
$0.4< r/D_j < 0.8$, which is a first indicative of a larger velocity 
spreading of the jet. The mean longitudinal velocity component profiles from
the S-2 simulation could provide a better agreement with experimental 
reference at this position. The behavior agrees with the postponed
transition presented in Fig.~\ref{res:velx_rms_ll}. The analysis of the
profiles of the mean longitudinal velocity component at station $x/D_j=5.0$, 
Fig.~\ref{res:radial_vxmean2}, indicates a larger jet spreading from the
simulations compared to the experimental data, characterized by the larger
slope profile in $0.3< r/D_j <1.0$ from the numerical simulations compared
to the experimental data. All the simulations present profiles with a similar
shape between themselves. At station $x/D_j=10.0$, 
Fig.~\ref{res:radial_vxmean3}, the differences in the mean velocity profiles 
are apparent. The S-1 and S-2 simulations present a larger velocity reduction
in the centerline than the S-3 and S-4 simulations. The region of small 
velocity values extends up to $r/D_j=0.8$, where the S-1 and S-2 velocity 
distributions reach the values of the other simulations. The mean velocity
profiles from S-3 and S-4 simulations present a larger velocity reduction in
the centerline than the experimental data. In the last section analyzed
at $x/D_j=15.0$, Fig.~\ref{res:radial_vxmean4}, the velocity profiles present
a similar behavior to those observed in Fig.~\ref{res:radial_vxmean3}, where
the largest velocity reduction in the centerline is observed from the S-1 and 
S-2 simulations, then the S-3 and S-4 present similar velocity levels and,
above them, the experimental results. Extending above $r/D_j=1.0$, all the
simulations and the experimental data present similar velocity levels.

The radial progression of the longitudinal velocity fluctuation, 
Figs.~\ref{res:radial_vxrms1} to \ref{res:radial_vxrms4}, indicate a clear
division in the velocity distribution from the S-1 and S-2 simulations and 
the S-3 and S-4 simulations. In the first section, 
Fig.~\ref{res:radial_vxrms1}, at $x/D_j=2.5$ close to the jet lipline, the 
S-3 and S-4 simulations' velocity fluctuation has a small peak compared to
the S-1 and S-2 simulations. The four simulations generally present a larger
fluctuation value than the experimental data. The velocity fluctuation in 
the second station, Fig.~\ref{res:radial_vxrms2}, shows that the peak values 
from the S-3 and S-4 simulations are similar to the peak values of the
experimental data. The S-3 and S-4 simulations have a wider high-velocity
fluctuation region than the experimental reference. The velocity fluctuation
distribution from the S-1 and S-2 simulations present larger peak values and
spreading than the other two numerical simulations and the experimental data.
In the section $x/D_j=10.0$, Fig.~\ref{res:radial_vxrms3}, the velocity peak
of velocity fluctuation and the velocity distribution above the mixing layer
have similar levels between all the numerical simulations and the experimental 
data. Close to the centerline, the values of velocity fluctuation differ 
between the S-1 and S-2 simulations with the S-3 and S-4 simulations and the
experimental data. In the last section, at $x/D_j=15.0$, 
Fig.~\ref{res:radial_vxrms4}, the levels of velocity fluctuation between all
the numerical simulations are very similar, with small differences to
the experimental data only in the jet lipline.

The radial velocity component of the Reynolds stress tensor distributions 
are presented in Figs.~\ref{res:radial_vyrms1}, \ref{res:radial_vyrms2}, 
\ref{res:radial_vyrms2}, and \ref{res:radial_vyrms4}. The profiles are 
similar to those presented by the longitudinal velocity component. In 
Fig.~\ref{res:radial_vyrms1}, the peak values are obtained close to the jet
lipline, $r/D_j=0.5$, with nearly twice the values reported by the 
experimental data. The peak values of all simulations are close in this 
station. In the second station analyzed, Fig.~\ref{res:radial_vyrms2}, the
simulations' results continue to be close between themselves, and they got
closer to the experimental reference. The velocity distribution in the
section $x/D_j=10.0$, Fig~\ref{res:radial_vyrms3}, presents some differences 
in the profiles of the simulations. The S-1 and S-2 simulations present larger
values in the centerline than the S-3 and S-4 simulations. The S-3 and S-4
simulations present larger radial velocity fluctuations larger than the
experimental reference. The profiles of radial velocity fluctuation of the
simulations match close to $r/D_j=0.2$ with larger values compared to
the experimental reference up to $r/D_j=1.0$. In the last section, 
Fig.~\ref{res:radial_vyrms4}, the radial velocity distribution of all the
simulations agrees with the experimental reference.

The last component of the Reynolds stress tensor evaluated in this radial 
basis is the shear-stress component, Figs.~\ref{res:radial_uvmean1}, 
\ref{res:radial_uvmean2}, \ref{res:radial_uvmean3}, and 
\ref{res:radial_uvmean4}. The radial distributions from section $x/D_j=2.5$,
Fig.~\ref{res:radial_uvmean1}, distinct the simulation results into two groups.
The peak values from the S-1 and S-2 simulations are nearly three times larger
than those of the experimental reference, while the peak values from the 
S-3 and S-4 simulations are nearly twice those from the experimental reference.
The shear-stress tensor profile from the numerical simulations indicates a
larger spreading when compared to the experimental reference, which presents
small values close to the lipline. The radial distribution in the section
$x/D_j=5.0$, Fig.~\ref{res:radial_uvmean2} show a better agreement with
experimental data from the peak values of the S-3 and S-4 simulations with 
a reduction in the differences between the S-1 and S-2 simulations. Even with
an agreement in the peak values, a larger spreading is observed directed to 
the jet centerline and the external flow. The shear-stress distributions from
the sections $x/D_j=10.0$ and $x/D_j=15.0$, Figs.~\ref{res:radial_uvmean3} 
and \ref{res:radial_uvmean4}, indicate an agreement of the numerical results
with experimental data concerning the peak values and the profiles.

The general behavior of the Reynolds stress components indicates more 
considerable differences in the initial development of the jet flows,
$x/D_j<5.0$, with larger peak values and increased mixing layer. In the 
section $x/D_j=10.0$, there are similarities in the shape of the radial
profiles far from the jet centerline, where the numerical simulations present
more significant velocity fluctuation levels than the experimental data. In 
the lastsection analyzed, $x/D_j=15.0$, all the components of the Reynolds
stress tensor agree with experimental data. The behavior of the mean
longitudinal Reynolds stress tensor component profiles is similar to the
experimental data in the first section, $x/D_j=2.5$. In the second section, 
at $x/D_j=5.0$, the derivative of the mean longitudinal velocity with the 
radial position is larger than the experimental data, which is associated 
with a larger jet spreading, resulting in a smaller core of the jet and larger
velocity values far from the jet lipline. In stations, $x/D_j=10.0$ and
$x/D_j=15.0$ similar mean velocity values are observed between the numerical
simulations and the experimental data for radial positions larger than
$r/D_j=5$. In contrast, in the centerline of the jet, smaller mean velocity
values are obtained.

\FloatBarrier

\subsection{Power Spectral Density}

\begin{figure*}[htb!]
\centering
%\subfloat[Mean longitudinal velocity component at centerline]{
\subfloat[PSD from the longitudinal velocity fluctuation.]{
	\includegraphics[trim = 0mm 0mm 10mm 5mm, clip, width=0.48\linewidth]
    {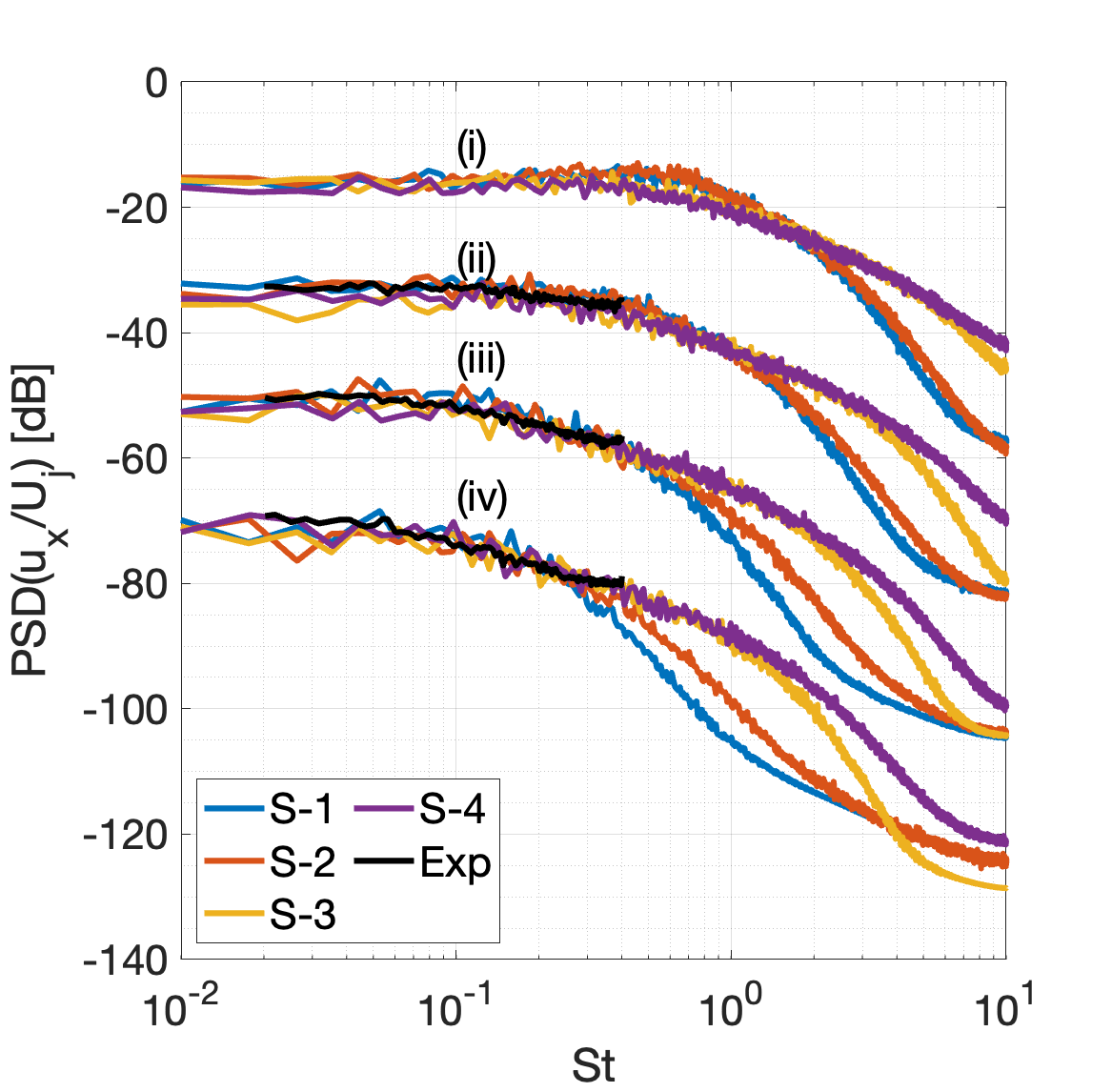}
	\label{res:psd_u}	
	}%
%\subfloat[RMS of longitudinal velocity fluctuation at centerline]{
\subfloat[PSD from the radial velocity fluctuation.]{
	\includegraphics[trim = 0mm 0mm 10mm 5mm, clip, width=0.48\linewidth]
    {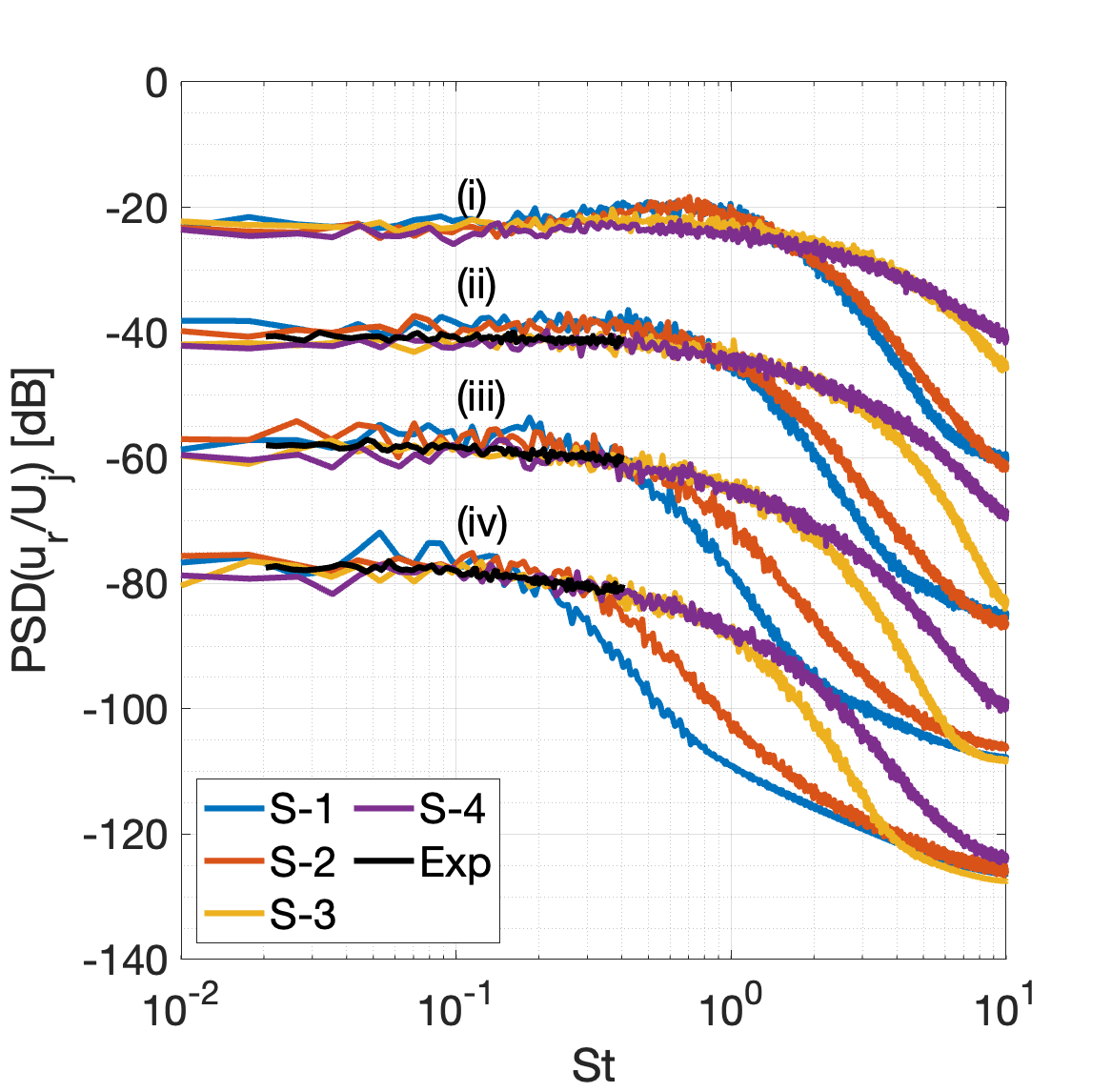}
	\label{res:psd_v}	
	}
\caption{Power spectral density (PSD) of the velocity fluctuation at the 
         lipline in four longitudinal positions: $x/D_j=2.5$ (i), 
         $x/D_j=5.0$ (ii), $x/D_j=10.0$ (iii) and $x/D_j=15.0$ (iv). 
         A cumulative value of $20dB$ is subtracted from the results of each 
         station, starting at $x/D_j=5.0$ to allow the distinction between 
         the velocity spectrum results in the same chart.}
\label{res:psd}
\end{figure*}

The velocity spectra analysis is performed through the power spectral density
(PSD), Fig.~\ref{res:psd}, of the longitudinal velocity component fluctuation,
Fig.~\ref{res:psd_u}, and the radial velocity component fluctuation, 
Fig.~\ref{res:psd_v}. The velocity spectra results are obtained in the same
four streamwise positions of the radial analysis, $x/D_j=2.5$, $x/D_j=5.0$,
$x/D_j=10.0$, and $x/D_j=15.0$, and they are all performed in the jet lipline
$r/D_j=0.5$. The experimental reference is available for the last three 
sections. The charts show a cumulative shift of $20dB$ starting in the second
position, $x/D_j=5.0$, to distinguish between the data in each section. Due to 
the limited acquisition frequency of the experimental apparatus, the
experimental reference is only available to a maximum Strouhal number of $0.4$.
It is possible to observe that for the positions $x/D_j=5.0$ and $x/D_j=10.0$,
the PSD from all simulations is similar to the experimental data in the range
of the experimental reference. In the last section, at $x/D_j=15.0$, for both
velocity components, the spectra of the S-1 and S-2 simulations detach from 
the experimental data inside the range of experimental reference. A clear 
trend is present when comparing the data of the four numerical simulations in
the available frequency range. The velocity spectra from the S-1 simulation
present an abrupt power reduction in frequencies inside the experimental range.
The S-2 simulation presents the same abrupt power reduction of the velocity
spectra in frequencies slightly higher than those of the S-1 simulation. The
power reduction of the velocity spectra from the S-3 and S-4 simulations occurs
at higher frequencies than the S-1 and S-2 simulations. For the first two
positions analyzed, in the range of the data, it is difficult to define
the initial position of the PSD reduction from the S-4 simulation, which
presents the best velocity spectra with larger power spectral values than
the other simulations at large frequencies. In the first position, at 
$x/D_j=2.5$, the S-1 and S-2 simulations present a peak value of the velocity
spectra close to a Strouhal number of $0.5$. This peak value can be associated
with the late transition in the two simulations.

\FloatBarrier

%-------------------------------------------------------------------------------
\section{Concluding Remarks}
%-------------------------------------------------------------------------------
\label{sec.conclusions}

In the present work, the impacts of {\it hp} resolution on large-eddy 
simulations of supersonic jet flows are investigated within the framework of 
the discontinuous Galerkin spectral element method, implemented in the 
open-source FLEXI solver. The simulations employed three computational meshes
with two polynomial options to represent the numerical solution, leading to 
schemes with second- and third-order accurate spatial discretizations. The 
ensemble of numerical meshes and polynomials are employed in four simulations,
with degrees of freedom from $50 \times 10^6$ to $410 \times 10^6$. The 
calculations are compared to experimental data. 

The qualitative investigation of the instantaneous velocity and pressure 
contours showed that the simulations with high resolution could reproduce
small flow features in the mixing layer and the region of the jet potential 
core. The highest-resolution simulation produced better-defined shock waves 
than the simulations with less resolution. Employing a third-order accurate
spatial discretization scheme led to improvements when employed in different
meshes with the same number of degrees of freedom and in the same numerical
mesh. The comparison of mean velocity profiles showed that the highly-resolved
simulations produced longer jet potential cores when compared to simulations
with smaller resolutions. The root mean square value of velocity fluctuation
indicates anticipation of the development of the mixing layer, which also
presents a small spreading rate. The mean pressure contours showed an improved
capacity to reproduce the shock structure in the jet interior with increased
resolution.

The numerical data are compared with experimental data in terms of the velocity 
profiles at the jet centerline, jet lipline, and radial profiles in four 
streamwise positions. The velocity profiles at the centerline indicate a 
monotone improvement toward the experimental reference with the increasing
resolution of the simulations. The monotone trend is observed for both mean
velocity and root mean square values of velocity fluctuation distributions.
In the jet lipline, the increase in the resolution of the simulations produced 
mean velocity values with a better agreement with experimental data. However, 
the root mean square values of the velocity fluctuation distributions distance
themselves from the experimental reference. This behavior is attributed to the
simplicity of the inlet profile imposed in the jet inlet condition, {\em i.e.},
that the flow entering the domain is inviscid. The larger resolution of the 
simulations produces an anticipation of the development of the mixing layer,
which is observed by a sudden change in the velocity fluctuation distribution 
slope. The highly-resolved simulation presentes the slope change occurring
closer to the jet inlet section than the simulations with lower resolution. 
This behavior is not observed in the experimental reference.

Power spectral densities are calculated  for the longitudinal and radial 
velocity signals in four streamwise positions in the jet lipline. The last 
three positions are compared with experimental data. The data acquisition
frequency of the experiment restricted the comparison to low-frequencies, in 
which there is a good match of all numerical results with the experimental 
reference. When the comparison is extended to the frequencies available only 
from the numerical simulations, a monotone improvement is observed with the 
increase in the resolution of the simulations, represented by the larger power
in the high frequencies. 

In general, the most refined simulation, which employed the mesh with improved
topology and a larger number of elements, $15.4 \times 10^6$ elements, with a
third-order accurate spatial discretization generated the best results when
compared to the experimental reference. Such simulation has reproduced, with 
the highest quality, the different structures present in the flow, visualized 
by a diversity of velocity and pressure contours. The differences between the
numerical simulation and experimental references are mostly correlated to the
simple inviscid profile employed in the jet inlet condition that could not
represent the aspects present in the experimental tests. The results obtained 
from the work show that the discontinuous Galerkin spectral element method is
an interesting tool to the simulation of supersonic free jet flows. The work
suggests a mesh topology, with a description of the element size distribution,
and indicates an adequate polynomial resolution that together are capable of
solving the large-eddy simulations of free jet flows with good accuracy. 
{The present research could also profit from data of direct
numerical simulation of the same supersonic jet flow configuration. Such 
effort could generate complete turbulent kinetic energy spectra which, in turn,
would provide reference results that would allow a more precise indication of
the portion of the spectrum which is actually being resolved by the present
large eddy simulations. Clearly, however, such an effort is a quite demanding
endeavor and requires adequate computational resources}.

\section*{Acknowledgments}

The authors acknowledge the support for the present research provided by 
Conselho Nacional de Desenvolvimento Cient\'{\i}fico e Tecnol\'{o}gico, CNPq,
under the Research Grant No.\ 309985/2013-7\@. The work is also supported by 
the computational resources from the Center for Mathematical Sciences Applied 
to Industry, CeMEAI, funded by Funda\c{c}\~{a}o de Amparo \`{a} Pesquisa do 
Estado de S\~{a}o Paulo, FAPESP, under the Research Grant No.\ 2013/07375-0\@.
The authors further acknowledge the National Laboratory for Scientific Computing
(LNCC/MCTI, Brazil) for providing HPC resources of the SDumont supercomputer.
This work was also granted access to the HPC resources of IDRIS under the
allocation A0152A12067 made by GENCI. The first author acknowledges 
authorization by his employer, Embraer S.A., which has  allowed his 
participation in the present research effort. Additional support to the second
author under the FAPESP Research Grant No.\ 2013/07375-0 is also gratefully
acknowledged. The authors acknowledge Dr. Eron T. V. Dauricio, for the support
on the development of the simulations within the FLEXI framework.

\bibliography{jabcm}% common bib file

\end{document}